%% file: ConventionalFacitlitiesYemilab.tex
\documentclass[12pt]{report}
\usepackage{graphicx}
\usepackage{amssymb}
\usepackage{epstopdf}
\usepackage{xspace}
\usepackage{xcolor}
\usepackage{multicol}
\usepackage{floatrow}
\usepackage{float}
\usepackage{dcolumn}
\usepackage{bm}
\usepackage[dvips]{epsfig} 
\usepackage[font=footnotesize,labelfont=bf]{caption}

\usepackage{xspace}
\newcommand{\degree}{\ensuremath{^\circ}\xspace}

\usepackage{etoolbox}
\makeatletter
\patchcmd{\@makechapterhead}{50\p@}{-10pt}{}{}
\patchcmd{\@makeschapterhead}{50\p@}{-10pt}{}{}
\pretocmd{\subsection}{\addtocontents{toc}{\protect\addvspace{-2\p@}}}{}{}
\pretocmd{\subsubsection}{\addtocontents{toc}{\protect\addvspace{-2\p@}}}{}{}
\makeatother

\begin{document}

~~~~

\thispagestyle{empty}

~~

\vspace{-0.5in}

\begin{center}
{\Large \bf IsoDAR@Yemilab:\\~\\  A Conceptual Design Report for the\\
  Deployment of the Isotope Decay-At-Rest Experiment in Korea's New Underground Laboratory, Yemilab \\~~\\ }
\end{center}


\begin{center}
\noindent {\it Editors: } J.R.~Alonso\footnote{Corresponding Author:
  Jose R. Alonso (JRAlonso@MIT.edu)}, J.M.~Conrad, D.~Winklehner, \\S.H.~Seo, K.M.~Bang, Y.D.~Kim, and K.S.~Park\\for the
IsoDAR Collaboration
\end{center}

\begin{figure}[h]       
\begin{center}
{\includegraphics[width=4.5in]{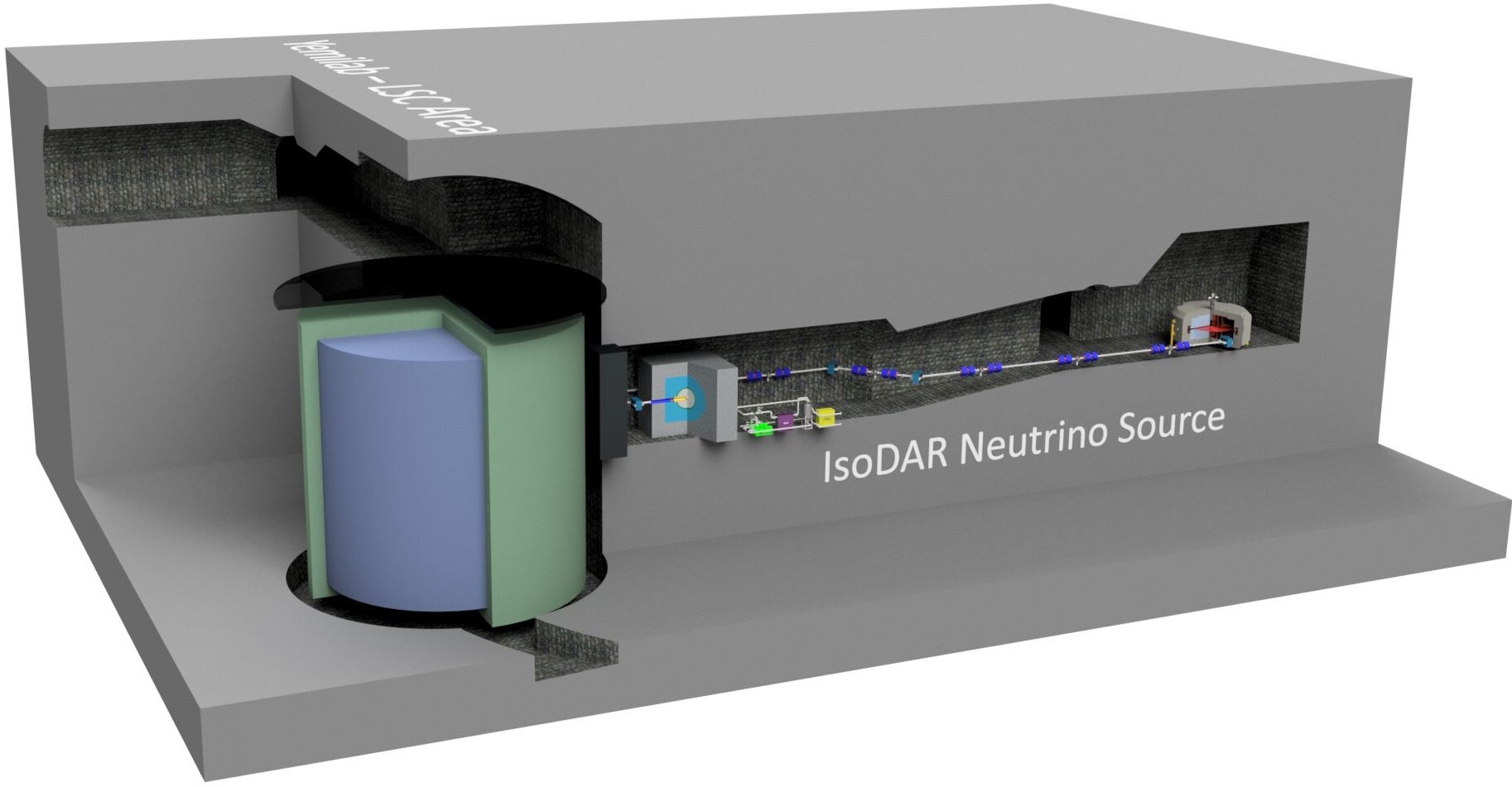}}
\end{center}  
\end{figure}


{ \it Abstract:}  

This Conceptual Design Report addresses the site-specific
issues associated with the deployment of the IsoDAR experiment
at the Yemilab site.   IsoDAR$@$Yemilab pairs the IsoDAR 
cyclotron-driven $\bar \nu_e$ source with the proposed Liquid Scintillator Counter (LSC) 
2.5 kton detector.
This document describes the proposed siting:   requirements
for the caverns to house the cyclotron, beam transport line,
and target systems; issues associated with transport and assembly
of components on the site; electrical power, cooling, and ventilation; 
as well as issues associated with radiation protection of the
environment and staff of Yemilab who will be interfacing
with IsoDAR during its operational phases.
The onset of construction of the IsoDAR area at Yemilab, in tandem with the release of this 
design report, represents a key step forward in establishing IsoDAR$@$Yemilab.


\newpage

~~~~~

\vspace{1.in}

\begin{center}
 {\it \Large Contributing Authors: }\\
~~~\\
~~~\\

J.R.~Alonso$^5$,
K.M.~Bang$^{6}$,
R.~Barlow$^{3}$,
L.~Bartoszek$^{1}$, 
A.~Bungau$^{3}$,
L.~Calabretta$^{4}$,  
J.M.~Conrad$^{5}$,
S.~Kayser$^{5}$, 
Y.D.~Kim$^{6}$,
K.S.~Park$^6$,
S.H.~Seo$^{6}$,
M.H.~Shaevitz$^2$, 
J.~Spitz$^{7}$
L.H.~Waites$^{5}$,
D.~Winklehner$^{5}$. 

~~~~~\\

{\it 
$^1$Bartoszek Engineering, Aurora IL, US\\
$^2$Columbia University, New York NY, US\\
$^{3}$University of Huddersfield, Huddersfield, UK\\
$^{4}$INFN  Sezione di Catania, IT \\ 
$^{5}$Massachusetts Institute of Technology, Cambridge MA, US\\
$^{6}$ Institute for Basic Science -- Center for Underground Physics, Daejeon, KR\\
$^{7}$ University of Michigan, Ann Arbor MI, US\\
}
\end{center}

\newpage 

\tableofcontents

\clearpage
\input{Ch1-Intro-physics_v1.tex}

\clearpage
\input{Ch1-Intro-equipment_v1.tex}

\clearpage
\input{Ch2-trans-instal-v1.tex}

\clearpage
\input{Ch2-rest.tex}

\clearpage

\input{Ch3-Utilities-Environment_v1.tex}

\clearpage

\input{Ch4-RadiationProtection_v1.tex}

\clearpage
\input{Ch6-Conclusion_v1.tex}

\clearpage

\input{ConventionalFacilitiesCDRbib_v1.tex}
\end{document}

%% file: Ch1-Intro-physics_v1.tex
\chapter{Introduction:  IsoDAR$@$Yemilab}

The IsoDAR (Isotope Decay At Rest) source
offers a pure $\bar \nu_e$ flux from decay of $^8$Li that, when paired with the Liquid Scintillator Counter (LSC) detector at Yemilab, will allow for state-of-the-art tests for 
oscillations involving sterile neutrinos \cite{PRL} and for
beyond Standard Model interactions \cite{elastic}.  

The source makes use of a cyclotron-accelerated beam delivered to a novel decay-at-rest target-system.
Specifically, a high-intensity H$_2^+$ ion source feeds  a 
60 MeV/amu cyclotron.   After acceleration,  the extracted H$_2^+$
ions are stripped to form a proton beam.
The proton beam is then transported to a
target of $^9$Be, cooled by D$_2$O, producing neutrons.  The neutrons enter 
a surrounding $\ge$99.99\% isotopically pure $^7$Li sleeve, where neutron
capture results in $^8$Li, that $\beta$ decay with a half-life of 839 milliseconds.  The resulting high-intensity decay-at-rest (DAR) $\bar \nu_e$ flux, peaking at $\sim 6$ MeV, interacts in the multi-kiloton liquid scintillator detector at Yemilab, the LSC, allowing physics searches 
exploiting the inverse beta decay (IBD), $\bar \nu_e
+ p \rightarrow e^+ + n$, and  $\bar \nu_e$-$e^-$ elastic scattering (ES)
processes.

This document briefly describes the components of the IsoDAR experiment: cyclotron, transport, and target systems; 
addresses the requirements and plans for the caverns that will house the experiment; and describes how components will
be transported and installed at the Yemilab site.  Utilities needs and environmental concerns will be addressed as well, including
shielding and radiation protection.  
In order to provide context, in the remainder of this chapter,  
we summarize the physics case and the introduce the technical components of the neutrino source to be housed within the infrastructure.

\section{Physics Summary}

A detailed description of the IsoDAR$@$Yemilab physics program will be published separately.  Here, we summarize highlights.   The physics capability and the design choices, including those for the conventional facility, are highly intertwined.   The implications for each design choice have been carefully studied, leading to the configuration reported in this document.

\subsection{Overview of the Physics Program}

The physics capabilities described below assume the parameters in Table~\ref{assumptions_table}.     

\begin{table}[h]
      \begin{tabular}{|c|c|} \hline 
Runtime  &  5 calendar years  \\ \hline
IsoDAR duty factor  &  80\%  \\ \hline
Protons on target/year  &  $1.97\times 10^{24}$  \\ \hline
$^8$Li/proton ($\overline{\nu}_e$/proton) &  0.0146  \\ \hline
$\overline{\nu}_e$/year  &  $1.15\times 10^{23}$  \\ \hline
1$\sigma$ uncertainty in $\overline{\nu}_e$ creation point  &  0.41~m  \\ \hline
IsoDAR@Yemilab mid-baseline  &   17~m  \\ \hline
IsoDAR@Yemilab baseline range  &   9.5-25.9~m  \\ \hline 
IsoDAR@Yemilab fiducial mass  &  2.57 ktons  \\ \hline
IsoDAR@Yemilab fiducial size (rad, height)  &  7.5 m, 17 m \\ \hline \hline
$\bar \nu_e + p\rightarrow e^+ + n $ (IBD) & $1.9\times 10^{6}$ events with cuts\\ \hline 
$\bar \nu_e + e^- \rightarrow \bar \nu_e + e^- $(ES) & 8000 events with cuts\\ \hline
\end{tabular}
\caption{Assumptions for the physics case.  ``Cuts'' included fiducial volume and muon-timing cuts. For ES it also includes $E_{vis}>3$ MeV.}
\label{assumptions_table}
\end{table}

\begin{figure}[b!]       
\begin{center}
{\includegraphics[width=7.in]{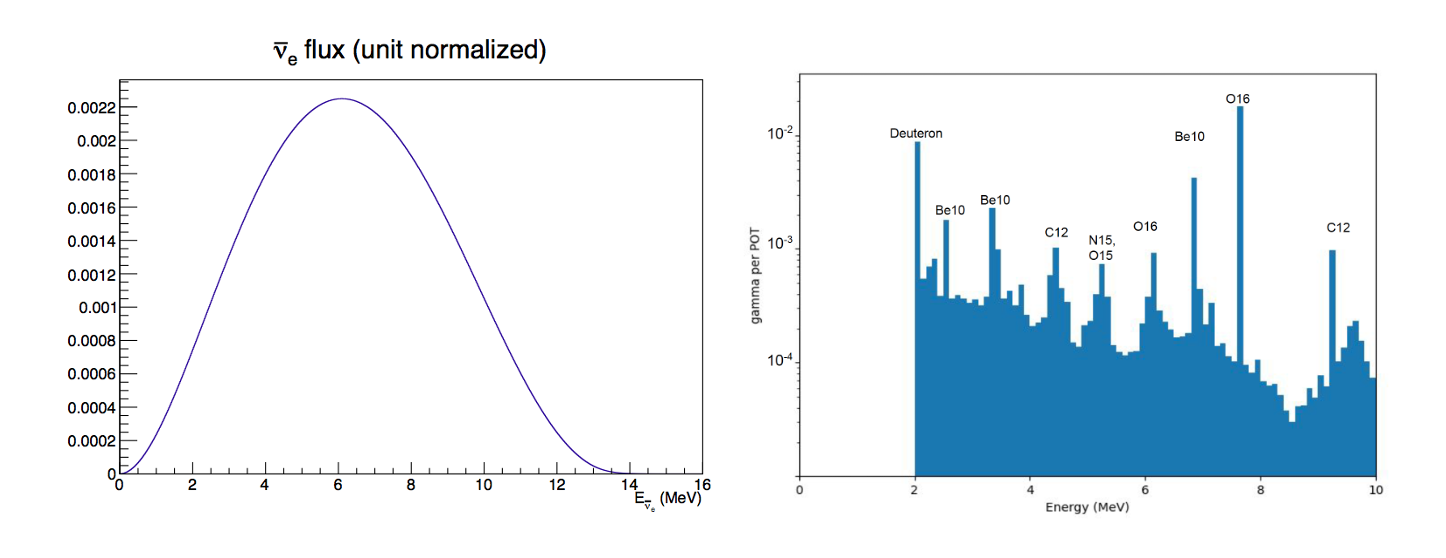}}
\end{center}  
\caption{\label{fluxes}  Left: Neutrino flux from the IsoDAR source, unit normalized.   Right: photons produced in the IsoDAR target/sleeve, normalized per proton on target.
        }
\end{figure}

The primary IsoDAR physics program is made possible by the unique, single-isotope-produced ($^8$Li), relatively high energy ($\sim 6$ MeV), pure $\bar \nu_e$ flux shown, unit normalized, in Fig.~\ref{fluxes}, left.   While the antineutrino flux provides the basis of the main physics program, interactions in the IsoDAR source also produce a high rate of neutrons and photons that are contained within the source region. These can be used for new physics searches involving conversion of photons and neutrons to non-standard particles.  
Ref.~\cite{MHneutrons} provides examples of dark sector searches using a ``neutrons-shining-through-walls'' method.
An interesting feature of the photon flux that can be exploited is the mono-energetic lines from decay of nuclear excited states, as shown in Fig.~\ref{fluxes}, right.

Given this $\bar \nu_e$ flux and the LSC detector design, IBD and ES will dominate the interaction types.   IBD has very low backgrounds because of the coincidence signal from the positron scintillation light followed by neutron capture, but environmental backgrounds to the single-flash ES signal are high below 3 MeV.   Given the parameters of Table~\ref{assumptions_table}.    
The layout at Yemilab yields very large samples:   Approximately 2 million IBD events and 8000 ES events with energy, timing and fiducial cuts.  
We provide two examples of the types of Beyond Standard Model (BSM) studies this opens up.  The program includes cross section measurements relevant to supernova physics studies, as well.

\subsection{IBD-based Search for Exotic Neutrinos Motivated by Very-Short-Baseline Anomalies}

IsoDAR was developed to allow a highly sensitive test of the source of anomalies observed in the very-short-baseline neutrino oscillation experiments.    If these anomalies are demonstrated to be BSM physics, this would revolutionize the field.    If these anomalies are instead due to Standard Model (SM) sources, the source must be identified, since these may affect other measurements important to the community, including that of CP violation. 

These anomalies are consistent with oscillations on very-short length scales ($\sim 10$ m).   Thus a picture involving a ``sterile'' neutrino ($\nu_s$) --a lepton that has no SM interactions itself,
but that can mix with the active neutrinos involved in the weak interaction-- has emerged.    The model that introduce one sterile neutrino in addition to the three active flavors is called ``3+1.''

A simple 3+1 explanation of $\bar \nu_e$ disappearance anomalies has become especially difficult in the past few years because of strong internal inconsistencies within experimental results.   The problematic picture is seen in Fig.~\ref{sterilesense}, which summarizes results within the 3+1 framework.
There is no single mass splitting, $\Delta m^2$, and mixing angle, $\sin^2 2\theta_{ee}$, that clearly explains all of the data.  
In short, the 3+1 model, alone, cannot resolve the present contradictions from the data sets. 

\subsubsection{The Electron Flavor Picture}

\begin{figure}[p]
\begin{center}
{\includegraphics[width=5.in]{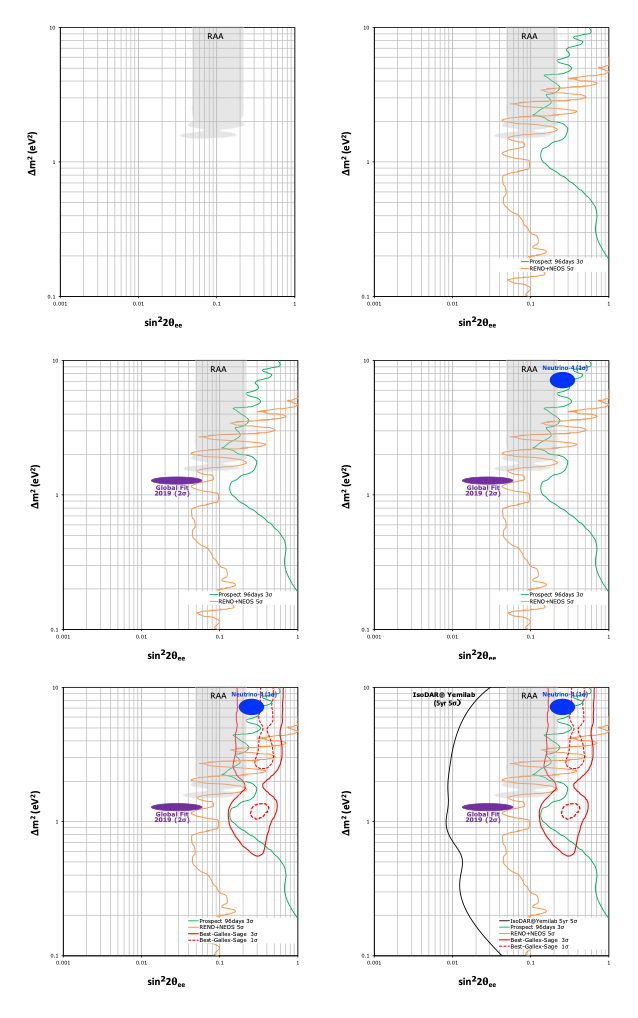}}
\end{center}  
\caption{\label{sterilesense}  Progressive plots summarizing electron-flavor disappearance results.    New information is added with each plot.   Top Left: Reactor Antineutrino Anomaly (RAA) allowed region (gray);   Top Right:    PROSPECT limit (green) and NEOS/RENO limit (orange) at 3$\sigma$ added;   Middle left: global fit allowed region (purple) added;  Middle right: Neutrino-4 (blue) added;   Lower Left: BEST/SAGE/GALLEX--red added;  and Sensitivity of IsoDAR at 5$\sigma$ added.}
        \end{figure}

At present, all $\bar \nu_e$ studies are reactor-based.   Nearly all
reactor experiments have reported an anomalously low normalization for the rate of $\bar \nu_e$ interactions compared to prediction \cite{RAA}, and that has been interpreted as a possible 3+1 oscillation signal (Fig.~\ref{sterilesense}, Top Left, gray, labeled ``RAA'').   
At the same time, these experiments have reported an unidentified peak in the prompt energy at with $\sim$5 MeV \cite{RENO, DayaBay, DChooz} 
that is not described by reactor flux simulations \cite{Huber}, leading to questions concerning the absolute flux predictions \cite{absflux, huberflux1, giuntiflux}.     

In response, a set of short-baseline (100 m to 1 km) reactor experiments were recently run that use energy-dependence rather than normalization in 3+1 searches.  Most of these short-baseline experiments set limits.  The result from PROSPECT (added in Fig.~\ref{sterilesense}, Top right, green)  \cite{Prospect} is a typical example.
Alternatively, data from very-short and short baseline reactor experiments can be combined, canceling most systematic uncertainties due to the absolute flux prediction. For example, the NEOS/RENO combination is added in Fig.~\ref{sterilesense}, Top Right, orange \cite{NEOSRENO}.   The 3$\sigma$ limits are shown,  
and one sees that a portion of the RAA is excluded at that level, but a great deal of parameter space remains uncovered.  

Although these experiments are presented as limits, in fact, they have allowed regions at 68\% to 90\% CL. This does not rise to the $2\sigma$-level that is usually used to define an anomaly, and so the allowed regions are not shown.    However, while each best-fit allowed region is statistically weak on its own,  the results cluster at about 1.3 eV$^2$ and $\sin^22\theta=0.03$.  As a result, when the collection of experiments is included in a global fit, which is described in the next section,  the best fit appears at roughly these parameters \cite{wherearewe}.  This is added in Fig.~\ref{sterilesense}, Middle left, purple.

Surprisingly, one recent reactor experiment, Neutrino-4 \cite{Nu4}, has reported a very large disappearance signal with associated oscillation parameters of $\Delta m^2\sim$ 7 eV$^2$ and $\sin^2 2\theta \sim 0.26$ (added in Fig.~\ref{sterilesense}, Middle Right, blue).  This result agrees with the RAA at the 1$\sigma$ level.    The PROSPECT limit at 3$\sigma$ does not cover the Neutrino-4 result, but PROSPECT, and other short baseline reactor experiments such as STEREO, do conflict with this result at $\sim 2\sigma$.   The Neutrino-4 result is far from the parameter space favored by global fits (Fig.~\ref{sterilesense}, purple), as discussed below.

 \begin{figure}[t]       
\begin{center}
\hspace{-0.3in}
{\includegraphics[width=6.9in]{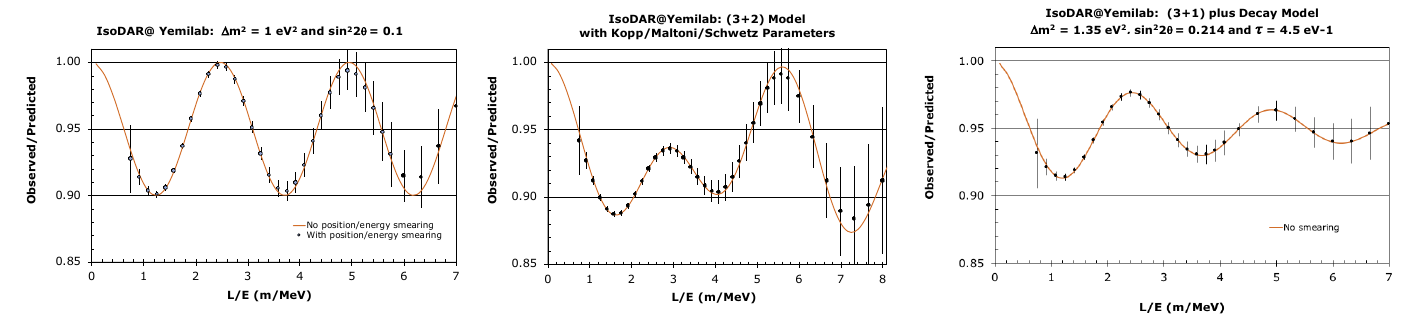}}
\end{center}  
\caption{\label{wiggles}  $L/E$ distributions that will be measured after a 5 year run at Yemilab for three models: 1)  Left:  3+1 global fit;  2) Middle: 3+2 global fit;  3) 3+1 with neutrino decay, consistent with a 95\% allowed region observed at IceCube \cite{MarjonThesis}.
        }
\end{figure}

If CPT is not violated, then $\bar \nu_e \rightarrow \bar \nu_s$ and $\nu_e \rightarrow \nu_s$ results will be in agreement,
 so $\nu_e$ disappearance searches involving Megacurie sources can add valuable information to this picture.  
Recently, the BEST $\nu_e$ disappearance search that makes use of a  $^{51}$Cr DAR source reported results \cite{2109.11482}.
These show a $\sim 4\sigma$ deficit in rate that, when interpreted as $\nu_e \rightarrow \nu_s$, yields a best fit parameters of $\Delta m^2 \sim 3$ eV$^2$ and $\sin^2 2\theta \sim 0.4$.   These results are in agreement with the SAGE and GALLEX experiments performed 25 years ago using the same type  of source and similar detectors \cite{2109.11482, Giunti}.  The BEST, SAGE and GALLEX combined results indicate the allowed region added in
Fig.~\ref{sterilesense}, Bottom Left, red.  One sees that BEST/SAGE/GALLEX overlaps with the Neutrino-4/RAA region within 3$\sigma$, and overlaps with Neutrino-4 alone at 1$\sigma$.    However,  most of the allowed range strongly disagrees with both PROSPECT and NEOS/RENO.   With that said, the lowest allowed region for BEST is consistent with the mass splitting favored by the global fits.

\subsubsection{The Global Picture}

A full 3+1 picture must also encompass results for 1) $\nu_\mu \rightarrow \nu_e$ appearance, 2) $\nu_\mu$ disappearance and 3) $\nu_e$ disappearance results \cite{wherearewe}.   Above, we have discussed issues that create tension when comparing the $\nu_e$ disappearance samples.  The other two categories also have samples with curious behavior also.
In the case of $\nu_\mu \rightarrow \nu_e$ appearance, the experiment with the most significant anomalous signal as a function of energy, MiniBooNE, has an angular distribution for the excess events signal is not consistent with arising entirely from SM $\nu_e$ interactions \cite{MB2020}.    This has led to extensions of the 3+1 model that also include photon signatures from heavy neutrino decay \cite{vergani}.   Related to this, in 
$\nu_\mu$ disappearance, the IceCube observes a signal at 93\% CL \cite{spencer} that is better fit, rising to 97.2\% CL, when neutrino decay is also included \cite{MarjonThesis}.    Thus, although the 3+1 model is the baseline used in most global fits, all three of the categories of oscillations have features that point to more complexity.

The global fit shown in Fig.~\ref{sterilesense}, purple, does not include RAA, Neutrino-4, BEST, or IceCube \cite{wherearewe}.   RAA was not included because of concerns about the poorly described  absolute reactor flux.  The other three experimental results are too recent to have been incorporated in the 2019 global fits.   The 3+1 fit prefers an oscillation solution with $\sim 2$ eV$^2$ and mixing of about 0.03 (Fig.~\ref{sterilesense}, purple) \cite{wherearewe}.   As pointed out above, this is far from the parameter range preferred by the BEST/Neutrino-4 combination.  However, the $\Delta m^2$ is consistent with allowed parameters from BEST.

\subsubsection{Why IsoDAR?}

How do we explain these results?  One explanation is that these experiments all suffer from unidentified systematic effects that unfortunately conspire to look like oscillations within the 1 to 10 eV$^2$ range.   Another is that Nature is presenting a more complex model than 3+1.  For example, there may be a second sterile neutrino involved in the oscillations--a so-called ``3+2 model.''  Alternatively, as explored by MiniBooNE \cite{vergani} and IceCube \cite{MarjonThesis}, sterile neutrino decay may be involved.
These possibilities are impossible to untangle with current technology.  Using standard methods seems to present new contradictions with each new experiment.

IsoDAR$@$Yemilab, with its innovative, powerful source and LSC detector, with its multi-kiloton target volume, can provide a definitive result.  The black line on Fig.~\ref{sterilesense} indicates the 5$\sigma$ sensitivity that can be obtained in 5 years of running.   The advantages of this program are:
\begin{itemize}
    \item Unlike the reactor experiments,  the flux is from a single isotope, $^8$Li, that is well measured and modeled.
    \item Unlike BEST, which is also a single isotope experiment, IsoDAR$@$Yemilab reconstructs the neutrino path length, $L$, and energy $E$, since the IBD interaction is used, rather than integrating 
    across these variables.
    \item Following from this, unlike all of the existing experiments, IsoDAR can trace the signal in $L/E$ across multiple oscillation lengths.  
\end{itemize}
These capabilities allow IsoDAR to resolve complex models for neutrino behavior.  This is shown in Fig.~\ref{wiggles}, which illustrates the capability to resolve the differences between three potential models that are often discussed in the literature.

\subsection{Precision Electroweak Tests through Electron Neutrino--Electron Scattering}

The 8000 $\bar \nu_e$-$e^-$ elastic scattering events produced in IsoDAR$@$Yemilab above the 3 MeV threshold that pass fiducial requirements and time-window cuts to remove spallation from muons
offer a valuable opportunity to test for BSM.   This is a very well-predicted lepton-lepton interaction. The ES rate can be normalized by the very high-statistics IBD sample to remove systematic uncertainties due to source and detector effects.  

An example of the power of this sample is a search for non-standard interactions (NSIs).   These arise from new physics that changes the Standard Model left- and right-handed couplings, $g_L$ and $g_R$ by effective parameters, denoted by $\varepsilon$.  NSIs are potentially flavor dependent and with differences in lepton-lepton versus lepton-quark couplings. Indicating the  electron-flavor neutrino-electron case.   NSIs may also allow for flavor change.   IsoDAR is sensitive to the flavor conserving NSI that couples electron-flavor neutrinos to electron targets.   Here, we will denote this by ``$ee$.''  Such NSIs modify the left- and right-handed couplings in the Standard Model cross section by: $g_R^\prime =g_R + \varepsilon_{eeR}$ and $g_L^\prime =1 +g_L + \varepsilon_{eeL}$.   With this replacement, the form of the Standard Model cross section remains the same.

\begin{figure}[t]       
\begin{center}
\hspace{-0.3in}
\includegraphics[width=5.in]{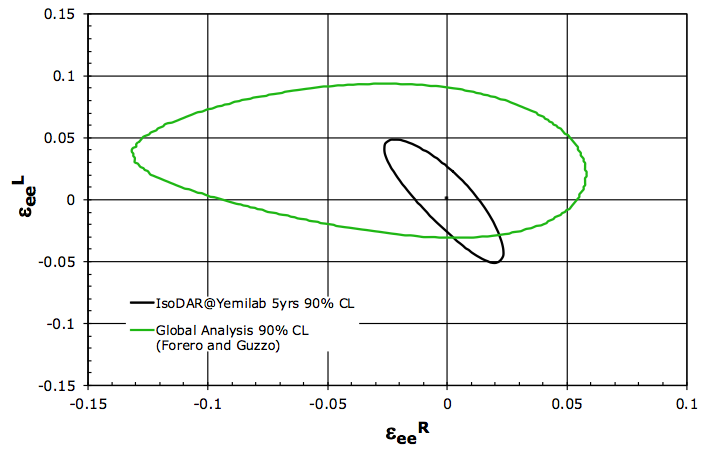}
\end{center}  
\caption{\label{epsilon}  
Expectation for the IsoDAR in the $\varepsilon_{ee}$ left-right parameter space (black) compared to world data (green) from Ref.~\cite{ForeroGuzzo}. The Standard Model predicts (0,0).        }
\end{figure}

IsoDAR measurements will substantially improve our knowledge of the NSI $ee$ coupling parameters.    The sensitivity is shown on Fig.~\ref{epsilon}, black, compared to the present measurements compiled by Forero and Guzzo\cite{ForeroGuzzo}, green.   The Standard model prediction is (0,0).
The present data includes reactor samples, but is dominated by electron-neutrino scattering from pion/muon decay at rest ($\pi/\mu$DAR) experiments at Los Alamos, leading to the rotation of the allowed region with respect to IsoDAR, which is purely antielectron neutrino scattering.   The high statistics of IsoDAR lead to far higher precision than has been achieved in the past.   There is no other experiment planned for the near future that competes with IsoDAR capability in the NSI $ee$ sector.

IsoDAR joins a vibrant program of experiments searching for complementary NSI effects that is reviewed in Refs.~\cite{NSIreview, DentDutta, Miranda}. Many neutrino-nucleon coherent scattering experiments are now proposed at $\pi / \mu$ DAR facilities. These can make precise measurements of lepton-quark couplings involving 
muon and electron flavors.  Also, Fermilab-based experiments using decay-in-flight beams now and in the near future, are improving the search for NSIs appearing in muon-flavor neutrino--electron scattering.

%% file: Ch1-Intro-equipment_v1.tex

\section{Technical Components of the Neutrino Source}

The steps in generating the antineutrinos are:

\begin{itemize}
\itemsep-0.1em
\item An accelerator that produces a beam of 5 milliamperes (mA) of H$_2^+$ ions
(which corresponds to 10 mA of protons) at 60 MeV/amu.
\item A transport line that includes a stripping foil close to the extraction point of the cyclotron
to remove the electron from the 
H$_2^+$ molecular ion, converting the H$_2^+$ ions to protons.
\item A layered beryllium and heavy-water target that is struck by the proton beam, producing large quantities of
neutrons.
\item A sleeve containing a mixture of highly-enriched $^7$Li and beryllium that 
is flooded by these neutrons, which are moderated and captured to make
the parent $^8$Li, which subsequently beta-decays releasing the electron-antineutrinos.
\end{itemize}

\begin{figure}[t]
\centering
\includegraphics[width=4.5in]{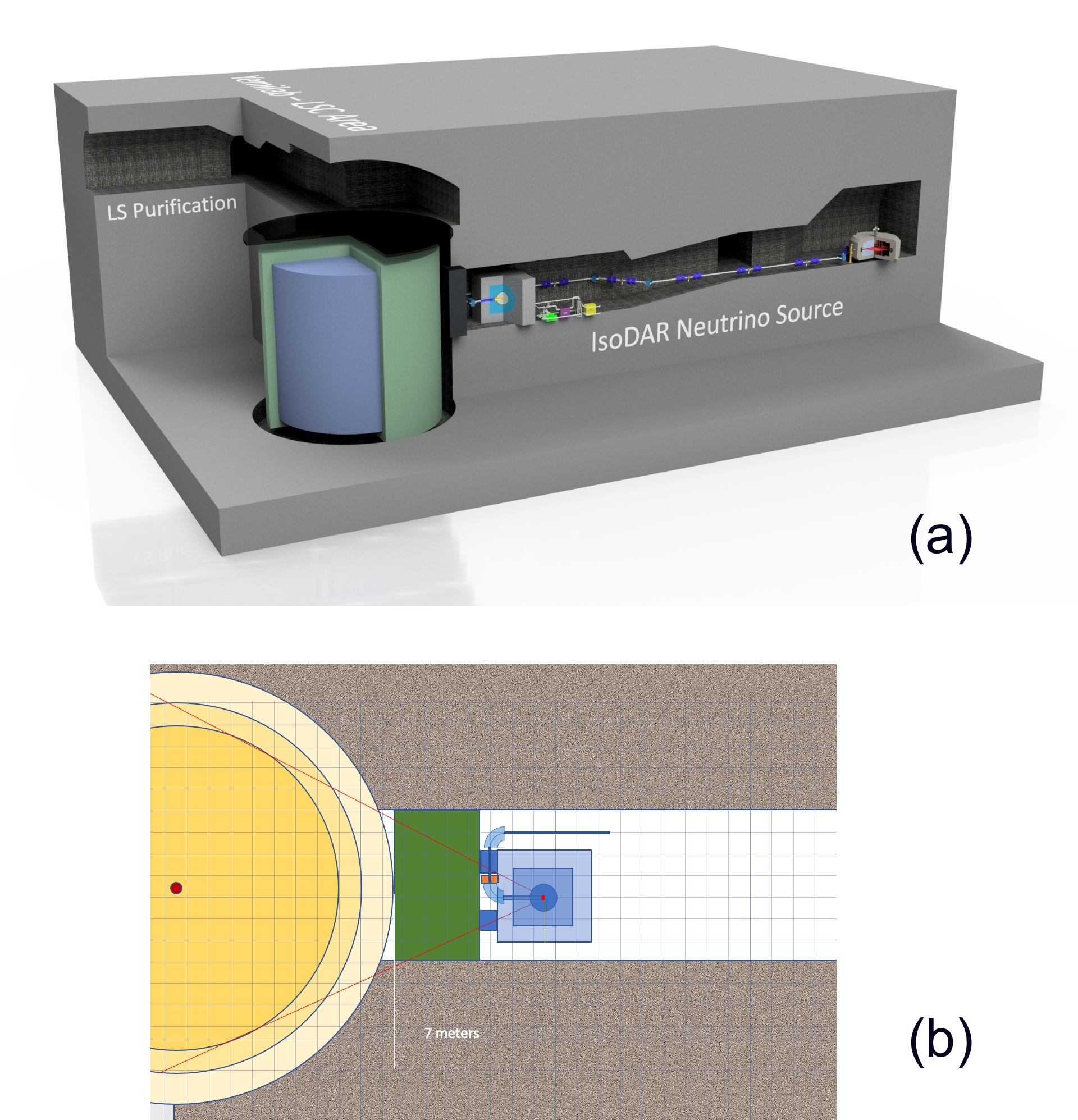}
\caption{
\footnotesize Schematic of the IsoDAR experiment deployed at the Yemilab site; (a) shows the cyclotron
at the far right corner and the transport line taking the beam through to the target area. The target assembly is represented by the blue cubes (steel and concrete) in (b), the target itself is the small red dot at the center.  The sleeve with the Be + $^8$Li where the $\bar \nu_e$ flux is produced, surrounds the target. The beam line comes in via two 90$\degree$ bends so the beam strikes the target pointing away from the detector. This greatly reduces the fast neutron flux directed towards the detector. The target volume of the detector is represented by the blue cylinder in (a) and darker yellow in (b), and the buffer and veto regions are shown in green (lighter yellow in (b)). }
\label{deployment}  
\vspace{0.2in}
\end{figure}

\begin{figure}[tb!]
\centering
\includegraphics[width=2.5in]{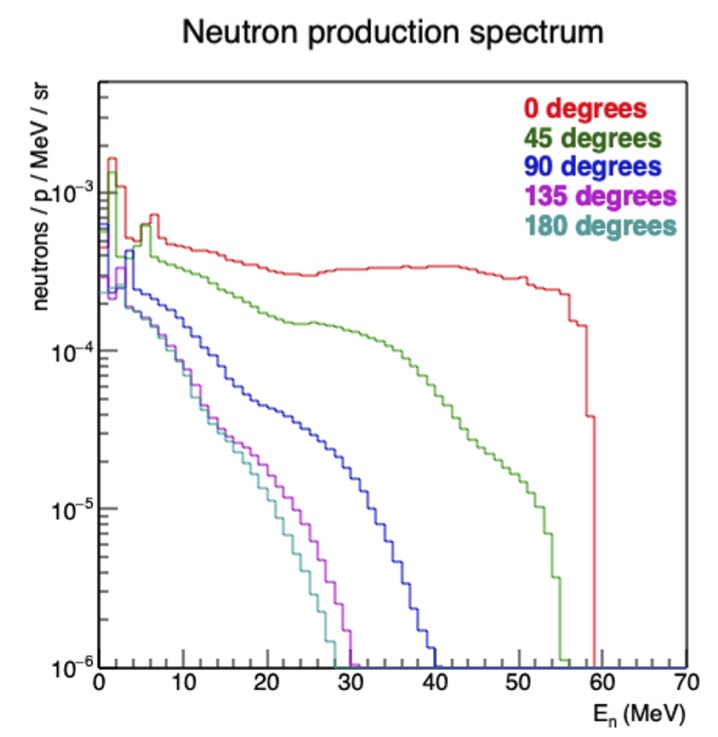}
\caption{{\footnotesize Angular distribution of neutrons emerging from the target.  High-energy neutrons in the 
backwards direction are greatly attenuated, reinforcing the value of the beam entering the target going away from
the detector.}
\label{neutron_spetrum}}
\vspace{0.2in}
\end{figure}

The layout of these components in the Yemilab setting is shown in Figure~\ref{deployment}. 
The cyclotron is placed in a dedicated room at a bend in the ramp from the main Lab passageway to the Target Hall.
The stripper and analysis magnet are close to the extraction point from the cyclotron.  
The transport line or MEBT (Medium Energy Beam Transport) brings the beam down the ramp and to the target.
A large steel block (shown in green in Figure~\ref{deployment}(b)) covers the opening to the LS detector, 
to shield the detector from gammas and neutrons
generated in the target hall.  The beam line undergoes two 90$\degree$ bends to orient the beam so it strikes
the target going away from the detector.  Figure~\ref{neutron_spetrum} demonstrates the value of this concept: 
high-energy neutrons emitted in the backward direction from the target towards the detector are greatly reduced
in both energy and intensity.
The target of nested shells of beryllium cooled with D$_2$O stops the beam and produces 
the neutrons shown in 
Figure~\ref{neutron_spetrum}.  These neutrons are moderated as they stream into the sleeve surrounding the target where
they are captured by the $^7$Li.  The high fraction $~$(75\%) of beryllium powder in the sleeve has been optimized to multiply the neutrons, and maximize the yield of $^8$Li. The target and sleeve are surrounded with highly efficient shielding material to minimize neutrons emerging from the shielding
structure.

As the purpose of this section is to provide context for the layout underground and the infrastructure requirements discussed in this CDR, the elements are discussed 
only briefly.  For more substantial information on the components, please see the 2015 Technical Facility CDR on the neutrino source \cite{CDR2015}.  Updates to that information and substantial detail on the most complex elements of the neutrino source are provided in references ~\cite{winklehner:nima, winklehner:mist1, winklehner:RFQDIP, bungau:optimization, bungau:shielding}

\subsection{Overview of the Accelerator System}

IsoDAR proposes to break new ground in the performance of cyclotrons,
 by delivering an order-of-magnitude more beam current than existing compact cyclotrons.  
 The standard commercial compact cyclotrons today are limited by space-charge forces, especially at injection energies.
One must compress the charged particles into a tight bunch for injection and acceleration into the cyclotron, 
but the charge in this bunch generates a strong repulsive force (referred to as ``space charge'') that is very large, 
and at high beam currents becomes too large to be overcome by the available focusing forces.  
A second limit for conventional cyclotrons,  all of which accelerate H$^-$, is the lifetime of the stripping foil
used for extracting the beam.

To reach our performance goal, the IsoDAR system must address both of these
challenges, and must pay great attention to efficiently bunching, injecting,
accelerating and extracting the very high-current beam.  
In all of this, it is imperative to minimize beam loss.
First of all, large beam losses require higher currents from the ion source to 
ensure adequate beam intensity for the purpose for which it is accelerated.  
The source must make up for all the beam losses.
Beam losses for high intensity beams can be extremely damaging.  
At the low energies at the start of acceleration in the central region of 
the cyclotron, beam loss causes sputtering of material and voltage breakdown; 
at high energies, beam loss into the walls of the cyclotron produces neutrons
that cause activation and severely limit the ability to perform maintenance 
on the cyclotron because of the high radiation fields.
 
Three important design breakthroughs have allowed us to 
make a convincing case that we can reach the necessary intensities.   

\begin{figure}[tb!]
\begin{center}
\vspace{-0.2in}
{\includegraphics[width=5in]{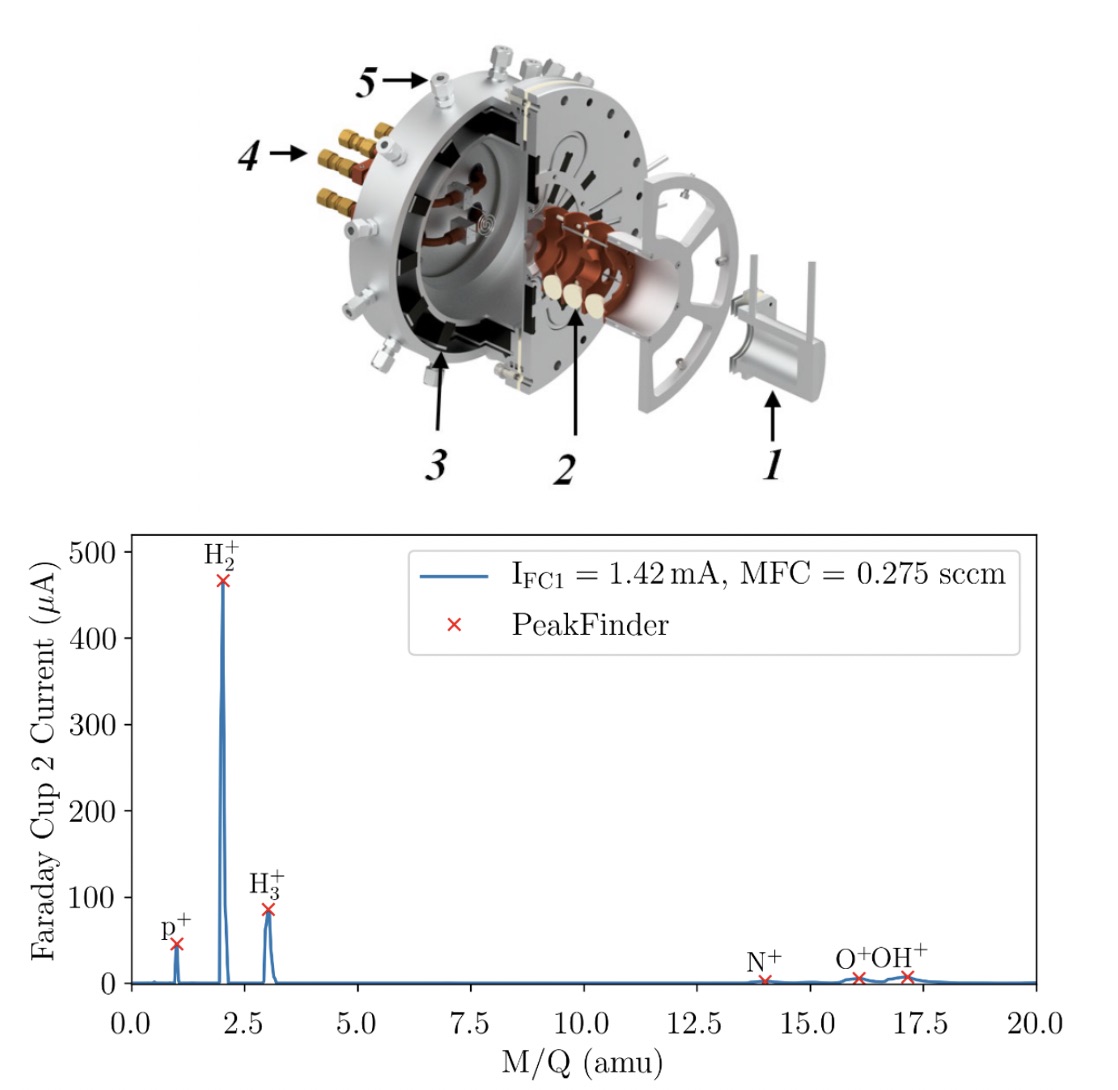}}
\vspace{-0.2in}
\end{center}  
\caption{\label{MIST-1}  Cutaway of MIST-1, filament-driven H$_2^+$ source, and recent spectrum.  H$_2^+$ peak dominates 
over proton and H$_3^+$, contaminant fraction is low.    See Ref.\cite{winklehner:mist1} for details.  
}
\end{figure}

First, we
use H$_2^+$ ions rather than protons, which has benefits at a number
of places in the acceleration cycle, not the least being a significant reduction in space-charge effects.   
We discuss the value of selecting H$_2^+$ as the accelerating ion over protons,
H$^-$ or deuterons in Sec.~\ref{costbenefit}. A prototype high-current, 
filament-driven multi-cusp ion source has been built, shown in Figure~\ref{MIST-1}
that also shows the results of first studies of the source~\cite{winklehner:mist1}. 
Development and optimization of this source is continuing; at this writing H{$_2^+$} current density and 
ionic fraction are moving well towards design requirements~\cite{winklehner:ICIS2021}.

\begin{figure}[tb!]
\begin{center}
\vspace{-0.2in}
{\includegraphics[width=5in]{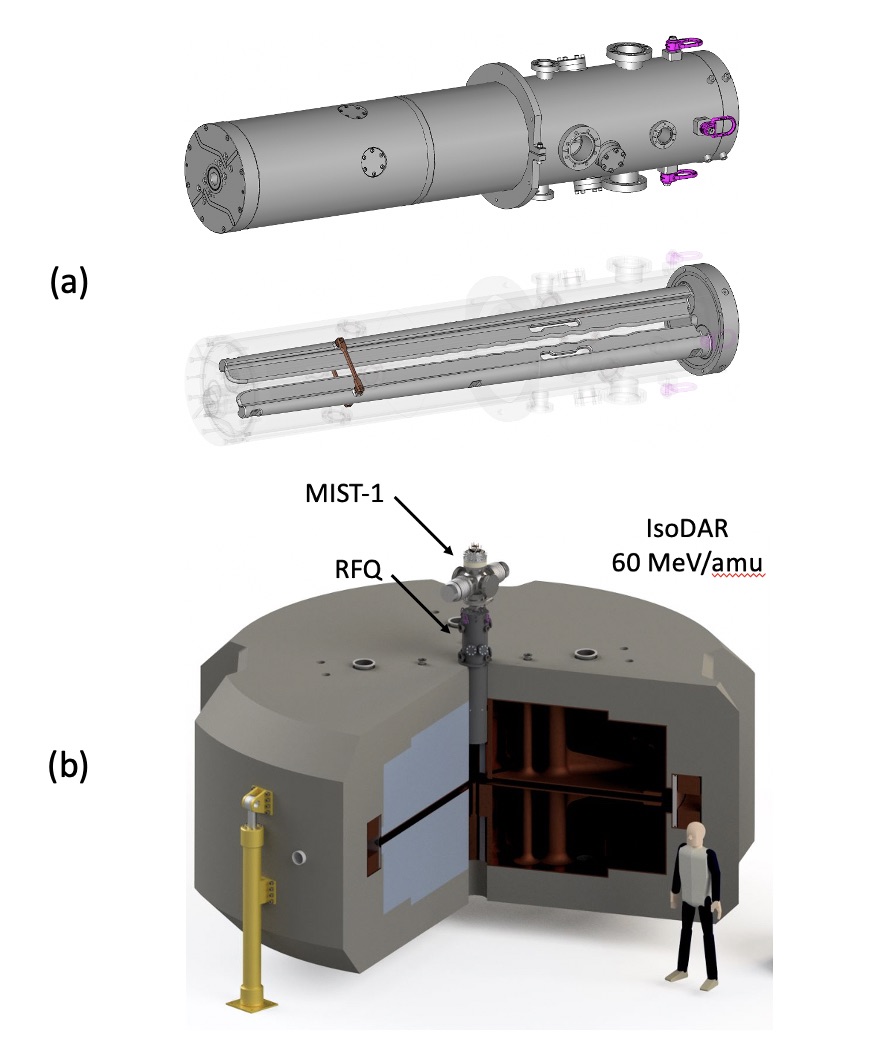}}
\vspace{-0.2in}
\end{center}  
\caption{\label{RFQ}  (a) Engineering model of RFQ designed by Bevatech GmBH, Frankfurt.  The bottom half of (a) shows two vanes of the split-coaxial structure mounted on one end plate.  The other two vanes are mounted on the opposite end plate. This geometry allows for the very low resonant frequency of 32.8 MHz, the frequency driving the cyclotron RF system.  The MIST-1 source is mounted on the right side, beam exits from the left side. (b) The RFQ is shown mounted along the central axis of the cyclotron.  A bit more than half of the structure is inside the steel of the cyclotron magnet.  The exit point is located 20 cm from the first accelerating Dee of the cyclotron.  This short flight path preserves most of the bunching provided by the RFQ. 
        }
\end{figure}

Second, we employ direct axial injection through an RFQ (Radio Frequency Quadrupole)
buncher for the first time in a compact cyclotron, allowing the high
efficiency beam capture that is necessary to reach high intensity~\cite{winklehner:RFQDIP}. 
This RFQ replaces the conventional ``low energy beam transport'' (LEBT) that 
transports the continuous stream of particles from the ion source to the point 
of cyclotron injection through the so-called ``spiral inflector''~\cite{winklehner:spiral}.
For efficient capture, the cyclotron can accept only beam 
that is within  $\sim \pm10^\circ$ of the synchronous phase
of the RF accelerating voltage.
As beam from the ion source is continuous, capture efficiency is less than 10\%.
This requires a factor of 10 higher current from the ion source, and leads to huge beam losses in the 
central region of the cyclotron.  As stated above, this beam loss leads to sputtering damage 
and high-voltage arcing problems.
Because of the high space-charge in low-energy high-current beams, ``classical'' double-gap bunchers, 
attempting to compress more particles into the acceptance window, are not at all effective.
The 1-meter long RFQ buncher, operating at the same frequency as the cyclotron, is capable of placing
over 90\% of the continuous beam from the ion source into the phase-acceptance window of the cyclotron
and so to increase beam intensity. The RFQ buncher is shown in Fig.~\ref{RFQ}. 
Construction of a prototype RFQ for IsoDAR is now underway \cite{winklehner:nima}.

\begin{figure}[tb!]
\begin{center}
\vspace{-0.2in}
{\includegraphics[width=3.in]{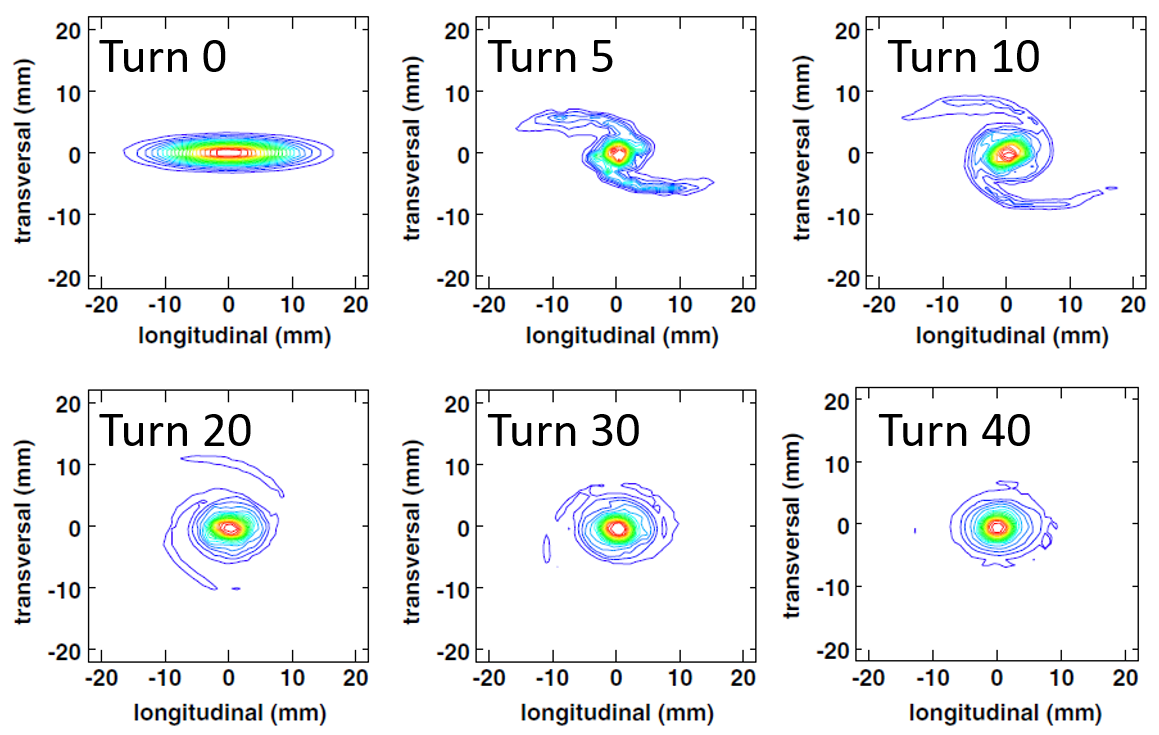}}
\vspace{-0.2in}
\end{center}  
\caption{\label{vortex}  Vortex motion in PSI Injector II; simulations using the OPAL code confirm the 
experimental observations \cite{yang:vortex}. In only a few turns the bunch evolves into a spiral, 
with the majority of the particles wrapped into a tight bunch.  Halo particles, amounting to 10-20\% of the total, 
are scraped off with judiciously placed collimators, all before the beam has reached 5 MeV, so causing no activation.  
From turn 40 onwards there is virtually no change in the bunch shape.
}
\end{figure}

Third, and perhaps the most important breakthrough, was the discovery 
that for high-current beams where space charge is a dominant force,
an effect called vortex motion could stabilize the beam and reduce growth 
during acceleration. This surprising finding was first observed at the 
PSI Injector II cyclotron \cite{stetson:vortex, stammbach:vortex}.  
As shown in  Fig.~\ref{vortex}, as the beam circulates,
the combination of space charge effects and external focusing forces 
induce bunches to curl tightly in phase-space.
This prevents the massive losses that some accelerator 
physicists had predicted would occur on the extraction septum,
due to radial overlap of bunches in adjacent turns close to extraction 
from the cyclotron \cite{yang:vortex}.

\begin{figure}[tb!]
\begin{center}
\vspace{-0.2in}
{\includegraphics[width=4.in]{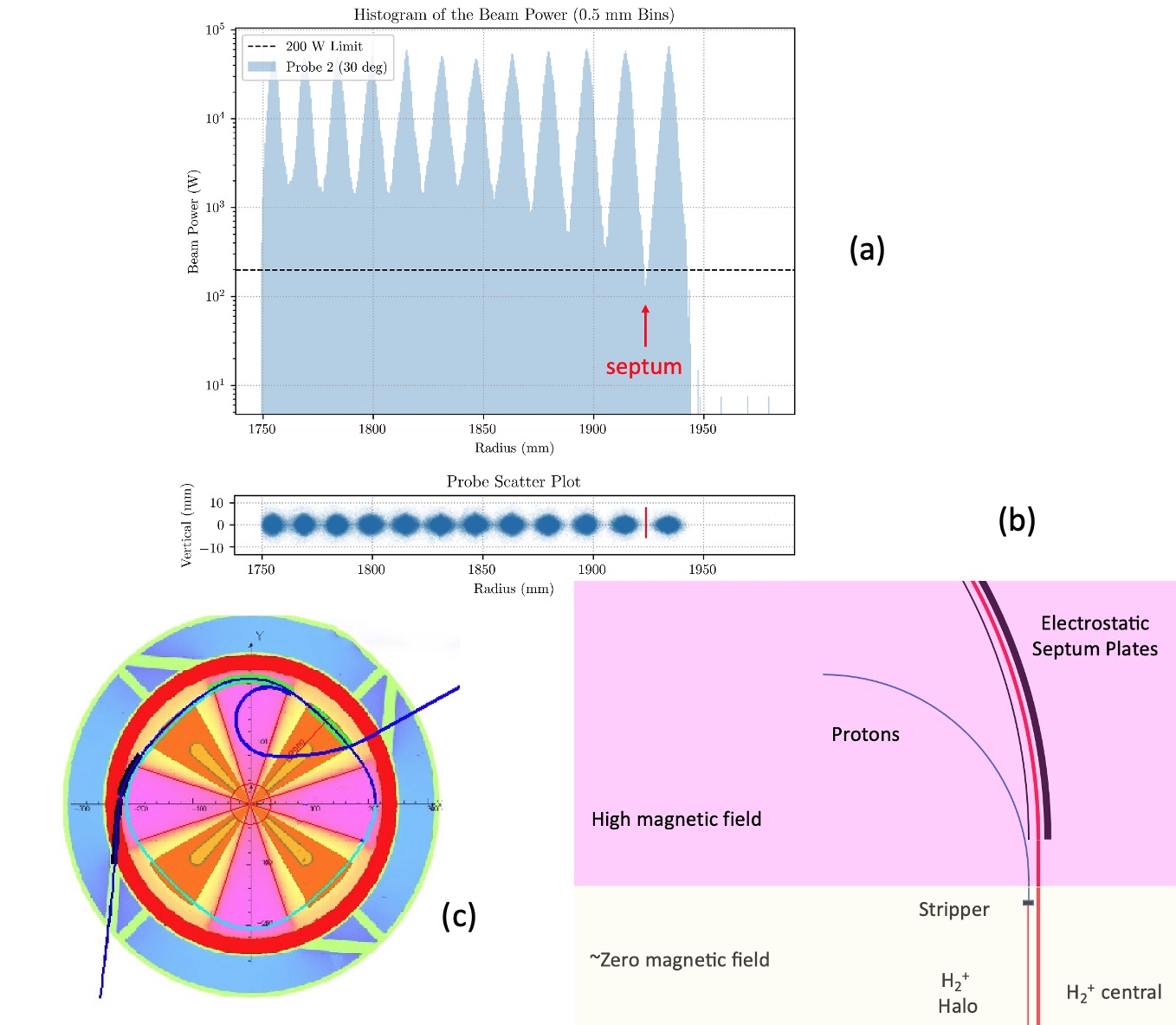}}
\vspace{-0.2in}
\end{center}  
\caption{\label{extract}  (a) Simulations of outer orbits in the cyclotron showing clean turn separation, particularly on the last turn.  Note the logarithmic scale.  (b) Schematic of the electrostatic extraction channel (black) and the H$_2^+$ beam being extracted (red).  Residual beam between turns, referred to as ``halo'' that would strike the inner septum plate is intercepted by the thin stripper.  These ions are converted to protons that are more sharply bent in the high ``hill'' magnetic field of the cyclotron, missing the septum plate.  (c) Plan schematic of cyclotron showing the four high-field ``hills'' (pink), the four low-field ``valleys'' (yellow), the orange RF accelerating Dees, the red coil surrounding the entire pole section, and the blue yoke return steel for the magnet.  The H$_2^+$ extraction channel is shown with two electrostatic deflectors followed by two magnetic channels.  The orbits of the stripped protons are also shown as the blue spiral exiting to the upper right.
        }
\end{figure}

Another feature of using H$_2^+$ ions relates to extraction from the cyclotron, which is described below.
Instead of extracting the beam with a stripping foil, we revert to
the early technique of an extraction channel employing a thin electrostatic septum for guiding
the beam out of the cyclotron.  
For this to work, one must have good turn-to-turn separation at the last orbits, and must
keep the size of the bunch small in order to have few particles at the radius where the septum
is located.  Introducing a structure resonance helps increase turn separation at the extraction
point ~\cite{seidel:extraction}, while the afore-mentioned vortex effect controls the bunch size.
The result, shown in Fig.~\ref{extract} shows that loss on the septum will be less than 100 watts which,
considering the total beam power is 600 kilowatts, is a remarkable accomplishment.  This level
of beam loss is well below the ``rule-of-thumb'' from PSI, of maintaining beam losses below 200 watts
to enable hands-on maintenance of accelerator components.

But there is a further step we can take to protect the electrostatic extraction septum.  
Placing a narrow stripping foil just upstream of the septum intercepts ions that would strike
the septum.  These H$_2^+$ ions are converted to protons which are then bent inwards by the 
cyclotron magnetic field, and miss the septum.  They orbit tightly in the strong ``hill'' 
section of the magnetic field, and emerge in the weaker ``valley'' on a trajectory that takes
them safely out of the cyclotron. The details of this extraction scheme are shown in Fig.~\ref{extract}.

The hill/valley design of the magnet may be unfamiliar to some readers.  This is employed in cyclotrons that are ``isochronous,'' {\it i.e.} the time for a particle to complete
one orbit (``turn'') is constant, and does not depend on the energy of the particle.
In such a machine, the magnetic field varies along the trajectory, adjusted 
by detailed shaping of the steel (``shimming'')  the main coil geometry, and ``trim coils.''
This allows for a single-frequency RF system 
to be used, and for particles of all energies, from injection to extraction, to 
reside stably in the cyclotron.  Thus, particles are continuously accelerated and
extracted. 

The accelerating system will be constructed in parts and assembled in the Yemilab space.   While this is true for all conventional cyclotrons, the space limitations of an underground laboratory require the sizes of the pieces to be smaller than usual.    The specific limitations on size are discussed later in this report. Refer to the Technical Facility CDR \cite{CDR2015} for an explanation of how the accelerator components can be broken down.  In fact, the limitations on size discussed in that CDR are more restrictive than what is required at Yemilab. In particular, we are confident that 
the cyclotron coil, which is 5 m in diameter, can be brought through all Yemilab passages if tilted on the diagonal.

\newpage				
\subsection{Frequently Asked Questions About the Accelerator System \label{costbenefit}}

As a neutrino source to be installed in an underground laboratory, IsoDAR has many unique issues that drive the design.  The questions that are most often raised focus on the choice of a compact cyclotron for the design.  We have published a detailed cost/benefit analysis of the accelerator options \cite{costeffective}.   Here we address the questions that are most often asked.

\subsubsection{What Metrics Were Considered During the Design Phase?}

The following metrics were considered when selecting the design:
\begin{enumerate}
\item {\it Cost:}~~Minimizing the total acquisition and operations costs is important to any project.   In the case of running at an existing underground lab, this includes the cost of the accelerator, the target ($\bar \nu_e$ source), electricity, and other relevant infrastructure--all of which are addressed in this Conceptual Design Report.

\item  {\it Rate and energy distribution of the $\bar \nu_e$ flux:}~~Maximizing the $\bar \nu_e$ flux is vital for sensitivity to sterile neutrinos as well as accomplishing the other physics goals. 
Note that a  mean $\bar \nu_e$ energy well above 3~MeV is important, as this requirement helps differentiate signal events from (usually less energetic) radiogenic backgrounds in the detector.

\item  {\it Rate and energy distribution of backgrounds:}~~The 
$\nu_e$ intrinsic background must be minimized with respect to the 
$\bar \nu_e$ flux, and limited to $<3$~MeV. 

\item{\it Low technical risk:}~~No existing cyclotron can meet our physics goals.
  Therefore, some R\&D has been required for the base design.  We consider the risks of each element of the design in our Technical Facility CDR \cite{CDR2015}.  We compare the risk
  involved to the risk of implementing alternative designs in Ref.~\cite{costeffective}.
  
\item {\it Compactness of both 
accelerator and $\bar \nu_e$ source:}~~Compactness of the accelerator is driven 
by cost and space considerations underground.   Compactness of the target/sleeve is driven by the precision required on knowledge of $L$, the distance that the neutrinos travel to the target, as well as by cost considerations.  The level of neutron reduction by shielding is set by regulation, but the choice of shielding material balances the cost of the shielding itself against both the added distance from target to detector and cost of excavation of space to accommodate the shielding.

\item {\it Simplicity and Reliability:}~~Overall cost, along with compactness, application underground, and safety issues all encourage simplicity of design.   The goal has been to simplify both the construction and operations phases, but where there were trade-offs, we have opted to simplify long-term operations and minimize downtime for maintenance. This will enhance reliability.

\item {\it Long-term value
to future physics programs and industry:}~~Developing engineering and
infrastructure for future physics programs is desirable.
Developing a
  design which is of interest to industry, for 
  possible medical or ADS (Accelerator-Driven systems) applications has 
  already led to strong industrial partnerships and sharing of development costs.
\end{enumerate}

For the full set of considerations, the reader should refer to  Ref.~\cite{costeffective}.   
The scope of this document is too large to summarize here.   
Instead, we now consider some specific value-engineering questions that often arise when we present the project.   
We highlight where the decisions on the issues are directly related to this CDR.

\subsubsection{Why the Choice of a Compact Cyclotron?}

 In Ref.~\cite{costeffective}, we consider many options for drivers, including accelerators, D-T generators and more exotic options.  In each case, we looked for a design that would match the flux specifications and design criteria described above.   Here is a brief summary of the options and the findings:
 
\begin{itemize}

\item {\it Alternative Cyclotron Designs:~~}    An RFQ injecting into a 4-sector ring cyclotron with conventional magnets was considered the best alternative design.   However, this would increase the cost of the cyclotron itself by an estimated 40\% and would occupy a substantially larger footprint, increasing the cost for cavern space.    An alternative to the conventional magnets in our compact design is to use superconducting magnets.  This has much higher up-front costs for the cyclotron and for the supporting cryogenic plant.  It also places much stronger constraints on losses and rapid removal of dumped-energy around the collimators and extraction region.   In addition, wall-power costs may not be lower during operation. As most of the electric power goes into the beam through the RF system, the magnet power is a small part of the total.  Furthermore, maintaining the necessary up-time is riskier and maintenance of the machine is more difficult.   Safety issues related to use of cryogens is also a concern.

    \item {\it Linear Accelerators:~~} Linacs could be either normal-conducting (copper) or
    superconducting.  
    A high-current normal-conducting linac 
    would not be a practical option for IsoDAR. For one, 
    the electrical efficiency of these structures is very low.  
    For example,  LANL built a CW copper RFQ demonstrator called LEDA
    as the ``front end" (low-energy portion) for the APT project in the 1990's.  
    This RFQ accelerated 100 mA of protons to 6.7 MeV and required 
    four 1.5 megawatt klystrons to drive the RF system.  
    That's 6 MW of RF power to produce 675 kilowatts of beam power.  
    The efficiency is about 11\%, with 89\% of the power going into heating the copper structure.
    The klystron itself can be quite efficient, about 70\%, but the total
    efficiency, wall-plug to beam, is no higher than 8\%.  By comparison, the
    RF system for the compact cyclotron has a wall-plug to beam-power efficiency of about 50\%.

    A superconducting linac would not have the structure-heating problems, and the coupling
    of RF power to the beam would be substantially higher.  
    But low beta superconducting structures are not as cost-effective as normal-conducting ones; for example the ESS low beta structures, up to 90 MeV, are all normal conducting while the remainder (up to 2 GeV) are superconducting. (The difference between ESS and LEDA is the duty factor:  for ESS it is only 4\%.)
    
    But in any event the hardware cost of a 60 MeV CW linac, whether normal or superconducting will be substantially higher compared to a cyclotron.  In addition, the cavern space required for a linac would be substantially larger than what is needed for a cyclotron.

    \item {\it Other Options:~~}  Ref.~\cite{costeffective} covers several other options,  but two that are often raised are worth discussion here.   The first is a state-of-the art D-T generator.   These devices are reaching rates as high as $2{\times}10^{11}$ neutrons/s.   However, they are large devices that produce neutrons isotropically.    To reach $\bar \nu_e$ fluxes close to that of IsoDAR, an extremely large surrounding sleeve is necessary. This is very high cost and presents a much less compact source to the detector than the default design.  The pulsed nature of these generators is not  an advantage because the $^8$Li lifetime is 839 ms.    
    
    Secondly, one might look into an FFA (Fixed Field Alternating Gradient) structure. 
    While these devices have a faithful and dedicated following in the accelerator community, 
    the technology has yet to demonstrate anything close to the required beam intensity for our application.  
    A device at our required energy would also occupy a significantly larger footprint than a compact cyclotron.

\end{itemize}

\subsubsection{Why the Choice of H$_2^+$?}

If one wants high-current proton beams to strike the target, 
the particles accelerated could be either protons, H$^-$ or H$_2^+$.  
Early cyclotrons all used protons from the ion source; 
however it proved difficult to cleanly extract proton beams 
without substantial losses on the extraction septum.  
This came about because of incomplete understanding of the beam dynamics, 
and inadequate control over the magnetic field quality to ensure the clean turn separation
needed to avoid the septum.  
The resulting losses at the outer edge of the cyclotron, 
at its highest energy, produced very high radiation levels from activated components.  
This limited the highest currents one could safely accelerate, and still allow
hands-on maintenance.  
Another limit was space charge of high-current proton beams: 
increasing the number of protons in a bunch caused the bunch to grow in size
because of Coulomb repulsion of the protons.  
The bunch size is determined by the balance between this Coulomb repulsion and
the focusing forces in the cyclotron; if the bunch is too large the separation
between turns will become blurred and much beam will be lost on the extraction septum.
Only recently have properties of space-charge dominated beams, and the ``vortex motion'' described above that stabilizes these beams, been discovered.  However, use of this effect requires careful collimation of the ``halo'' of the beam bunches at low energies to realize the beam size control properties of this effect.

A major advance was the use of H$^-$ ions, injected through an axial channel 
and bent into the plane of the cyclotron by means of a spiral inflector.  
These ions behaved like protons (with the opposite charge, of course), 
except that to extract them all one needed to do was insert a thin carbon foil 
that stripped the two electrons off of the ion.  
The resulting bare proton, having the opposite charge, 
was bent away from the center of the cyclotron, and could be cleanly extracted.  
Activation levels in H$^-$ cyclotrons are considerably less.
All commercial cyclotrons today that are employed in fabricating radioisotopes accelerate H$^-$ ions.  
However, there is a limit to the maximum current in an H$^-$ cyclotron, due to the lifetime
of the stripper foils.  The problem is heating of the foil; not caused by the protons, but by the
electrons that are stripped off.  These ``convoy electrons" (so called because they have the same velocity as the
proton) are tightly bent in the magnetic field and pass through the foil again.  And again and again, 
until all their
energy is lost.  The electrons deposit many thousands of times more energy into the foil than do the protons.  
At beam currents of around 1 milliampere, the foil temperature rises to over 2000 degrees; at higher
currents the foils vaporize instantaneously.
H$^-$ ions have another problem, called Lorentz stripping.  The binding of the second electron is very weak
(around 0.7 eV), and if the velocity of the ion is very high the relativistic conversion of the 
magnetic field in the rest frame of the ion produces an electric field
that can cause the dissociation of the H$^-$ ion.  This becomes a problem for the ``normal" cyclotron fields 
of about 1.5 to 1.8 T and for beam energies above 70 MeV. (Higher energy cyclotrons, such as TRIUMF, must lower the magnetic field.  TRIUMF's is about 0.5 T.) 

H$_2^+$ ions, offer a significant advantage in space-charge mitigation.  
The ion has two protons bound by a single electron, so effectively, 
10 milliamperes of protons have only the charge of 5 mA.  
This makes it much easier to keep the individual bunches of particles small, 
and reduces loss from this cause.
The charge-to-mass ratio is 1/2 (compared to protons or H$^-$ which both have a Q/A of 1),
which means the particles are stiffer to bend in a magnetic field.  This requires a larger 
diameter of magnet, but this is a small price to pay for the factor of 10 higher beam
intensity.  
The binding energy of the ground state of the H$_2^+$ ion is about 2 eV, so Lorentz stripping 
is not an issue.  It is true that ions extracted from a conventional ion source have a 
population of vibrational states, but the fraction that is dissociated in the IsoDAR cyclotron field,
at 60 MeV, is less than 1\%.
It is surprising, but stripping extraction is still possible with these ions as well.
The protons emerging from the foil have lower rigidity so are bent towards the center.
However, the magnet geometry, as shown in Fig.~\ref{extract}, puts these stripped protons
on a path that cleanly exits the cyclotron.  Note that the ``convoy electrons" are not a problem 
as they are not bent back into the foil.
Nevertheless, the preferred extraction technique employs the original thin electrostatic septum.
This is made possible by a much better understanding of beam dynamics and control of the 
shape of the magnetic field,
by the space-charge mitigation using H$_2^+$ ions that limits beam growth from space charge, and by the 
aforementioned ``vortex effect" demonstrated in Fig.~\ref{vortex}. As stated also above, a narrow
stripper foil can be used to shadow the septum, further limiting beam losses on the septum.
For all these reasons we strongly prefer the H$_2^+$ ion over the other two.

A note should be added about the option of using deuterium beams.  While ions of this nuclear
species already carry a neutron that we are trying to make in our target, these neutrons
are easily released, and carry the very high beam velocity, whenever the deuteron strikes any material.
Practical experience is that the fast neutron fields around deuteron accelerators from beam losses
is very high
and preventing activation is extremely difficult.  
In general deuteron beams are avoided as much as possible.

\subsection{Beam Transport (MEBT) System}

In Figure~\ref{deployment} one sees the transport line connecting the cyclotron to the target referred
to as the MEBT (Medium Energy Beam Transport).

The H$_2^+$  ions extracted from the cyclotron are passed through a thin
carbon stripper foil that removes the electron, leaving two bare protons.
It is easier and safer to transport protons: beam losses can be better controlled.
The stripper and an analyzing magnet are placed close to the extraction point.  The analyzing magnet  
steers protons into the transport line, and detects any
H$_2^+$ passing unstripped through the foil, as a sure sign the foil is failing and needs to be replaced. 
This area will be the highest beam-loss point
in the MEBT, and sufficient shielding will be provided.

A note about shielding:  This question will be addressed in much further detail later; 
however, it should be made clear that while high-energy beam is being produced by the cyclotron 
there will be no access allowed anywhere
within the rooms housing the IsoDAR cyclotron, transport lines or target.  
When beam is off, the radiation level in these areas is determined by the amount of activation 
of beam-line elements and material in the the caverns from neutrons produced by interactions of errant beam particles.
Steady-state beam loss in the transport line, from beam halo and interactions with residual gas, can be calculated, and 
shielding, whether localized to specific areas or distributed along the beam line can be provided.  
Prior to access for maintenance, these areas will be surveyed by radiation safety personnel 
to ensure safe radiation levels exist.

Beyond the stripper stage, the proton beam is transported down the ramp to the target area.  
Standard transport elements are used:  quadrupole magnets for focusing and dipoles
for bending the protons. Beam position and profile are monitored by standard beam
diagnostic devices designed for high-current beams, and sensitive area monitors will pick up any unexpected
beam loss through rise in the ambient radiation level, so corrective actions will be taken by the operations staff.

Gas scattering is a source of beam loss that can be mitigated by maintaining the transport
line at a very high vacuum.  Though this source of beam loss is substantially reduced by
transporting protons instead of H$_2^+$ ions, it will still be the primary source of beam loss in the
transport line.  Once the design vacuum level has been specified, calculations will be performed to determine
if added shielding is required along the beam line, and if so this shielding will be provided.

Beam halo, a known phenomenon in transport of high-current beams, can also be calculated. 
Collimators can be introduced an appropriate locations to strip particles that would be lost
in the beam line.  Such collimation stages provide what is called ``controlled beam loss" points
that can be shielded efficiently, and avoid the loss of these particles in areas that might 
be more difficult to shield.

The beam line goes as close as possible to the large steel neutron-shielding plug and, as seen in Fig.~\ref{deployment} (b),
makes a 90$\degree$ bend towards the target, then a second 90$\degree$ bend into the target shielding.
The beam is now traveling away from the detector.  As seen in Figure~\ref{neutron_spetrum} this orientation
is highly advantageous for reducing the flux of the highest-energy neutrons directed towards the detector.
The beam is spread out with a wobbling magnet to paint the beam over the face of the target, making as even as possible the distribution of power over the target.

\subsection{Target System}{\label{target-section}}

\begin{figure}[tb!]
\centering
\includegraphics[width=3.5in]{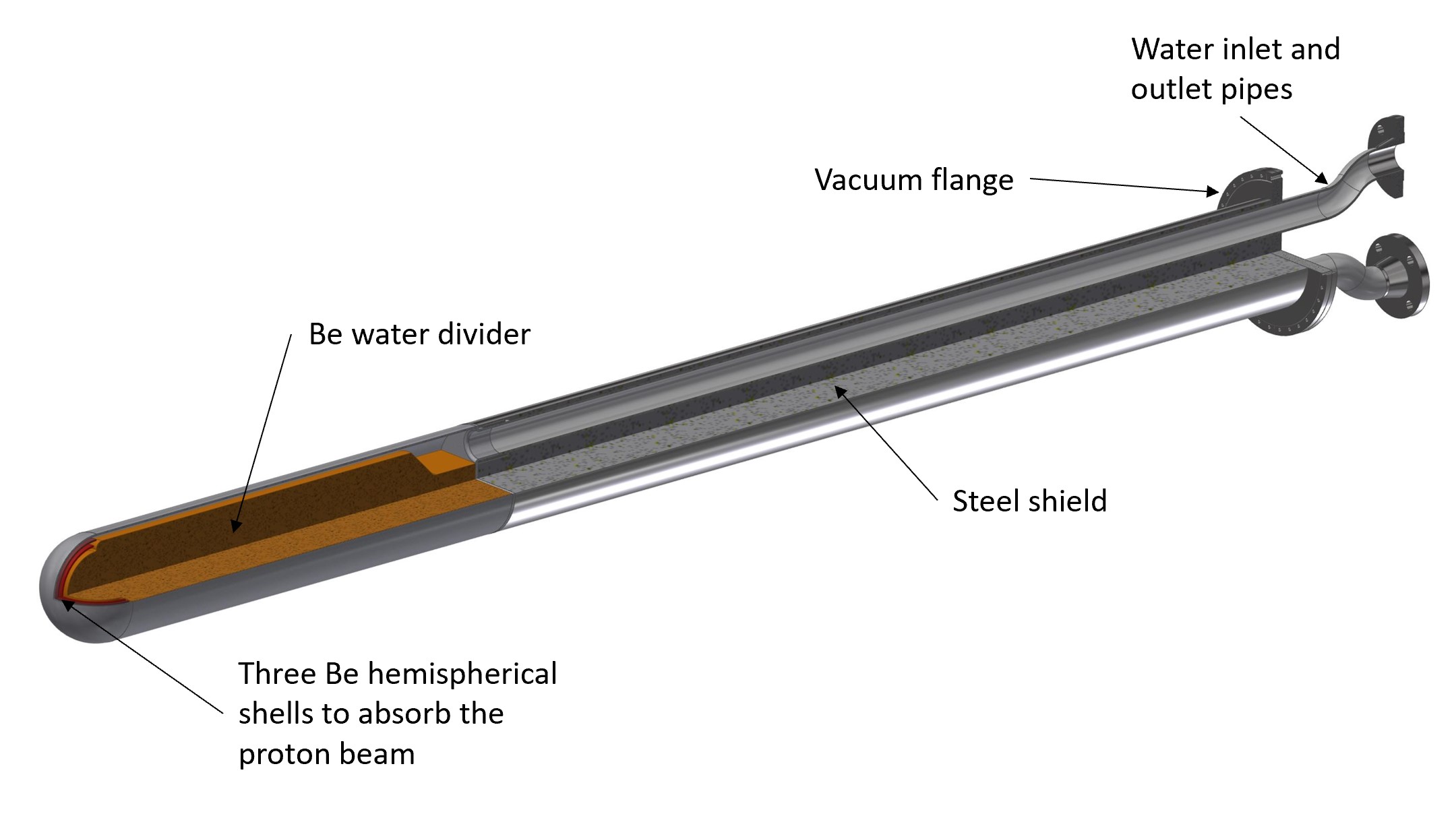}
\caption{{\footnotesize  Referred to as the ``Torpedo,'' the target assembly consists
of the beryllium surface hit by the beam, as well as the cooling system.  The entire assembly
can be easily detached from the cooling lines and removed into a shielded storage area.}
\label{torpedo}}
\vspace{0.2in}
\end{figure}

\begin{figure}[tb!]
\centering
\includegraphics[width=3.in]{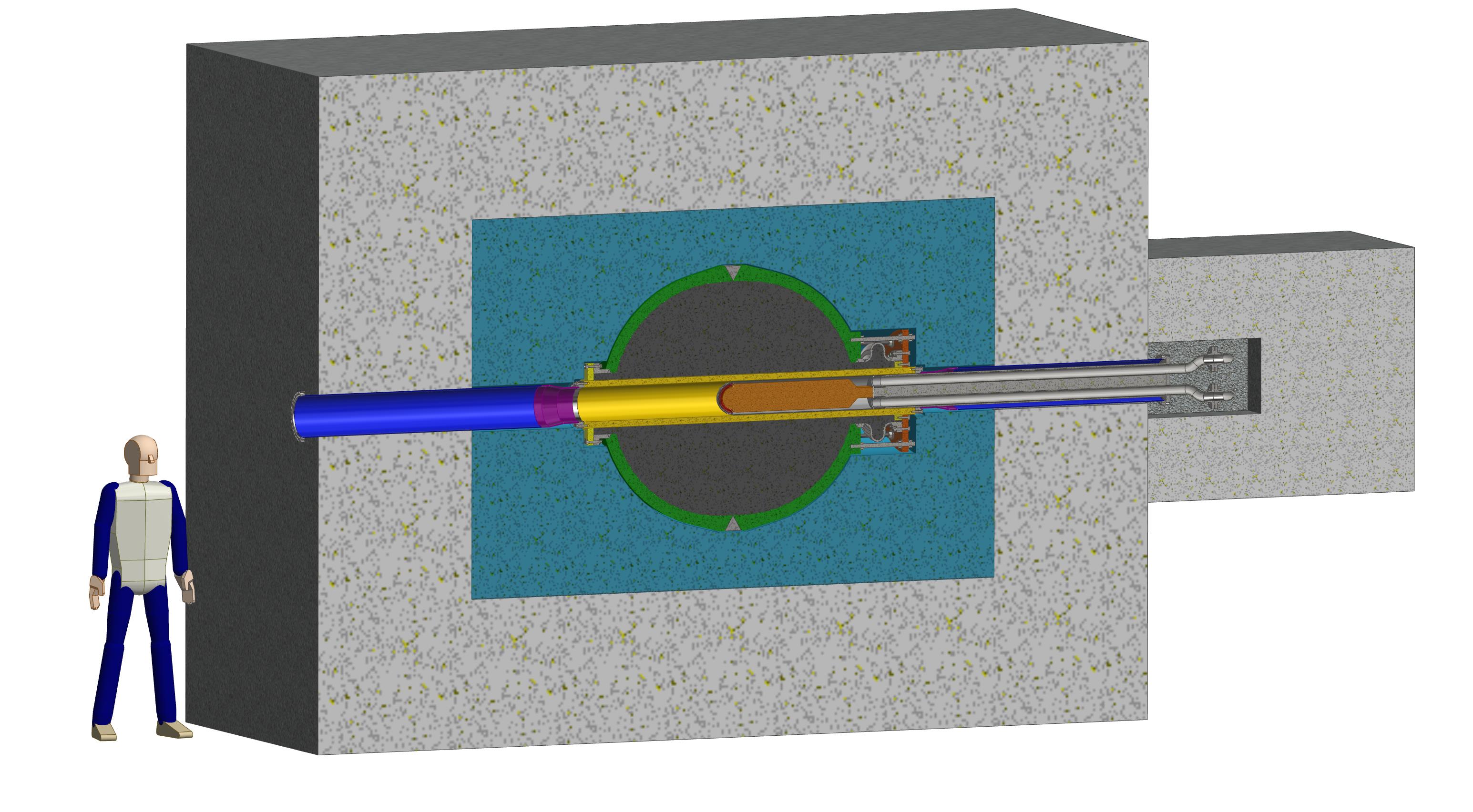}
\caption{{\footnotesize Section through the target system and the surrounding shielding. The target torpedo is surrounded by a Li/Be  sleeve pressure vessel. The shielding consists of inner layers of steel (shown in blue) and outer layers of boron rich concrete (shown in grey).}
\label{shielding}}
\vspace{0.2in}
\end{figure}

The target assembly, referred to as the ``torpedo'' (see Fig.~\ref{torpedo}), 
is designed for easy removal and replacement from the back of the target shielding
block.  Figure~\ref{shielding} shows the shielding structure surrounding the target 
and sleeve, with a 20 cm (inner diameter) vacuum pipe running the full length.  The 
inlet side brings the beam onto the target face.  The back side allows the torpedo
to be slid out.  
As mentioned earlier, the orientation of the target assembly in the hall faces away
from the detector (See Fig.~\ref{deployment} (b)), 
and so provides ample room for removal of the long torpedo assembly.
This will probably be done robotically because of high radiation levels.
The spent targets will be stored in bore holes drilled into the walls of the
target hall, as described in subsection 2.1.3 describing the Target Hall, 
and shown in Fig.~\ref{target_room}. 

\begin{figure}[tb!]
\centering
\includegraphics[width=3.in]{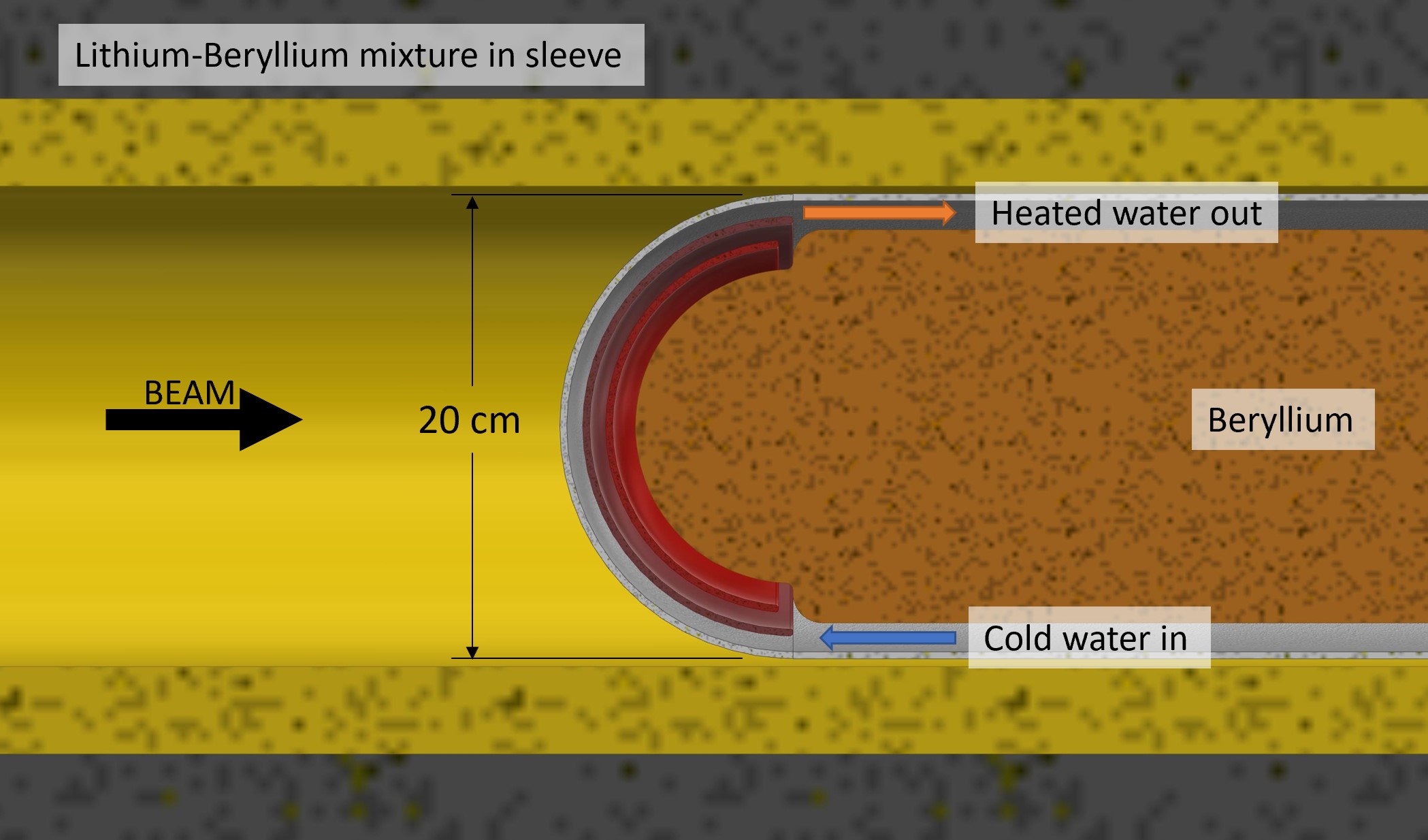}
\caption{{\footnotesize  Inside of the target vessel.  The beam is spread out over the face of the target by upstream wobbler magnets running at least at 50 Hz. Cooling is provided by circulating heavy water that also serves to generate more neutrons. 
  Neutrons stream
into the sleeve surrounding the target, consisting of a mixture of highly enriched ($>$99.99\%) $^7$Li 
and beryllium.  $^8$Li is produced by neutron capture.}
\label{target-01}}
\vspace{0.2in}
\end{figure}

The head of the torpedo, the actual target struck by the protons,
consists of three nested beryllium hemispheres
each having a thickness of 3 mm with a gap of 7 mm of heavy water between them, 
shown in Fig.~\ref{target-01}. 
Heavy water is used as the cooling fluid, as the absorption cross section for neutrons is
 lower, and there are added neutrons from the breakup of the deuteron.  
 These increase the neutron flux reaching the $^7$Li in the sleeve.
GEANT4 calculations indicate an almost 40\% improvement in the $^8$Li yield for
D$_2$O over H$_2$O.
Using heavy water does introduce tritium issues; however, the completely-sealed design of the primary cooling loop assures that tritium will not be released to the environment.  Mitigation of tritium will be considered as part of the decommissioning plan, where recovery and purification of the heavy water will be addressed. 

Water is circulated onto the target using a central divider plate made of 
beryllium, to separate the water flow.  
Water flows smoothly through the gaps between the beryllium hemispheres, 
Computational analyses of heat flow, temperature profiles and 
thermal stresses at the boundaries 
indicate acceptable performance of this target design, and that it is 
capable of handling the 600~kW of beam power.
The target geometry is still being optimized, specifically to address the known issue of
blistering of beryllium at the stopping point of the proton beam under 
intense bombardment due to the low solubility of hydrogen in beryllium~\cite{LENS}.

The target torpedo is surrounded by the sleeve, a roughly spherical pressure vessel
with an outer radius of 84 cm, and an inner radius of 74.4 cm.  
 Optimization studies have shown that a mixture of lithium and beryllium 
 with a beryllium fraction mass of 75\%  considerably increases
 the neutron yield and therefore the neutrino production.
 The sleeve is filled by pouring liquid lithium into the sleeve  pre-loaded
 with beryllium powder.  An experiment is currently underway to study
 the wetting properties of liquid lithium on beryllium surfaces.
 The sleeve vessel is rated at 
 2500 psi, 
 owing to the need to load the liquid lithium
 under pressure because of the significant decrease
 in volume between liquid and solid phases of the lithium.  Not doing so 
 would introduce unacceptable voids in the sleeve material.
 The $^{7}$Li isotopic purity assumed in our experimental rate determinations
 is 99.99\% although simulation studies 
 have shown that 99.995\% enrichment substantially improves the $^8$Li yield. 
 Even a very small amount of $^6$Li is damaging to $^8$Li production 
 because its thermal neutron capture cross section is a factor of 10$^3$ higher.
 We are told that the higher enrichment level is in fact achievable, and is the
 goal of several ongoing large-scale enrichment projects.

Beyond the sleeve is the shielding enclosure Figure~\ref{shielding}, consisting of steel 
cylinders and plates, with a nominal thickness of 30 cm, and boron rich concrete of minimum
thickness of 90 cm. Specifications for this shield are described in more detail in the next chapters.

%% file: Ch2-trans-instal-v1.tex
\chapter{Site Description of IsoDAR$@$Yemilab}

This chapter describes the location of IsoDAR within the Yemilab complex, addresses
the layout of the experiment and the caverns necessary for this.  It also
addresses many of the considerations for rigging and installation, and environmental
concerns, such as dealing with neutrons and activation.  Personnel safety is addressed,
and considerations related to access and maintenance.

\begin{figure}[t]
\centering
\includegraphics[width=5in]{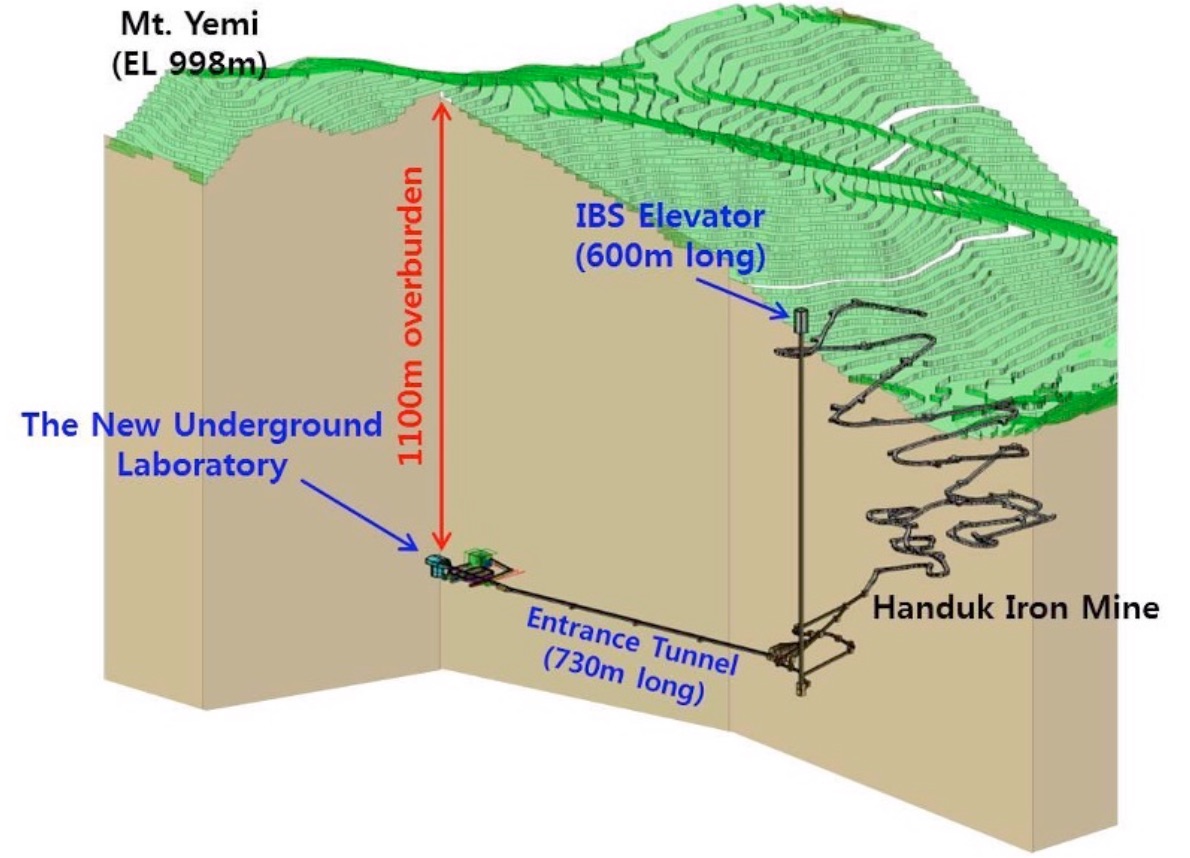}
\caption{{\footnotesize The location of Yemilab underneath Mount Yemi, adjacent to the Handuk Iron Ore Mine.  Access to the Laboratory level is via the 6.6 km mine ramp.  Personnel access is also available using the elevator shaft.}
\label{YemiLayOut}}
\vspace{0.2in}
\end{figure}

\begin{figure}[t]
\centering
\includegraphics[width=5in]{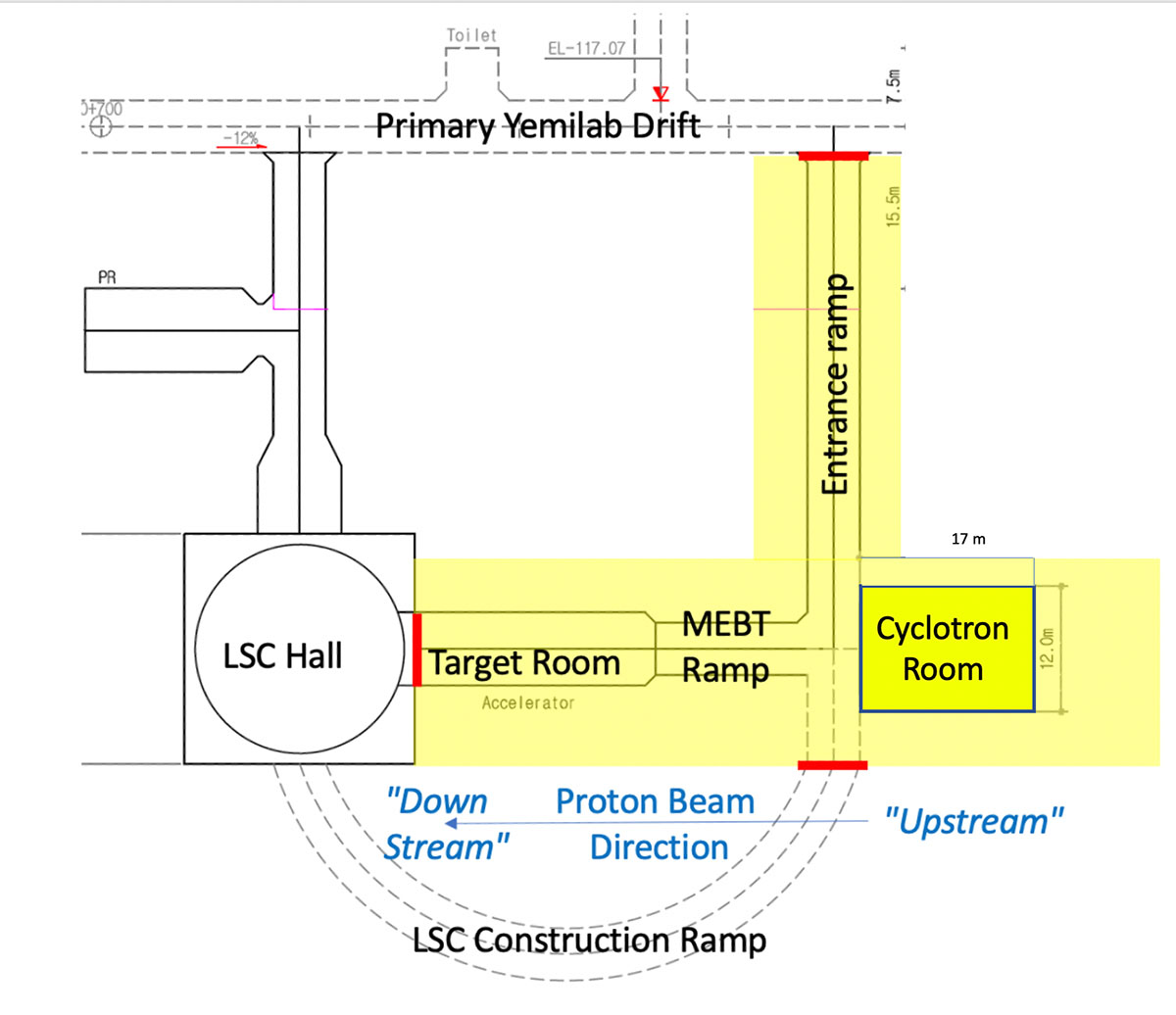}
\caption{\footnotesize  Layout of IsoDAR caverns at Yemilab.
These areas include the Entrance ramp, as the principal access to the IsoDAR experiment; this ramp drops about 6 meters
to the Cyclotron Room, then another 2 meters via the MEBT Ramp to the Target Room.  A large steel wall, for neutron shielding,
separates the end of the Target Room from the LSC Hall containing the 2.5 kiloton liquid scintillation detector.
\label{IsoYemiLayOut}}
\vspace{0.2in}
\end{figure}

The Yemilab complex is shown in 3-D in Fig.~\ref{YemiLayOut}.   
We will make substantial use of the mine ramp for transport of equipment.
This ramp is  6.6 km with slopes ranging from 
12\% to 15\%.  
Access to the Yemilab area from the base of the mine ramp and the elevator 
is via a drift 730 meters long with a 12\% slope.  
All drifts and ramps in the mine and Yemilab areas are nominally $5 \times 5$ meters, 
though some areas have more restricted cross sections 
because of ceiling-mounted ventilation trunks 
or other infrastructure needs.

\section{The IsoDAR Caverns}

The primary IsoDAR cavern spaces, shown in Fig.~\ref{IsoYemiLayOut}, consist of 1) the entry ramp; 
2) the cyclotron room; 3) the MEBT ramp; and 4) the target room.    
The ramps have $12\%$ slopes, while the floors of the rooms are flat.  
The floor of the cyclotron room is about 6 meters below the level of the
primary Yemilab drift, the target room is another 2 meters lower.
The target room connects to the LSC hall at the vertical mid-point of the LSC detector.

\begin{figure}[t]
\centering
\includegraphics[width=5in]{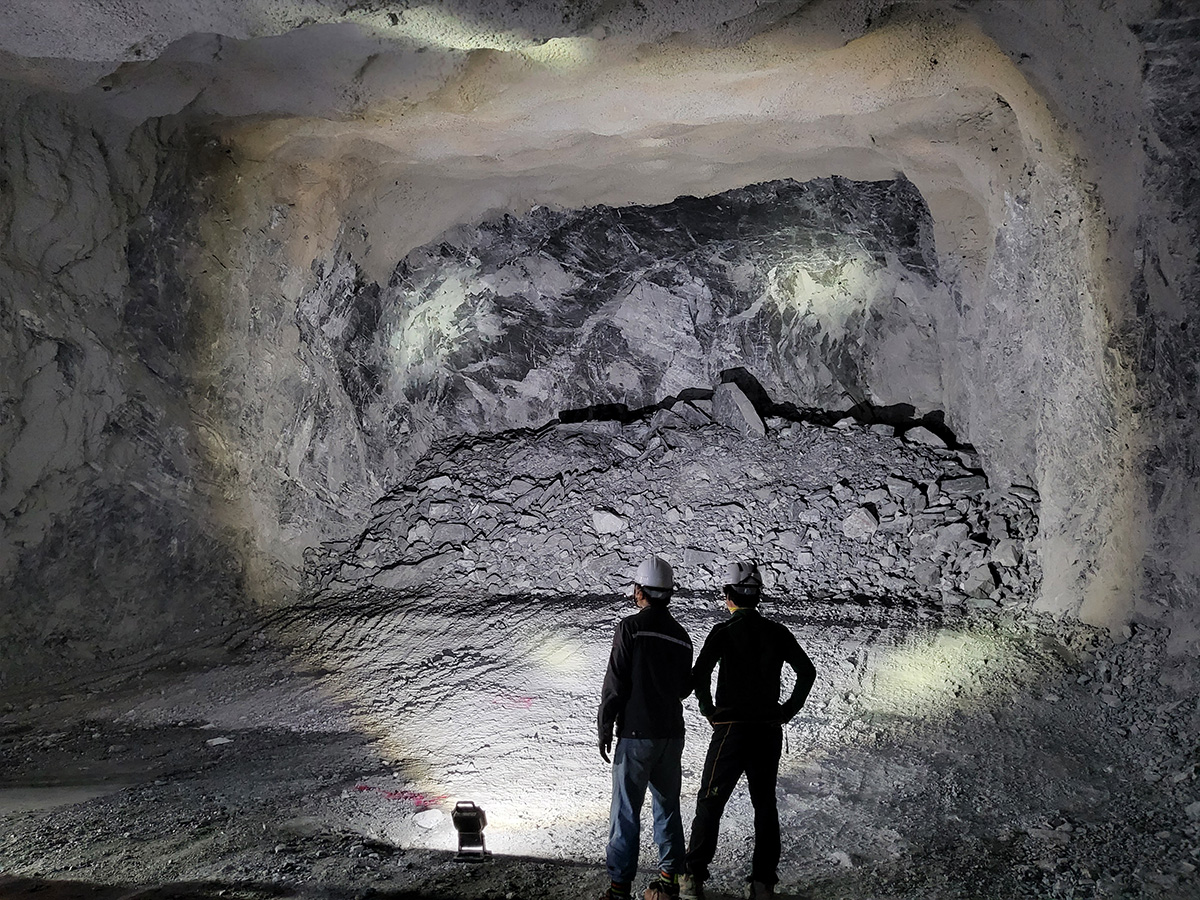}
\caption{{\footnotesize Photograph taken December 16, 2021 showing the status of the cyclotron room.  Blasting is almost complete.}
\label{cyclotron-room}}
\vspace{0.2in}
\end{figure}

The construction of the LSC Hall and the IsoDAR caverns are already well-underway, 
and will be finished this winter.  
This will complete all major construction at the Yemilab site
before the first of the experiments 
are installed at Yemilab, thereby minimizing exposure of these experiments
to dust and vibrations.
The connected drifts that form the Target Room,  
the MEBT Ramp and the Entrance Ramp are already in place, 
and are in use for removing rock from the top half of the LSC Hall.  
A semicircular ramp is being constructed to connect to the bottom level of the
LSC Hall, and will be used for removing rock from the bottom half of the LSC Hall.
The cyclotron room, indicated in deeper yellow, is almost complete.  Figure~\ref{cyclotron-room} shows status of this excavation in mid-December 2021.

Once construction of the LSC Hall is completed, the semicircular construction ramp
will remain open with access restricted.  
A stub of this ramp will be retained in the IsoDAR space 
and the floor will be leveled making this space useful for power supplies or other items.   
It should be noted, incidentally, that large mining trucks are being used for removing
rock from the construction sites, and that these trucks can navigate all the turns and 
sharp corners in the ramp and tunnel system.  This is 
good, because it is these
same trucks that will be bringing cyclotron components into the IsoDAR area for assembly.
As we will see, though, there are still some components that will not fit on the trucks, for which
special transport arrangements will be necessary. 

A shielded, and hermetically-sealed door will be installed at the top of the entrance ramp.  It will be
part of the radiation interlock safety system, and will allow the IsoDAR areas to be maintained at a slightly negative pressure to ensure airflow is always into the area.  Opening this door will
be disabled while beam is on.  However if the door IS opened (overriding the interlock), 
this action will
automatically interrupt the beam.
There will be other labyrinth baffles and blocks installed in the Entrance Ramp
to ensure no migration of
neutrons into the body of the Yemilab area.

Above the IsoDAR cavern space is the Yemilab primary drift, which has a nominal 5~$\times$~5~m cross section.  
The Entrance Ramp to the IsoDAR area joins this primary drift at a 90\degree angle.
This ramp is
45 m in length, with also a $5\times 5$ m cross section. It descends at a 12\% slope
placing the Cyclotron Room about 6 meters below the level of the main Yemilab drift. 
The entrance ramp ends on a 5 m (w) $\times$ 5 m (l) $\times$ 5 m (h) flat space
that meets with the cyclotron room and the MEBT Ramp.  
This area is the principal
staging area where components of the cyclotron are delivered, and are then rigged into
their proper places for the assembly of the cyclotron.
Components destined for the Target Room area are brought around the corner
and are staged at the base of the MEBT Ramp.
The current design of the 
Cyclotron Room to the left at the bottom of the ramp 
is 12 m (w) $\times$ 17 m (l).  

In this design, the ceiling height is 10 m for the
back 12 meters, and slopes up from 5 meters at the front.
The beam will travel through the MEBT (Medium Energy Beam Transport) line 
from right to left in Fig.~\ref{IsoYemiLayOut}, 
hence the Cyclotron Room is referred to as ``upstream'' while the Target Room is ``downstream.'' 
The MEBT ramp is 5 m (w) $\times$ 5 m (h) $\times$ 16.5 m (l).  
It descends at a 12\% grade so the floor of the target room is 2 meters below the floor 
of the cyclotron room.  The MEBT will require vertical bending magnets to follow the contour
of the spaces.
The target room is 7 m (w) $\times$ 22 m (l) $\times$ 7 m (h).     
The features of the Target Room are described in more detail below.

The finish of the walls, ceiling and floor of the IsoDAR area will be planned
to coordinate with plans for the rest of Yemilab.  However, the requirements for 
the IsoDAR are somewhat different.  While the main laboratory areas stress 
cleanliness and low-radioactivity materials, IsoDAR areas require careful
choice of materials to minimize induced radioactivity from the large neutron
backgrounds generated by the experiment.  One known bad-actor is sodium.
As will be noted below, the sodium content of the rocks in the Yemilab area is very low.
This is excellent, as it minimizes both short-term induced activity from 15-hour
$^{24}$Na (from slow neutron capture on the only naturally-occurring isotope of sodium,
$^{23}$Na, and also the long-lived (2.5 year) $^{22}$Na isotope produced by fast
neutrons via an (n,2n) reaction.  But it is important that materials brought in to
the caverns are also low in sodium.  This relates to finishes on the cavern walls
and floor in particular.

We will discuss staging during construction at several points in this CDR, but here we briefly consider the general order of component installation.   

\begin{figure}[t]
\centering
\includegraphics[width=4in]{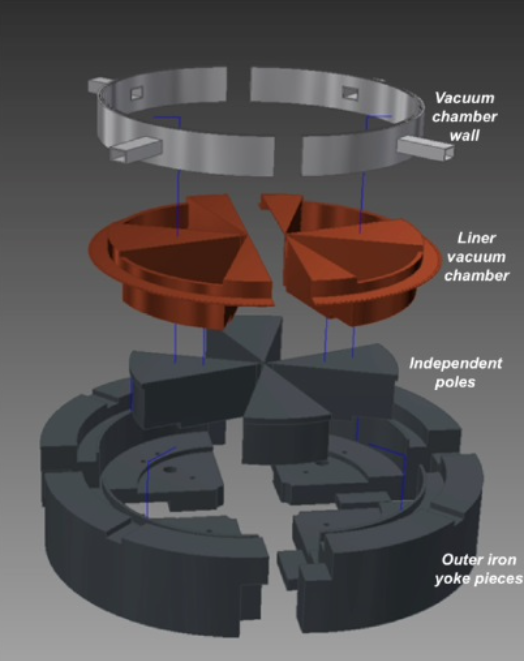}
\caption{{\footnotesize  Bottom half of cyclotron, split into parts.  An equal number
of parts are added to the top.  As the total weight of the cyclotron is over 450 tons,
the cyclotron is assembled in its designated operations location.}
\label{CyclotronParts}}
\vspace{0.2in}
\end{figure}

The first pieces to be installed will be the large iron shielding wall, 
shown as the large green block in Fig.~\ref{deployment} at the end
of the target hall.  This wall provides shielding to prevent neutrons and gammas
from reaching the detector.  
As will be discussed later, the maximum weight of the blocks making up this wall will
be determined by conveyance equipment and rigging strategies.  
The pieces will be stacked and welded in place.  The total volume of
this wall is about 200 cubic meters, and has a weight of about 1,600 tons.
It will be important to ensure the rock at the lip of this cavern is properly reinforced
to support this weight.

Next will be the cyclotron.  The cyclotron itself has a 6 meter diameter,
is about 2 meters high, and weighs about 450 tons. It must be
brought in pieces and assembled at the site.
Here too, the maximum size and weight of individual pieces will be
determined by the manufacturer,  by constraints in the shipping
and transport to the site, and by the rigging capabilities in the 
Cyclotron Room itself.  
Fig.~\ref{CyclotronParts} shows just the bottom half of the cyclotron.  
Each piece will be brought down to the staging area at the base of the
Entrance Ramp, and lifted into its assembly place.  
Referring to Fig.~\ref{RFQ} one can see the fully-assembled cyclotron.
Note the yellow jack on the side of the cyclotron steel.  
Six of these spaced around the outside circumference of the cyclotron
steel are capable of lifting the top half of the cyclotron up to a full meter,
allowing for access to the interior of the machine for maintenance.

Following the cyclotron will be the shielding for the target.  As seen in Fig.~\ref{shielding}
this structure is made up of large blocks of
special concrete and steel pieces, specified to prevent neutrons from the 
target (as many as we can make to maximize the $\bar \nu_e$ flux)
from leaving this shielding, to minimize activation of components in the cavern.
Then follow the beam line components, target and sleeve, power supplies, and support equipment.  
The MEBT will probably be last to be assembled, as it crosses the Entrance Ramp
and so blocks convenient access to the area.

Utilities for IsoDAR will probably be located in a common area, 
designed to service the needs 
of the whole laboratory.  Details of size and location of this area have not yet 
been specified.

\section{The Cyclotron Cavern Layout}

\begin{figure}[t]
\centering
\includegraphics[width=5in]{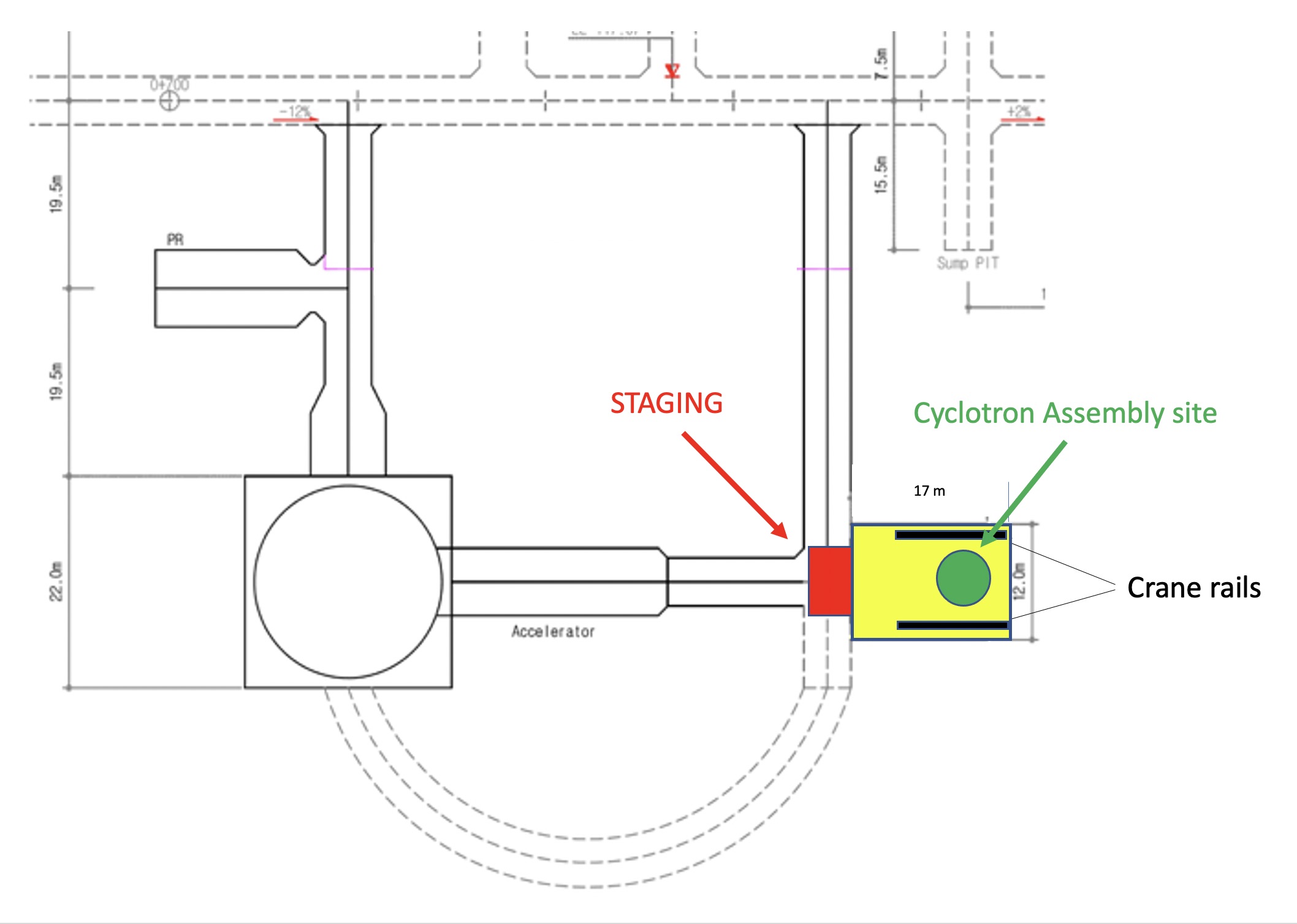}
\caption{{\footnotesize  Plan view of the Cyclotron Cavern.  Note the staging
area at the base of the Access Ramp, and the approximate location of the cyclotron.}
\label{Cyclotron-staging}}
\vspace{0.2in}
\end{figure}

\begin{figure}[t]
\centering
\includegraphics[width=5in]{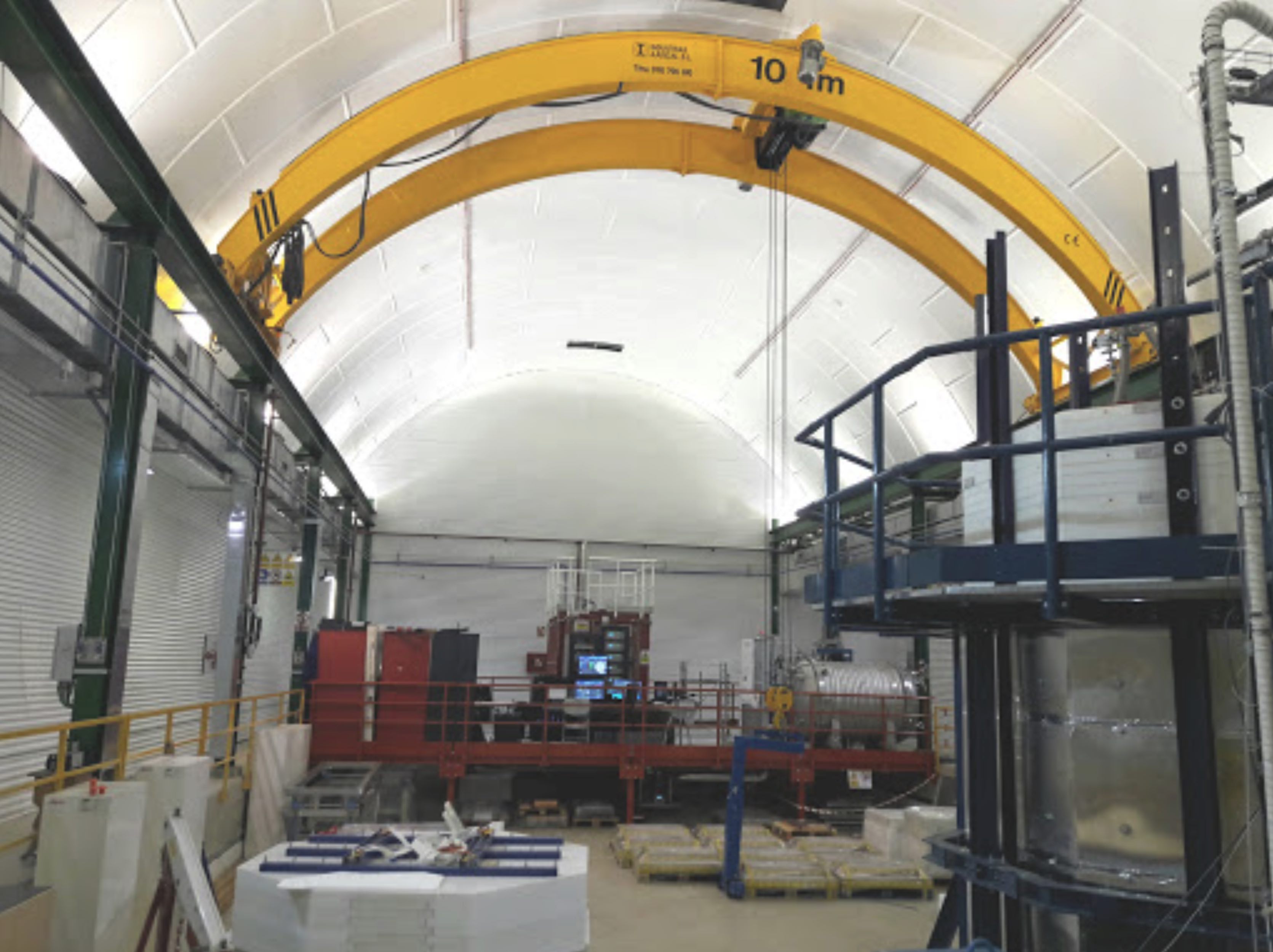}
\caption{{\footnotesize  Photograph of the interior of the main hall at Canfranc, the underground laboratory in Northern Spain
underneath the Pyrenees.  The crane, believed to have a 10 ton capacity, elegantly serves the rigging functions in
this laboratory.  Such cranes would serve well for the rigging needs in the IsoDAR cyclotron and target rooms.}
\label{Canfranc}}
\vspace{0.2in}
\end{figure}
The size of the Cyclotron Cavern must be adequate both for operations needs 
as well as for assembly and installation of all the technical components.

For operations, the cyclotron is quite compact, at 6 meters' diameter.
There are some critical elements, such as the final RF amplifiers that need to be relatively
closeby as the high-power lines between these and the cyclotron Dees must be as short
as possible.  Vacuum pumping equipment must also be nearby. 
Other elements can be located elsewhere, possibly even outside
the radiation control areas for easy access and maintenance.  Cable trays will provide
the pathway for connection of these units and the cyclotron.  The MEBT runs between
the cyclotron and the target, its components require relatively little space, 
with the exception of the stripper area that will require substantial shielding.
This will be the highest beam-loss point in the transport line.
However, this too will probably be the only area within the cyclotron cavern requiring
shielding.  Concerning shielding in the Cyclotron room, personnel protection is provided by exclusion
from the area during operation, and environmental protection requirements are
quite modest; specifically, we will see in a later chapter that
rock activation and ground-water management are almost
surely not issues of concern.
In addition, the bulk of the cyclotron itself provides a very large
amount of shielding from beam lost inside the machine. 
This topic will be revisited in a later chapter.

However, the largest constraint on the Cyclotron Cavern will be the needs for assembly
of the cyclotron.  Pieces of all sizes and weights must be transported to a staging area
in this cavern, and lifted into position for assembly of the 3-D jigsaw puzzle that is
the cyclotron.  The size and weight of these pieces is a major concern, and at
the present time we do not know what these will be.  In fact, the manufacturer will need
to establish these parameters based on many factors, 
of which limitations from shipping and transport are probably the major contributors.  
What we know of these will be discussed below.
The elements affecting the extremes of these limits are the cyclotron magnet coils: 
two rings each
4.95 meters in diameter and 20 cm high; and the magnet yokes and poles, totaling over 400 tons
of steel in total.  We know the magnet steel must be made in pieces, but the maximum size
of these pieces is the question.  This again is determined by manufacturing efficiency 
and cost
(calling for fewer pieces, but heavier) or by shipping limits and 
rigging capabilities (size of crane).
Bringing down the magnet coils may present a challenge because of aperture restrictions along the access route from the surface. If they cannot be transported in a single piece, they will need to be wound underground, or 
shipped in two halves.  Either of these would present very difficult problems and
would add significantly to the cost of the project.
Every effort
should be made to find a way of having the coil wound in the factory and shipped as 
a single piece.

Considering these uncertainties, specifying the dimensions of the cyclotron cavern 
becomes a challenge.  We have picked dimensions which we believe are adequate for
all the above requirements:  17 meters deep, 12 meters wide, and 10 meters maximum height.
The floor-space provides at least 3 meters of space all around the periphery of the cyclotron,
and the ceiling height could accommodate a bridge crane, or a reasonably-sized 
gantry crane.  In addition there is a substantial space adjacent to the cyclotron
where large (but light) parts, such as the vacuum liner, or the RF Dees, could be 
assembled, then moved to and installed into the cyclotron.

The top of the room will not be flat.  For rock stability, roofs of caverns are always
dome-shaped, and are reinforced with rock bolts driven many meters into the rock to 
maintain compression in the rock and prevent collapse.  Figure~\ref{Canfranc} shows
an elegant solution to installing a crane in such a cavern.  The rails are at 6 meter height
along the edges of the domed cavern, and the bridge is shaped to follow the curve of the dome.
The structure of the bridge could be lighter because of the strength of the arch shape. 
It is also possible that the trolley motor may not have to work harder:  if the hook is
kept at the same height by letting cable out while the trolley is moving up the arch, in principle the system 
is doing no work.

Such a crane could prove of immense value in assembling the cyclotron, and
for maintenance activities.  
However, this crane will not be able to reach to the staging area where the trucks bring loads.  
This requires the full ceiling height to extend over the staging areas, which was not an
option for the design of the cyclotron room.
As will be discussed in the transportation section, a separate hoisting system will be used
to offload the pieces off the truck bed and transfer them to a fork lift or similar
device to move them under the crane.

Another definite requirement for the cyclotron cavern will be a very strong floor.
The cyclotron weighs about 450 tons, and its 8 support legs must rest on a solid
foundation.  The manufacturer will specify this base, probably a very large-area, and very thick
steel plate that must be set in the floor of the cavern.

\section{The Target Cavern Layout}

\begin{figure}[t]
\centering
\includegraphics[width=5in]{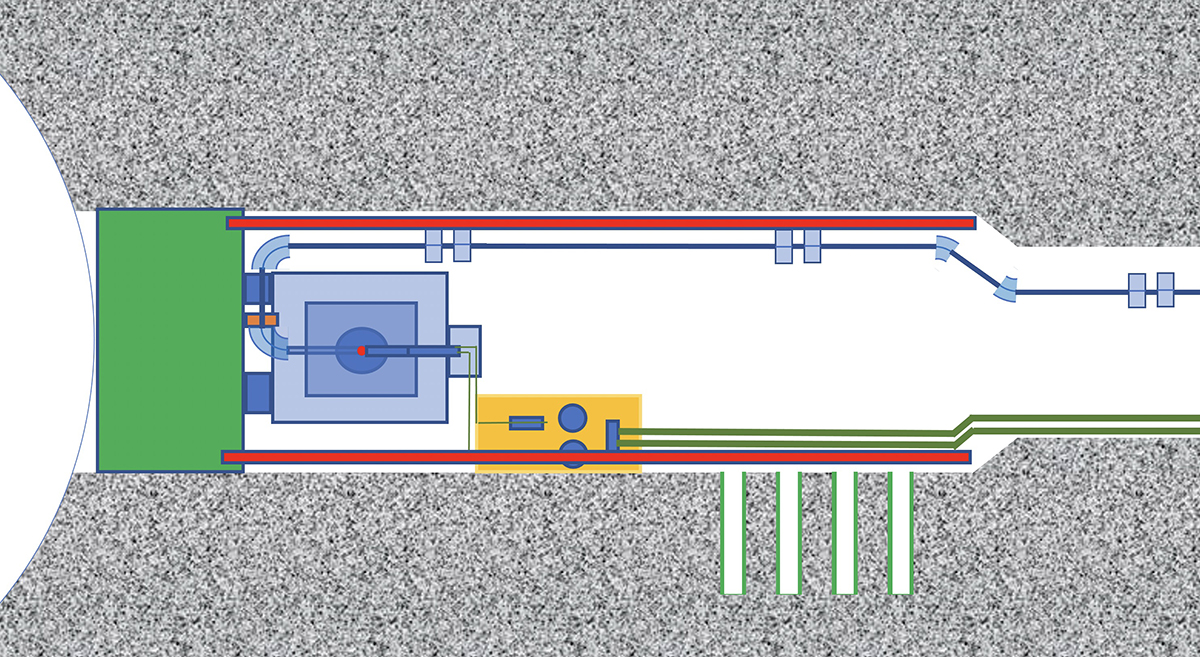}
\caption{{\footnotesize  Plan view for the conceptual layout of target. The MEBT beamline is seen along the top edge of the plan.  Beam enters from the right and is directed to the target (with shielding) region at left, in blue, next to the shielding wall indicsated in green.  The orange rectangle indicates a utilities-skid.  Red lines indicate crane rails. Bore-holes for spent targets are indicated at lower right.    See text for further explanation. 
 }
\label{target_room}}
\vspace{0.2in}
\end{figure}

A conceptual layout of the target cavern is shown in Fig.~\ref{target_room}.  This
figure shows the beam line bringing the 60 MeV protons to the target in the center of the large shielding block
of iron and boron-loaded concrete.  The protons produce neutrons that are moderated as they flood into the roughly-spherical sleeve
surrounding the target.  Neutrons are captured by the $^7$Li to produce $^8$Li that beta decays in less than one
second to produce the isotropic $\bar \nu_e$ flux.  The geometry pictured provides a solid angle of $\sim$0.2$\pi$ steradians,
or roughly a 5\% efficiency for the $\bar \nu_e$'s produced to reach the fiducial volume of the detector.
This is the downside of the`Decay-at-Rest" concept, that the desired particles are emitted isotropically.
It also indicates that one must place the source as close to the detector as possible.

The MEBT beam line provides dipole magnets to bend the beam (around corners and down the ramp) and quadrupole magnets
to focus the beam.  Detailed beam-optics calculations will identify the growth of beam halo, which will lead
to beam loss along the transport line.  This halo comes from gas scattering as well as from 
non-linearities in the transport itself.  Locations will be identified (not shown on the figure) where collimators
will be placed, these collimators will be surrounded by shielding structures to capture the neutrons produced
from cleaning up the beam halo.  This process is known as ``controlled loss" because one provides localized shielding 
for absorbing the neutrons produced in eliminating these halo particles.

Fig.~\ref{target_room} provides details of the MEBT layout, where the beam enters from the right.    The beam line (dark blue) with transport magnets is shown along the ``near side" (side closest to the main Yemilab drift); where the rectangular pairs are quadrupole focusing magnets.  The 45\degree wedge-shaped bending magnets 
 are canted to accommodate the 12\% slope of 
 the target-room access ramp.  The two 90\degree bends closest to the target area bend the beam so it strikes the target
 going away from the detector, decreasing the fast neutron flux aimed at the detector.  The dark orange element is 
 a wobbler magnet to sweep the beam over the target surface.  
 
 The MEBT brings to beam to 
 the red half-circle in Fig.~\ref{target_room} that is the beryllium + D$_2$O target in the
 center of the  ($\sim$spherical)(Be + $^7$Li) sleeve producing the $\bar \nu_e$ flux.  
 The large green block in Fig.~\ref{target_room} is the $\sim$4 meter steel shield blocking neutrons and gammas from the LS detector.

Cooling for the target is provided by the pumps, filters and heat exchangers shown on the yellow skid in Fig.~\ref{target_room}. As the 
(heavy) water
circulating here comes directly from the very high radiation environment of the target area, 
it must be treated as RAW (RadioActive Water) with suitable isolation and QA procedures to
ensure complete containment and zero leaks.  
If major maintenance of these components is required,
the entire skid can be placed in a shielded container and removed to the surface to a suitable hot cell area
where repairs or component replacement can be safely conducted.
Pipes with water from the secondary cooling system transfer heat from the primary heat exchangers; eventually
this heat is transferred to cooling towers on the surface.  (This is discussed in a later section.)  
The heat exchangers between primary and secondary systems provide complete separation between the two systems, 
so there is zero contamination of any radioactive material into the secondary system. 
If there are concerns from the Yemilab management in this regard, it is certainly possible to place more
water circuits in series to provide further isolation of the cooling system for the IsoDAR target and the
main cooling system for the Yemilab experiments.

A very important element of the Target Cavern is provision for target changes. 
Over the 5 years of IsoDAR operation, targets will need to be replaced.   We are drawing on
experience at high-power-beam installations (Fermilab, SNS, JPARC, ISIS
for example) in developing target-changing procedures.  Space within the hall is adequate for this operation.   
Spent targets must remain within the target space for several years to cool before transport to the surface.
For storage, we are proposing that a series of $\sim$25 cm diameter bore holes be drilled in the wall of the cavern.  These are
shown schematically in Fig.~\ref{target_room} as the four perforations in the lower wall, right.  
It is important that a sufficient number of these storage repositories are provided, and
we make a preliminary recommendation of 12 (3 vertical layers of these 4 holes shown), but this will be the subject of further review.
The inner surfaces need to be smooth, to ensure that a spent torpedo will not snag while being
inserted.  For this reason we suggest that the bore holes should be lined with a ($\sim$2 cm) steel pipe.  
This pipe will also serve as 
added gamma shielding.  As the hottest part of the torpedo is at the deepest end, this pipe and the roughly 2
meters of rock will provide the shielding needed to maintain adequate radiation levels in the Target Cavern.
For added protection, it is suggested that the bore holes be made deep enough to accommodate a steel plug of maybe 
half a meter that blocks
the entire opening of each bore hole.

A medium-capacity bridge-crane system will ease construction and can be used during target-changes.
 The capacity of this should probably be in the 10-15 ton range.  
 The crane rails are shown in red in Fig.~\ref{target_room}.  The ends closest to the detector can be supported on the steel wall, but will need columns
 along the walls of the cavern to provide further support.  The crane shown in Fig.~\ref{Canfranc} is an example.

\section{Environmental Considerations and Personnel Safety}

Many of these topics will be considered in greater depth in later sections, but we will summarize
these here for completeness. Several other topics can be addressed here and need no further discussion beyond what
is presented here.

The first point is an overall assessment of the IsoDAR environment in the context of Yemilab.
Yemilab is a modern, state-of-the-art underground laboratory, the hallmarks of which are cleanliness
and exceedingly low background radiation levels. The IsoDAR spaces are certainly not a low-background area,
so it is important to ensure that the IsoDAR spaces are suitably isolated from the clean Yemilab areas.
This relates to personnel access, barriers for leakage of neutrons, and environmental management of air
and water to prevent migration of any activated material out of the IsoDAR areas.

\subsection{Cleanliness Boundary}  

It will be necessary to establish where the principal entrance to the Yemilab area is, 
that is, the
area that is the interface between ``dirty" and ``clean."  It would probably be best if this dividing
point were to be located inside the main steel door entrance to the IsoDAR area.  There is no need to have
special cleanliness conditions for access to IsoDAR and, on the contrary, personnel exiting the IsoDAR
area should take special precautions before being allowed into the main Yemilab ``clean'' area. 
To take the example of SNOLAB, entry into the Yemilab clean areas might include requirements for 
showering and changing clothing into protective outer ``bunny suits'', gloves,
hair nets and masks.  Certainly changing footwear is the most basic precaution.

\subsection{Personnel Access to IsoDAR} 

There is only one way for ingress and egress from the IsoDAR area, via the Access Ramp.
The entrance to this Ramp, at the location of the main Yemilab backbone drift, will be blocked by a Radiation
Safety gate that cannot be opened when the beam is on.  Or said another way, it is interlocked so that
if it is opened the beam will be shut off.  This will ensure that there will be no personnel exposure to 
``prompt'' radiation.  

\subsection{Personnel Radiation Protection}  

If interlocks protect personnel from ``prompt'' radiation, there remains
``residual background'' radiation that must be mitigated.  First of all, all personnel accessing the IsoDAR area 
must be trained and classified as ``radiation workers'' and carry proper dosimetry devices for monitoring exposure.
As residual activity will all be gamma radiation, there is no need to monitor for neutron exposure.
The normal procedure for access will be that a Radiation Technician will be the first person into the area
following beam shut-off, to monitor the radiation environment in the areas.  This will include not only the
general ambient radiation level, but also the identification and characterization of 
specific areas that may be hot, from stray beam particles striking
vacuum pipes, diagnostic devices or collimator slits.  The technician will mark these areas and record levels
observed, and place barriers or portable shields around these places to limit exposure to personnel that will
be working in the area.  The technician may also impose limits on the time that personnel are allowed to remain inside
the IsoDAR areas, or may mandate time necessary for the radiation levels to drop to levels where access can be allowed.
Personnel entering these areas must be cognisant of the radiation levels, and plan their work to minimize
the time they are in the radiation fields.

Note that while the ``cleanliness'' criteria for access to the Yemilab clean areas may not apply to IsoDAR, 
it may still be necessary to wear protective clothing, masks, hair and shoe coverings to enter the areas, to ensure 
that no radioactive contamination is brought out of the IsoDAR areas.

\subsection {Personnel Protection:  Underground Environment}  

As noted above, the beginning of the Access Ramp is the
only ingress and egress point to the IsoDAR area.  It is a mandatory requirement for underground spaces
that either there be always a second ingress-exit path to all occupied underground spaces, or there must
be a ``refuge chamber'' where personnel can be protected from fire, or air deprivation in the event of
an accident between them and the only egress point.  Space must be allocated in the IsoDAR area for 
such a refuge chamber.  This chamber must have communication capabilities, as well as emergency food, water
and air supplies, and be isolated from the external air in the event of fire or oxygen deprivation in the
cavern air supply.  In fact, another restriction on access to IsoDAR areas must be that no more persons
are allowed in the area than there are spaces for in the refuge chamber.

\subsection {Neutron Shielding} 

While the shielding around the target, and in other areas where neutrons might be produced,
is designed to minimize the number of neutrons escaping to the environment, the attenuation factors from
the shielding in the immediate vicinity of the places where neutrons are generated 
will almost surely not be at the level needed to meet the stringent standards established for the main Yemilab labortory spaces.  
It is possible to bring these levels down; for instance the added $\sim$4 meter steel shield between the target
and the LS detector is calculated to keep the neutron flux in the fiducial volume to less than a few fast
neutrons per year of running.  This is below the level calculated for fast neutrons from cosmic ray muon
spallation at the Yemilab depth.

Similar calculations and measures must be taken to ensure that no neutrons from the IsoDAR area reach the very sensitive
experiments deployed at Yemilab.  This must be done carefully, but is not difficult to assess.  For starters, the smallest
thickness of rock between any of the IsoDAR caverns and the nearest experiment is about 50 meters.  Attenuation
of neutrons in this thickness of rock certainly places neutron transport through the solid rock
at below the cosmic-ray background.
But neutrons do have a nasty habit of working through cracks and air spaces. However, the flux is
attenuated every time a neutron encounters
a ``bounce'' surface, as seen in the effectiveness in labyrinth designs placed between neutron sources and 
outside environments.  There are very good codes for calculating neutron transport, MCNP, PHITS, etc.  that will 
be used to evaluate the location of labyrinth blocks in the Access Ramp to ensure that neutrons cannot reach
the main Yemilab drift, and bounce into experimental areas.  If need be, movable barriers can be inserted in
the Access Drift to totally block the path.  We are confident the most stringent neutron-attenuation goal can be met.

\subsection {Air Management} 

In areas where high beam currents are transported, a concern is air activation from fast neutrons.  
This is much more of a concern where beam energies are in the GeV range or higher; however, in the interest of complete
safety an analysis of air activation in IsoDAR from the 60 MeV protons must be conducted, 
and mitigation measures proposed to
ensure this is not an issue for the Yemilab environment. 
Typical products are light isotopes of carbon, nitrogen and oxygen that all have very short halflives.  
The typical mitigation measure is to transport the air to the outside, and provide long paths where transit times 
are longer than the activity halflives, so when the air reaches the outside environment
these activities have decayed.  
But in our case we wish to make sure that any air-activation products do not reach the Yemilab low-background
air supply.  

While mainly the air supply provided to underground areas brings clean air into the area with ventilation shafts
or trunks, and allows exhaust air to be pushed back out to the outside through corridors and shafts, 
the IsoDAR air system should be designed
such that the air drawn out of the area is captured and pushed through conduit trunks to the surface, and specifically
prevented from circulating through other Yemilab experimental areas.  In addition, maintaining a differential pressure, so the air inside the IsoDAR area is always at a slightly lower pressure than in the main laboratory, will help ensure that air from the IsoDAR area will not escape to the main Laboratory area.

One concern of air activation is $^7$Be, which has a long (53 day) half life, and condenses out on 
surfaces, so is picked up on clothing.  This is perhaps the principal reason why personnel entering
the IsoDAR area after a period of beam running might need to wear disposable clothing, gloves, hairnets, etc. 
and that these items are disposed of prior to re-entering the main Yemilab area.
To assess the extent of a potential $^7$Be risk, careful assays need to be
performed following the first few sessions of high-power running, at which time appropriate safety and containment strategies can be developed. 
The possibility exists that there will be {\em no} $^7$Be detected.
However, it is necessary to measure this beforehand, and not risk the
consequences of overlooking this potential hazard.

\subsection {Water Management}  

In many underground environments, ground water is a major concern, to control 
humidity levels, and prevent damage to equipment from high moisture levels. 
Extensive pumping systems are needed, as well as measures for channeling 
this water away from experiments.  For example Gran Sasso has had to line all of its caverns with heavy
polyethylene sheets to channel water seeping through the ceiling from dripping onto their experiments.  Similar
measures have had to be taken in sensitive areas of the Kamioka Observatory.  SURF in South Dakota has a
well-documented inflow of groundwater to the extensive Homestake mining complex of over 2500 liters per minute.
In areas where neutron fluxes 
are present, this ground water can be activated, which could prove to be a serious environmental hazard if this
water eventually finds its way into a drinking water aquifer.

At Yemilab, the absence of ground water in the immediate surrounding area 
has led to very dry limestone formations, alleviating this concern to the
point where ground water management will not be an issue.


However, cooling water is used to dissipate the heat deposited in the IsoDAR components. Two very different
cooling circuits are needed.  The first is for the target, which generates about 600 kW of power.  This circuit uses
heavy water as a cooling fluid, and as it is exposed directly to the proton beam, it amplifies the neutrons available
for the neutrino experiment.  But it also produces tritium, and will likely become activated due to trace impurities
in the water.  So this water loop must contain special filters and tritium-mitigation stages.  It is 
 a true ``RAW"---RadioActive Water---system.  
The second system cools the cyclotron and beam-transport magnets, and the RF amplifiers and Dees. These 
components dissipate between 2 and 3 megawatts.  As they are cooling electrical components, there is a strong
requirement for low electrical conductivity, or an ``LCW'' system.

Both of these ``primary'' systems must be coupled through heat exchangers to a (possibly common)
``secondary'' system that either
goes directly to a cooling tower on the surface, or to a central area where water loops from other experiments are
collected to run (in parallel) to a central heat exchanger.
As the IsoDAR secondary system will be running inside the IsoDAR caverns, there is a potential that its water
may become very slightly activated.  This should be evaluated carefully, and if necessary another heat-exchange
station be installed outside the IsoDAR areas to guarantee that water running outside the IsoDAR area
carries absolutely no activated material.

\section{Principal Shielding Considerations} 

The purpose of shielding in the IsoDAR area is two-fold: protection of
a) personnel  and b) the environment. 
Shielding must address both prompt radiation, mainly neutrons, but also x-rays and
gammas that are present when the beam is on; and residual background 
radiation from material that is activated by neutrons or beam particle interactions.
In fact, as personnel are excluded from the area while beam is on, shielding design should really
focus on ensuring that background radiation levels are not high when access is required for maintenance
activities.  
This is a complex problem, involving first containment to the greatest extent possible 
of neutrons produced in the target and from
beam loss during the acceleration and transport stages, but also management of materials that are
exposed to neutrons that are not contained.  For example, it is well known that sodium content in materials
such as concrete, should be kept as low as possible.  $^{22}$Na and $^{24}$Na, the former produced from high energy 
(n,2n) reactions (11 MeV threshold) with a 2.6 year halflife and the latter with a 15 hour halflife produced by slow
neutron capture, are both responsible for significant contributions to background activity.  $^{22}$Na plays 
an important role in long-term management of radioactive contamination, sodium content in rocks contributing
to problems during the decommissioning phase of an underground experiment such as IsoDAR.  $^{24}$Na produces
a gamma background that may limit the ability to rapidly access the IsoDAR caverns for maintenance activities.

We are truly fortunate that the limestone rock in which Yemilab is located is extremely low in sodium content,
and as we will see in a later section, this allows us to reduce the shielding bulk around the target to manageable
levels.  We must further analyze the details of shielding around the cyclotron and beam lines with regards to 
neutron production from beam loss in these areas.  In addition we must take great care in specifying the shielding
material used to ensure that this shielding material is not itself contributing to the background gamma levels.

Let us look a little more closely at the areas we must consider, following the beam from its generation to the target.  

\subsection {Cyclotron}

The cyclotron itself is a huge block of iron
that provides a significant amount of self-shielding.  First of all, note that no neutrons are produced
until the beam has at least 5 MeV/amu of energy.  So the ion source, RFQ, and first stages of acceleration
are of no concern regarding neutron production.  There will be beam losses inside the cyclotron; the design of 
the beam optics plans for the greatest part of the losses to be at low energies, below the Coulomb barrier, close
to the center of the machine.  A series of collimators will intercept particles that have orbits that would lead
to their losses later in the accelerating cycle, where such losses would produce neutrons.  Stopping them at lower
energies prevents neutron production.  However, there will be  losses throughout the accelerating cycle
due to gas scattering, present even
though we will have the highest-possible vacuum inside the machine.  Extraction will also generate losses, 
though these can be kept very small by careful design and management of the beam dynamics. As stated,
the cyclotron steel will provide good shielding for neutrons produced from these losses inside the machine.
However, there are perforations in the steel, for the RF dee stems, for vacuum feedthroughs, for the injection and
extraction ports, possibly for radial probes and other diagnostics.  These openings must be analyzed one by one
and shielding covers designed that can limit the leakage of neutrons to the outside.

\subsection {Stripper}

The stripper is the next major source of neutrons.  The beam is extracted from the cyclotron as H$_2^+$ ions
but it is much more efficient to transport protons.  The H$_2^+$ ions will easily dissociate with any residual
gas in the beam pipes, leading to beam losses throughout the transport line.  Protons will still scatter on the 
residual gas molecules,
but the amount of particle loss from protons is many orders of magnitude less than for the H$_2^+$  ions.
The stripper stage will be located very close to the extraction point, so will be in the Cyclotron Room.  It
consists of a thin carbon foil, usually 200 $\mu$gm/cm$^2$ ($\sim$1 $\mu$meter) thick placed in the beam.  This is thick
enough to dissociate 100\% of the ions, but causes almost no disruption to the beam quality or extra losses.  Experiments
at PSI have demonstrated that such foils in a beam line with no magnetic fields can survive several hours in our
10 mA proton beam ~\cite{PSIstripper}.  This foil is placed just upstream of an analyzing dipole magnet, that
will bend the protons into the entrance of the MEBT (Medium Energy Bean Transport) line where they are conducted
to the neutrino-producing target system.  If there are any unstripped H$_2^+$ ions exiting the foil these will
be bent less in this magnet and can go into an instrumented beam dump.  Any ions reaching this point will indicate 
the imminent failure of the stripper foil, and will notify the operators that it is time to slide a new foil into
the stripper position. (Usually there will be foil holders with as many as 50 foils on a ladder so they can be
changed remotely, and very quickly.)  There will be some losses in this area, at the stopper, for instance, though 
the current here will only be high for a brief period when the foil is failing.  Any material in the beam will
involve multiple
scattering, and will lead to some growth in the beam size.  The entrance to the MEBT will have collimators that scrape
the halo produced by this multiple scattering.  Thus, the areas around the beam stop, and the collimators must be
provided with shielding to absorb neutrons produced from the interactions of the stopped particles.

\subsection{Transport Line}

The transport line consists of focusing quadrupole magnets and bending dipoles, so the line can follow the contour
of the caverns and be directed to the target.  As indicated before, transport of high current beams is always accompanied
by the growth of halo particles around the core of the beam.  These particles are at a much higher risk of having their
orbits grow to where they will strike the walls of the vacuum tube.  The customary procedure is to place collimators
along the beam line to intercept these particles before they hit the walls.  These collimators constitute what 
is referred to as ``controlled'' loss points, and can be surrounded by (localized) shielding, thus preventing
the need for shielding along the entire transport line. On approaching the target region at the back of the Target Room,
the beam line takes two 90\degree bends and is directed towards the target.  
Of necessity the target shielding has a hole of 20 cm diameter necessary for the beam to reach the target.  This hole
is also a direct path for neutrons produced in the target directed backwards to escape.  As much shielding material as
possible needs to be placed around this hole.  The primary shield here will be the large steel wall plugging the entire back of the Target Room, and designed to be thick enough so that the neutrons leaking into the Liquid Scintillator detector are below background levels from cosmic and other sources.

\subsection{Target}

The  target shield must attenuate as much as possible the neutrons produced in the target.  As the
entire experiment is predicated on maximizing this neutron flux as it floods the $^7$Li sleeve, we must
design the shield around the target and sleeve to contain all those neutrons that do not contribute to
$^8$Li production.  The detailed design of this shield is described in a following chapter. The bottom line
is that a structure with a minimum of 30 cm of iron and 90 cm of a special boron-containing concrete 
will most likely provide adequate containment of these neutrons.

%% file: Ch2-rest.tex
\chapter{Transport and Installation}

This chapter will address the steps necessary to bring the cyclotron, other technical
components, as well as shielding materials to the Yemilab site, and the
process for bringing these components underground to the IsoDAR area.  It will
also address the steps necessary for installing these components and possible means for doing this.
The following represents a preliminary plan, and further study of transportation on rail, highway, Handuk Iron Ore Mine rampway, and within Yemilab, to understand all possible limitations, is planned.

\section{Transportation to the Handuk Iron Ore Mine}

Technical components that cannot be produced or acquired in Korea will arrive
via shipping at a yet-to-be-designated harbor.  Transport will be arranged to the 
Yemilab area, located in the vicinity of the Handuk Iron Ore Mine in the Gangwon province of Eastern Korea
(about  200 km ESE of Seoul).  
Transport could either be by rail or by public highways.  
Many components can be procured within Korea, and shipped in a similar fashion.
Bulk materials such as shielding blocks of concrete or iron
will also be procured within Korea.

It would be very convenient to identify a possible warehousing facility with easy access to the mine 
for temporary storage of components as they arrive, to allow
flexibility in planning for moving components underground. 

While most of the units to be shipped fall within normal size and weight
limits established for shipping, the IsoDAR cyclotron has several specific
components with unusual
requirements that must be accommodated for shipping and handling: 
the magnet steel, the magnet coils, and the possibly the vacuum liner.

\subsection{Magnet Steel Transport}

\begin{figure}[t]
\centering
\includegraphics[width=5in]{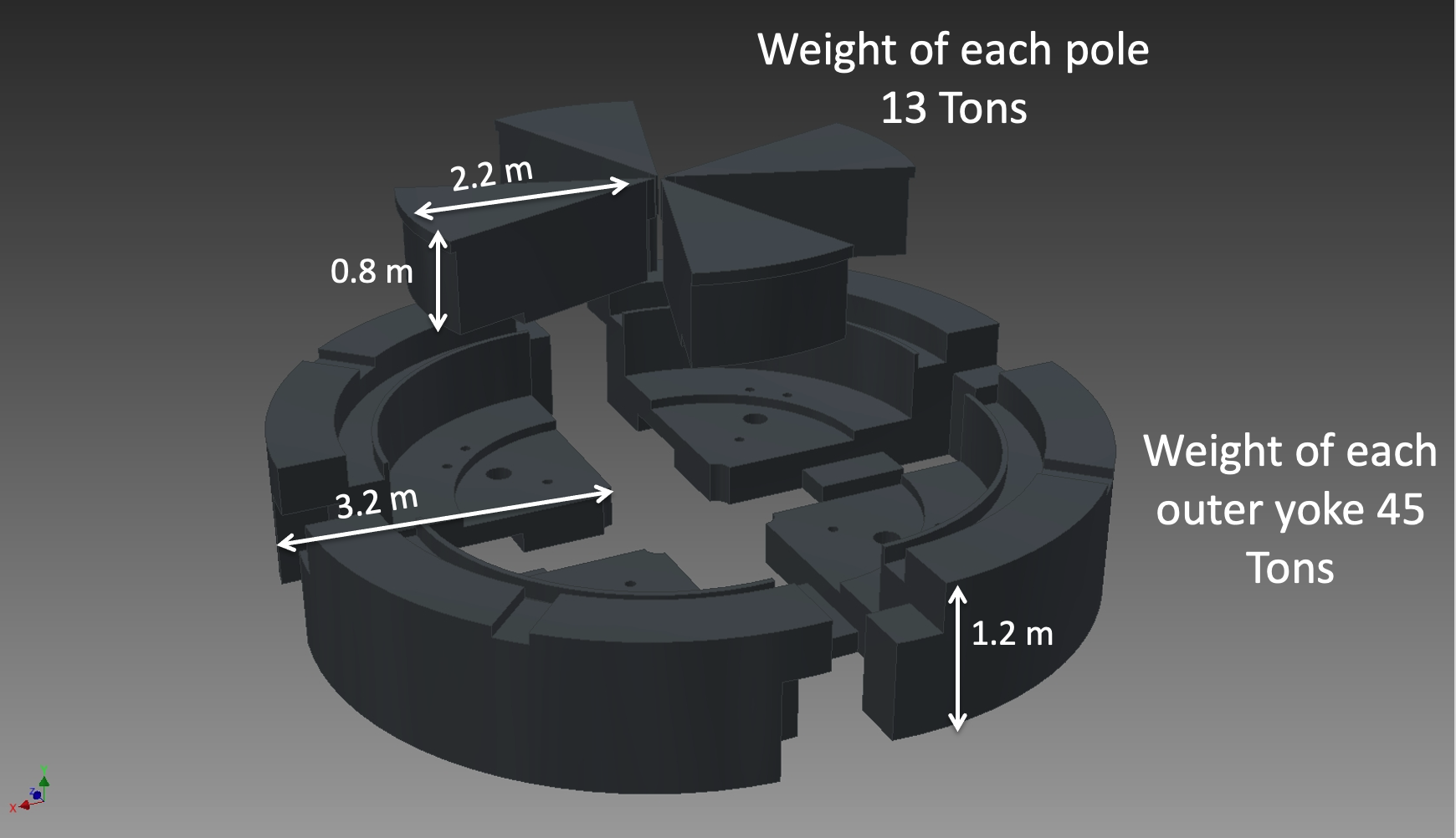}
\caption{{\footnotesize Bottom half of cyclotron steel yoke and pole pieces showing sizes and weights.  The top half
of the cyclotron is a mirror image of the bottom.}
\label{CyclotronSteel}}
\vspace{0.2in}
\end{figure}

The magnet steel challenges the weight limits.
The total magnet weight of steel is in
excess of 400 tons and is 6 meters in diameter (by 2 meters high).  
The magnet will be built in several pieces. Each will be transported separately,
and will be assembled in the location where the cyclotron will rest permanently during operation.  
Once assembled, the cyclotron 
will not be moved. 
Fig.~\ref{CyclotronSteel}
identifies the various pieces of the (bottom half of the) magnet steel.
The number of magnet steel pieces is established in the design and manufacturing process. 
The pieces must be assembled into the final structure with extremely precise tolerance---a
real 3-dimensional jigsaw puzzle.  There is a tradeoff between making a smaller number
of larger pieces, or many smaller ones.  From informal communications with IBA (in Leuven-la-Neuve, Belgium), 
one of the leading cyclotron builders, we learn that their largest milling machines can
accommodate pieces as large as 40 tons.  On the other hand, they can make pieces smaller,
but the cost may be higher because of the larger number of precision surfaces, and the assembly process 
may be more difficult.  The most realistic alternative from the preferred (40 ton maximum)
configuration would probably be to have no piece
larger than 10-15 tons.

Transportation of heavy steel pieces may not present a problem.  The limit
for rail loads appears to be 50 tons.  Truck limits are probably less, but one way
or the other the steel pieces can be shipped to the entrance to the mine.
If rail transport is used, one problem that must be faced is off-loading of the rail pieces at the rail station
closest to the Handuk Iron Ore Mine (about 5 km away).  There are no crane facilities with
this capacity available---a crane would need to be provided.  
We have been assured that it is possible to use the mining trucks that are used to bring ore 
to the rail station.  
These trucks could transport the pieces to the afore-mentioned warehouse for temporary
storage, or directly to the underground site through the Handuk Iron Ore Mine Ramp Way.

As we will see later, though, the weight limit for steel pieces will almost surely be 
determined by the 
underground rigging capabilities.  This will be discussed at length.

\begin{figure}[t]
\centering
\includegraphics[width=5in]{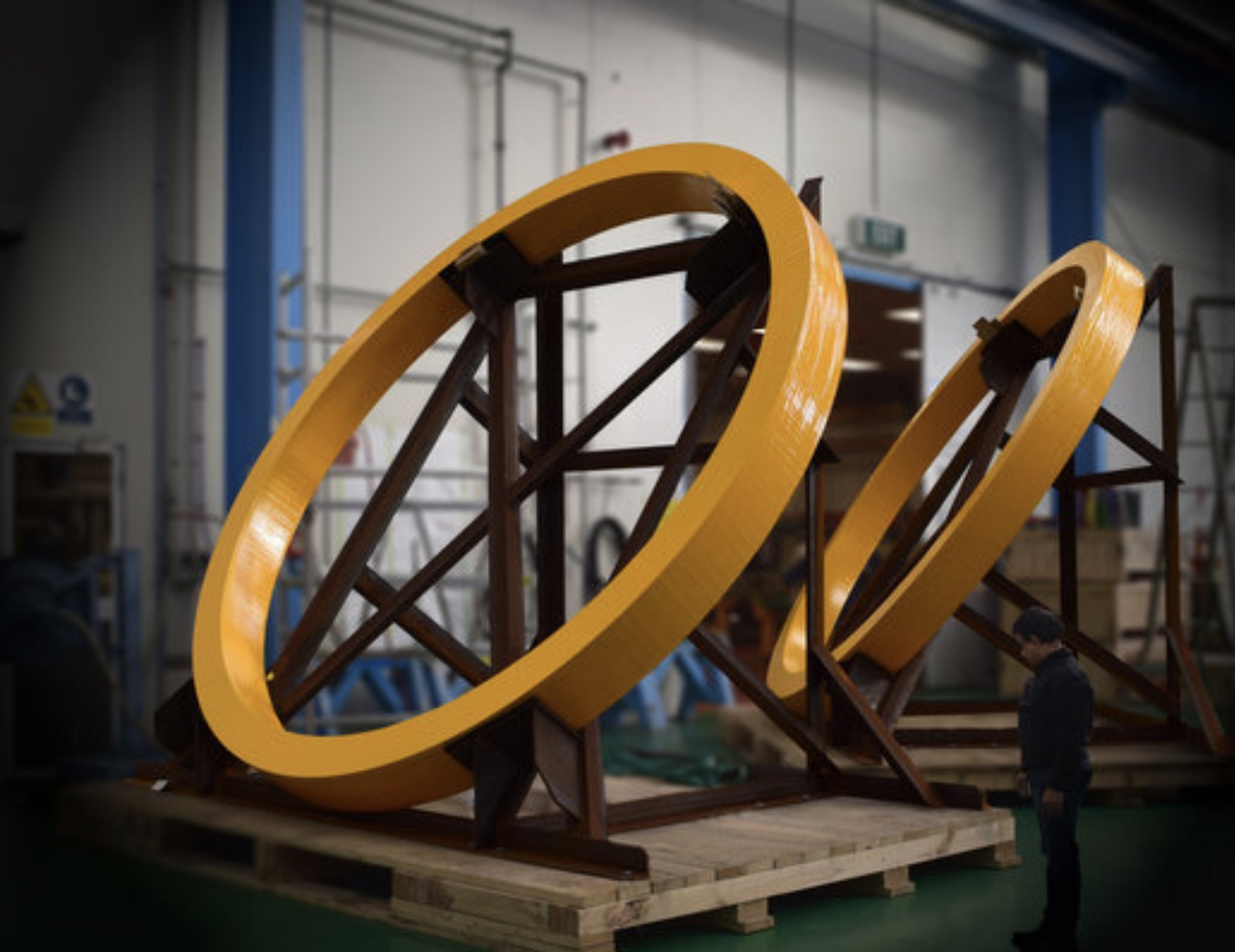}
\caption{{\footnotesize Photo of cyclotron coils built by Buckley Systems, Auckland NZ.
These coils are 3.33 meters diameter, ours are 4.95.  The proportions are about the same as ours. Notice the man in lower right.}
\label{coils}}
\vspace{0.2in}
\end{figure}

\subsection{Magnet Coil Transport}

A greater challenge will be posed by the magnet coils.
The two magnet coils are toroidal assemblies each 4.95 meters outer diameter, 20 cm high, and
4.40 meters inner diameter.  For approximate scale, see Fig.~\ref{coils}. Each of our coils weighs just over 1 ton.  
We are told the maximum load
size for a rail car is no more than 3.4 meters wide by 3.5 meters high.  The diagonal of
such a box is 4.88 meters, just not quite enough. And, if you include the 20 cm
coil thickness, and the need for a container, which will increase the width of the package even more,
it is clear the coil will not fit on a rail car.

On the other hand it may  be possible to ship the coil by truck over public highways.
It would be a ``wide load" and would require special permissions, but it may be feasible.
However, as is the case with the magnet steel, more severe limits may be imposed in transporting
the coils underground.  This too will be addressed below.

\begin{figure}[t]
\centering
\includegraphics[width=5in]{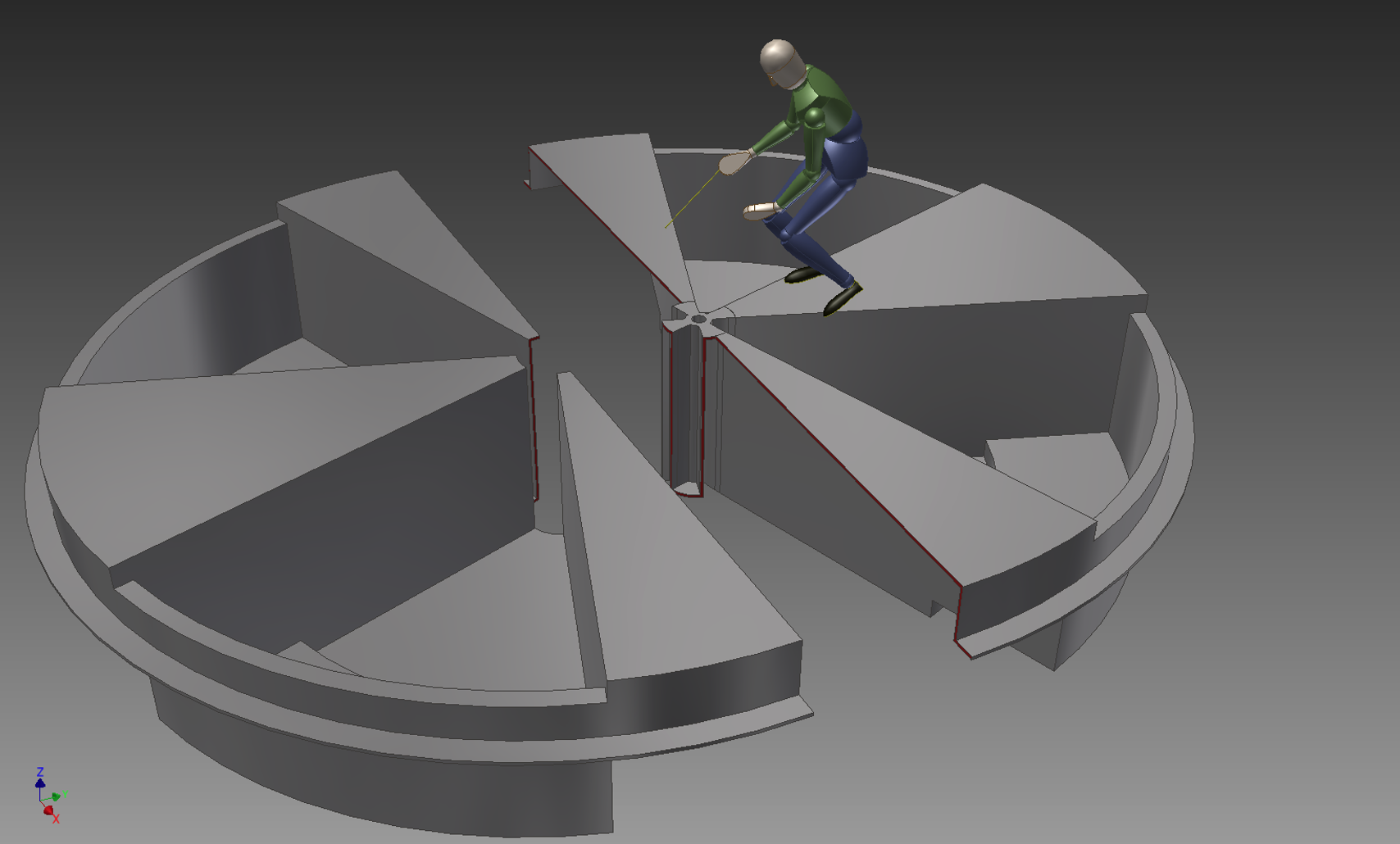}
\caption{{\footnotesize Bottom half of the vacuum liner (top half is a mirror image), made from stainless steel
sheets.  One option is to bring pieces and weld them on site, but it may be more efficient to do at least some
of the welds at the factory and ship larger sub-assemblies.}
\label{vac-liner}}
\vspace{0.2in}
\end{figure}

\subsection{Vacuum Liner Transport}

The vacuum liner halves are large and somewhat delicate, not heavy, pieces of welded stainless steel 
that are  smaller in diameter than the coils (just over 4 meter) but are about 90 cm high.
Fig.~\ref{vac-liner}  The manufacturer may suggest the stages of assembly that should be
performed in the factory, but this too must be folded into shipping possibilities and access size limits.

\section{Transportation from Surface to the Yemilab site}

This section will cover the process for delivering pieces from the surface warehouse or
staging area through the entire underground transport process, including
the mine ramp, and Yemilab access drift, to the most relevant staging area in
the IsoDAR caverns.

\begin{figure}[t]
\centering
\includegraphics[width=5in]{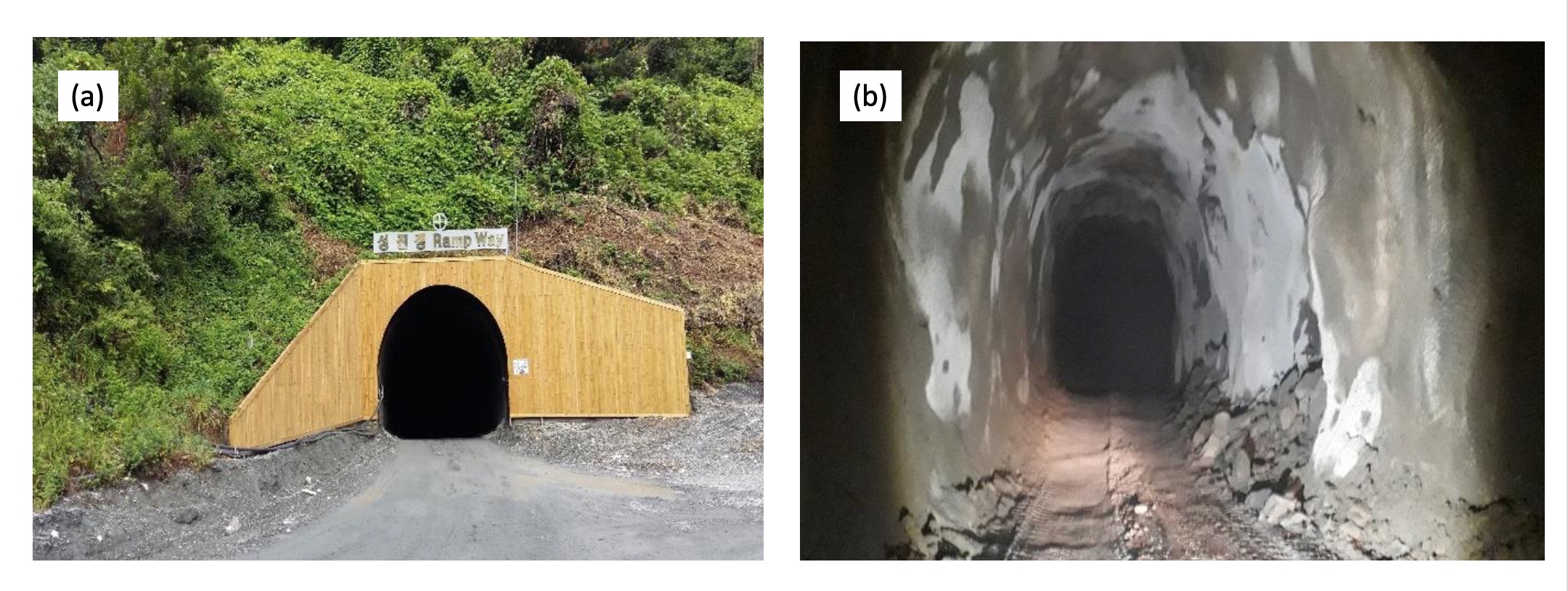}
\caption{{\footnotesize Photos of the Handuk Iron Ore Mine ramp.  (a) the adit, entrance way, (b) internal ramp section.  
Nominal cross section is $5~\times~5$ meters, but is highly variable along the 6.6 km length. Most important
is that the mining trucks can pass through.}
\label{adit}}
\vspace{0.2in}
\end{figure}

Figure~\ref{adit} shows the adit (entrance) to the Handuk mining ramp (a) and a photo
of one of the internal cross sections of this ramp.  It is rough, but certainly functional.

The Handuk Iron Ore Mine has a fleet of mining trucks, used primarily for transferring ore to the
rail outlet. 
The mining truck beds are 
7 m long, 3 m wide at base, and have 3 m clearance to the ceiling. 
These trucks fit through the $\sim$ 5 m $\times$ 5 m cross section of the drifts and can round all corners
of the very circuitous mine ramp shown in Fig.~\ref{YemiLayOut}.
Important too is that these trucks can navigate the 90\degree bends in 
the Yemilab tunnel system, so can deliver pieces directly to the unloading points in the IsoDAR areas
The weight capacity is $\sim$150 tons and 
the volume of the truck bed is 37 cubic meter.
This means of transport will be sufficient for nearly all pieces within the IsoDAR design.  
The exception is the cyclotron magnet coils, which will be discussed shortly.

As indicated before, though, the cyclotron coils will provide a particular problem.
They will certainly not fit in the truck bed.  
A detailed examination of the entire length of the transit path between the warehouse and the
Cyclotron cavern will need to be performed, to ascertain whether these coils will pass at
some diagonal angle, and then to see if a special cart or conveyance means can be designed
and built that will carry these coils through these restrictive points.

Possible alternatives for the cyclotron coils are discussed in a separate section.

\section{Handling Components in the IsoDAR Areas}

Once the various pieces arrive at the IsoDAR areas, they must be offloaded from the conveyances that brought them
to the site, and must be moved and oriented for proper assembly into whatever structure they were
meant for:  cyclotron, beam line, support stations, electronics, or shielding.  These pieces are all
heavy and some awkwardly shaped, and will require very specialized equipment for rigging and handling.

By far the most flexible and convenient is an overhead bridge crane, as seen in Fig.~\ref{Canfranc}.  We can envision such cranes 
in both the cyclotron room and the target room.   Capacity for this type of crane can vary from a low of 10 tons up to
50 tons or more, however space requirements, reliability and maintainability and other factors need
to be considered before finalizing a decision on capacity, and overall rigging solutions.

A most useful piece of equipment is a compact fork lift, 
with 
a 10 ton capacity.  This is a highly utilitarian device could be the workhorse for moving pieces around.

\section{Alternatives for the Magnet Coil}

We expect to be able to transport the magnet coils into the lab, but 
should it prove impossible, there are two other
alternatives that we have considered:  building segmented coils or winding the coils underground.
\begin{figure}[t]
\centering
\includegraphics[width=5in]{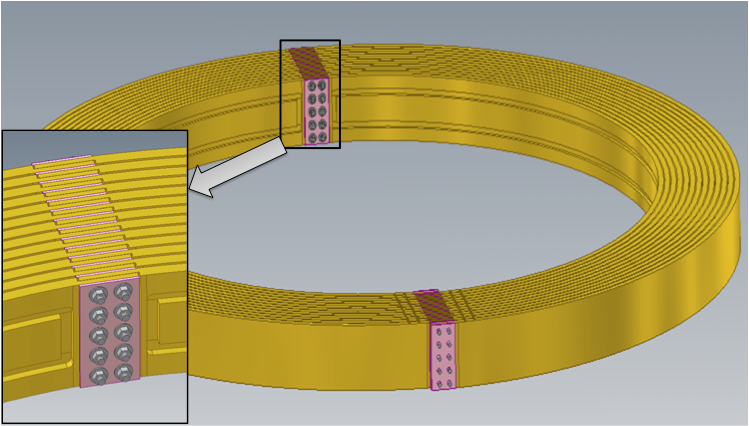}
\caption{{\footnotesize Magnet coil design splitting the coil into two parts.  Each coil consisting of about 15 plates bent into
half-circles.  Each end is bolted or welded to form a continuous spiral.}
\label{SplitCoil}}
\vspace{0.2in}
\end{figure}

\subsection{Segmented coils}

TRIUMF (in Vancouver) pioneered the concept of segmented coils~\cite{TRIUMF}.  Their enormous cyclotron has a coil diameter of almost 16 meters.  
An integral wound coil was just not possible.  Their coil was divided into six segments made up of 15 slabs of aluminum, 
each slab having a cross section of 2.5 cm by 50 cm.  
Each of these 8-meter-long slabs constituted a part of one turn of the coil.  
The ends of each slab were welded to the end of the corresponding slab in the adjacent segment, 
in the end making up a continuous spiral coil.  This required a total of 90 welds.  Cooling channels were added between the plates.
As the magnetic field is determined by the number of ampere-turns, if one has fewer turns then one needs more amperes.  
The TRIUMF coil requires 27,000 amperes.  The coil of a more conventional cyclotron may have 500 turns, dropping the current
to a value under 1000 amperes.  The power supply for this type of current is much more 
efficient and less costly than the TRIUMF configuration.

We have gone through the exercise of designing a split coil for the IsoDAR cyclotron.
Fig.~\ref{SplitCoil} shows schematically how a split-coil assembly might be constructed for IsoDAR. 
This concept would consist of aluminum or copper sheets bent into half-circles, and overlapped at the junction points. 
Insulated sheets between turns prevent shorts. The assemblies are tightly clamped at the junction points. 
Open cooling channels are milled into the sheets; the insulator seal must be good enough to prevent leakage. 
This concept would be implemented with as few turns as possible, to minimize the number of joints. 
The required ampere-turns are then provided by a very high-current, low-voltage driver, in this case 12 kiloamps at 30 volts 
(360 kW).

\subsection{Underground winding}

Winding the coil underground
requires transporting and assembling a coil-winding machine, bringing in the rolls of conductor and insulation, 
designing and constructing a collapsible potting fixture and mold for the wound coils, 
providing the appropriate epoxy-mixing, vacuum and ventilation capabilities. 
All of these components fit in the mining-truck bed.
The winding facility could be set up in the Target Room before any of the 
target or beam line components are moved into the area.    Alternatively, another room in the underground laboratory space may 
be available.   The coils would need to be made, transported and installed into the cyclotron, then the coil-winding
machinery dismantled and removed to the surface again.

%% file: Ch3-Utilities-Environment_v1.tex
\chapter{Utilities}

This chapter will cover the required utilities, electrical power and cooling water, as well as provisions
for management of ventilation air at suitable temperature and humidity levels.
The management of radiological aspects of air, water and material
leaving the experimental areas will be  addressed in the following chapter.
The plans for utilities presented here are preliminary and will be subject to reviews as the IsoDAR progresses.

\section{Electrical Systems}

The requirement for electricity for IsoDAR is 3.5 megawatts.  A good fraction of this needs to be
relatively ``clean" without large variations or spikes, so equipment should be installed for 
controlling noise and transient voltage changes.  Much of this will be included in the IsoDAR power distribution
and conditioning equipment; however, specifications for this equipment will require knowledge of the
condition of input power.

The situation at Yemilab is as follows:
\begin{itemize}
\item There is a contractual limit with KEPCO, the Korean Electric Power Corporation, to deliver 10 MVA to
the site of the Handuk Iron Ore mine. 
\item The mine is assigned 7.5 MVA of this, while 2.5 MVA is reserved for Yemilab.
\item Current projected need for Yemilab is 1.5 MVA.
\end{itemize}

This accounts for 9 MVA of the contractual limit, leaving a deficit of 2.5 MVA
to fully meet the requirements for IsoDAR.  To meet these extra needs, the following steps will be needed:
\begin{itemize}
    \item Negotiate a new contract with KEPCO to increase the power feed.
    \item Establish whether the lines to the substation are capable of supplying the extra power.
\end{itemize}

Further information about the power system is that:
\begin{itemize}
\item Power is fed to the Handuk Iron Ore Mine via 22.9 kilovolt lines.  
\subitem The full 10 MVA will draw 450 amps over these wires.
\subitem If this is increased to 12.5 MVA the current will increase to 550 amps.
\subitem It must be established whether the 22.9 kV lines can carry the extra 100 amps
\item High voltage transmission lines of 154 kV are not far away, so obtaining extra
power if needed should be not difficult.
\end{itemize}

The Handuk mine substation receives the 22.9 kV power, a transformer
will drop this voltage to 3.3 kV and one or more 3-conductor, heavy-duty
cables will be strung down the elevator shaft to bring the power to the Yemilab
substation.  This cable is a very common item, used for example for bringing power
to offshore platforms and lying on the sea bottom.

The details of the Yemilab substation are not known at this time, nor is the
detailed inventory or layout of the electrical equipment 
required for IsoDAR. However, we have estimated that the footprint of this equipment
will be about 100 m$^2$.

Power distribution within the Yemilab complex will be accomplished via cable trays or maybe, in usual
mining practice, via cables attached to wall hangers.

Electrical power requirements for IsoDAR include the power supplies for all the magnets, as well as vacuum systems and other ancillary devices.  The principal power requirement is for the RF system that accelerates the beam.  The beam power is 600 kW, the typical efficiency:  wall-plug to beam, is about 50\%, so at least 1.2 MW will need to be supplied to RF amplifiers.  The details of voltages, currents, cleanliness requirements for power conditioning, will all be developed as part of more detailed engineering studies for the technical components.

\section{Cooling Water Systems}

With only 1 to 1.5 megawatt planned usage, the original plans for Yemilab did not provide for an extensive cooling
system.  It was calculated that simply passing the heat from this usage to the air circulating system, with its continuous
flow to the surface, would be sufficient to maintain stable temperatures in the laboratory.
However, adding 3.5 MW from IsoDAR renders this solution unworkable.

As described earlier, IsoDAR will have two primary cooling loops, one for RAW (RadioActive Water) for the D$_2$O
circulating through the target (the heavy water is actually part of the neutron-producing target), the other a 
low-conductivity water system for the accelerator and beam-transport equipment.  
These loops will pass through heat exchangers to a second system that brings the heat out of the IsoDAR area.
The hot water must eventually reach the surface. Whether there is a dedicated loop just for this, or if the
afore-mentioned secondary loop carries the water to the surface is yet to be decided.
In any event, steel pipes, probably 15 or 20 cm diameter, capable of withstanding the pressure 
of a 587 meter head (the height of the shaft) will need to be installed in the vertical shaft.  
As water is flowing both up and down, 
it is only pipe friction that must be overcome by the pumps.  Because of the pressure at the base
of the shaft, it would probably be wise to have a separate system only for this circuit.

\begin{figure}[t]
\centering
\includegraphics[width=5in]{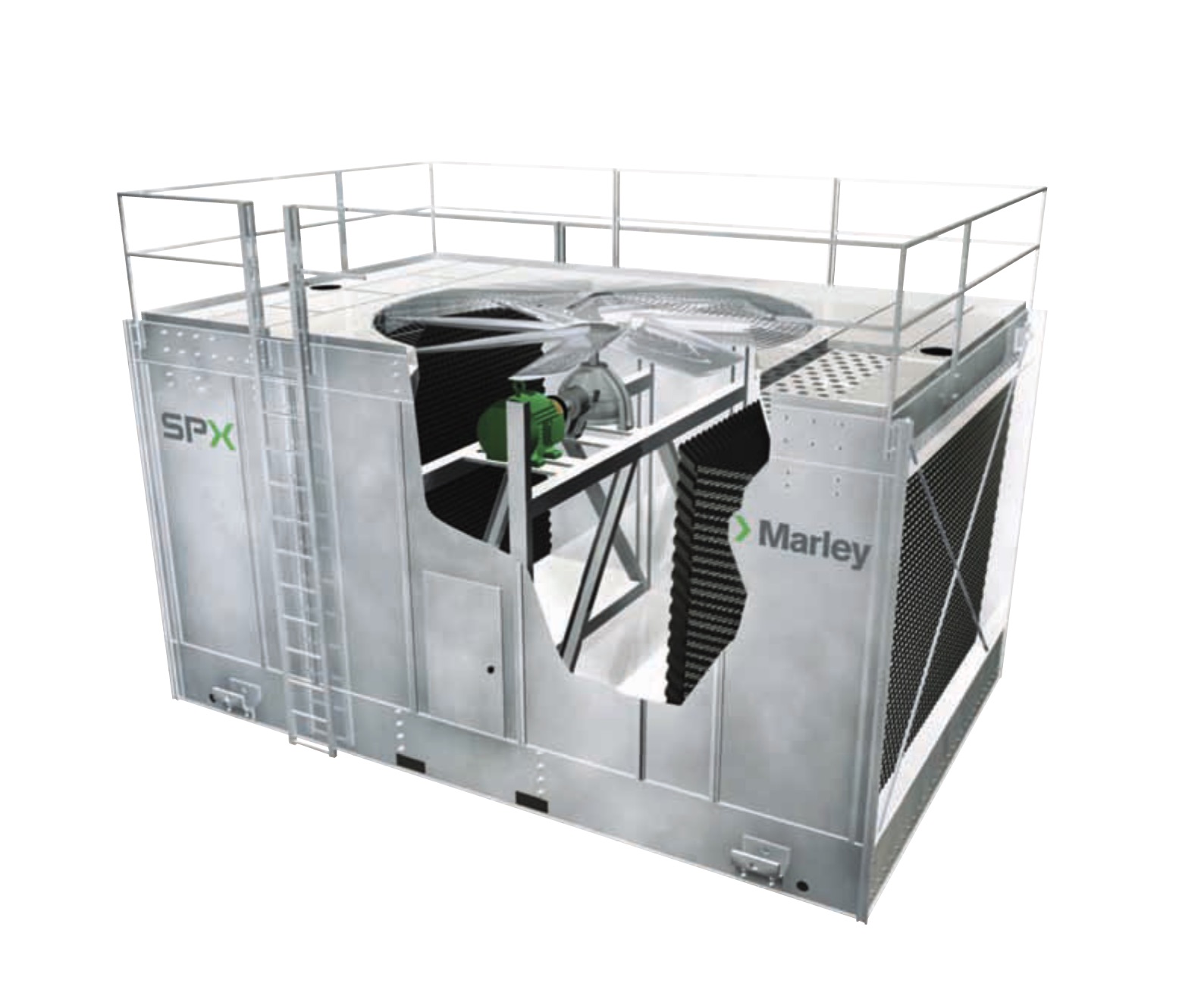}
\caption{{\footnotesize  1000 ton cooling tower, its footprint is 
 4.2 $\times$ 6.8 $\times$ 5.7 meters. This heat exchanger would be adequate for all the Yemilab needs.}
\label{coolTwr}}
\vspace{0.2in}
\end{figure}

Once on the surface, a cooling tower is required. 
A cooling tower that is sized for 1000 tons of refrigeration capability, would handle up to 4.5 megawatts, adequate
for all the Yemilab cooling needs.  An example is Marley Model NC8412V-1, with a footprint of 4.2 by 6.8 meters and a height of 5.7 meters.

\section{Input and Exhaust Air Management}

\begin{figure}[t]
\centering
\includegraphics[width=6in]{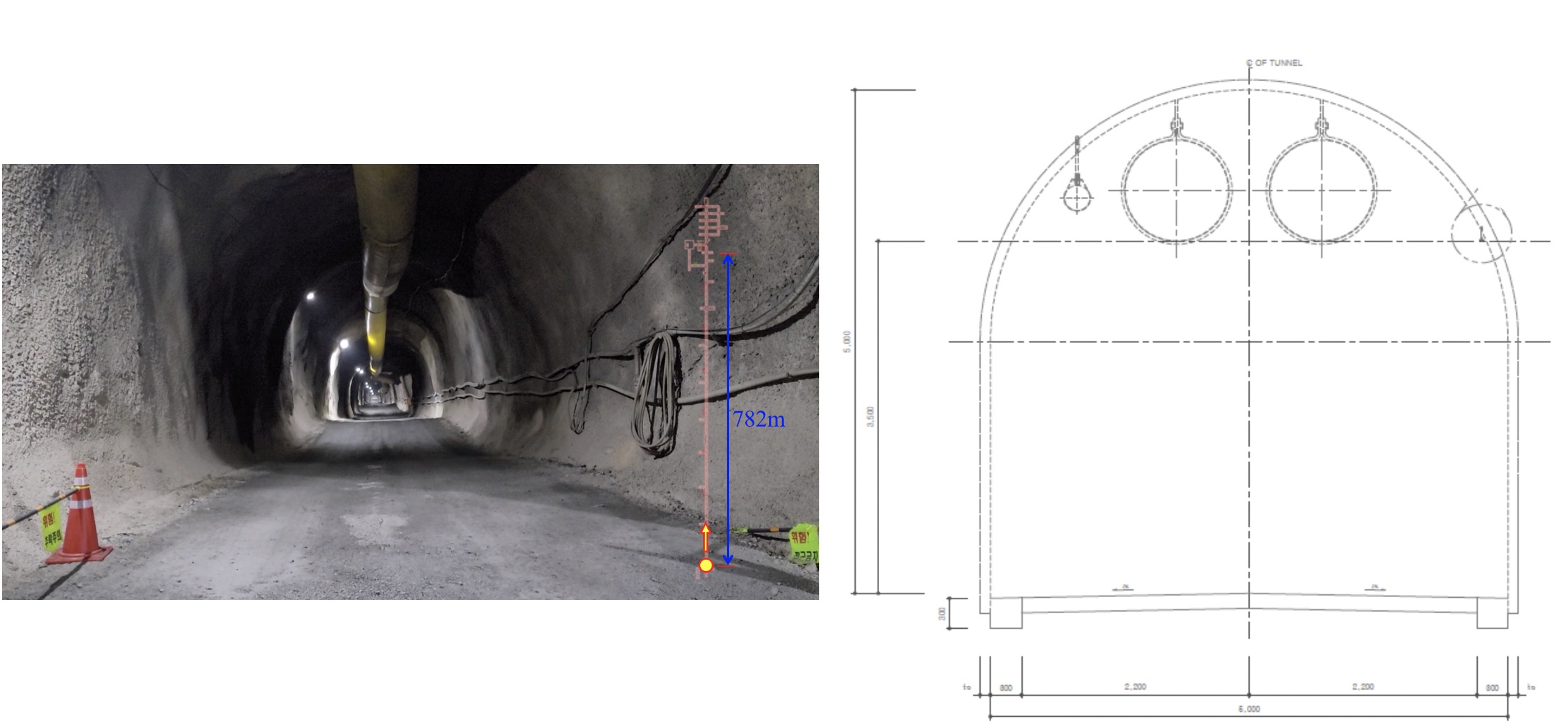}
\caption{{\footnotesize  Photo and cross section of the Yemilab access drift.  The drawing shows two ventilation
ducts strapped to the ceiling. Both of these are inlets, bringing fresh air to the Yemilab area from the surface.
It may be necessary to add a third, an exhaust duct from IsoDAR, as a means of removing air that may contain trace
amounts of activation from the neutrons in the IsoDAR area.  This would prevent any possibility of contamination 
from reaching the low-background Yemilab experiments.}
\label{airLine}}
\vspace{0.2in}
\end{figure}

Figure~\ref{airLine} shows a cross section of the Yemilab laboratory access ramp that runs from the base of the elevator 
shaft, as well as the end of the Handuk Iron Ore Mine Ramp, to the laboratory itself.  The drawing shows two large, 1 meter diameter, 
ducts strapped to the ceiling,  These had not been installed at the time the photo was taken.  These ducts bring 
fresh air into the Yemilab area, air that has been filtered and adjusted for temperature and humidity requirements.  
This air is released in several areas of the laboratory, and spent air is forced back to the surface through the shaft and mine
passages.  

Air handling in IsoDAR will require some special attention.  As there is a possibility of a small amount of 
contamination from neutrons interacting with air, no air from the IsoDAR area must be allowed to migrate to the clean 
areas of the Laboratory.  Controlling the air flow is actually fairly straightforward.  As there is only one ingress
to the IsoDAR area from anywhere in the Laboratory, one needs only to monitor and control the airflow through the  
door between the 
IsoDAR Access Ramp and the Yemilab main drift to ensure adequate control.  
The flow must be maintained always inwards, by maintaining a slightly negative relative pressure inside the IsoDAR area.  
A special duct line should be added to the two inlet lines that is dedicated to exhaust air from IsoDAR.  
This air is transported to the surface and released a suitable distance from the inlets for fresh air to the Laboratory.

\section{Moisture and Environmental Water Management}

We are most fortunate in that there is no ground-water seepage through the limestone into the
Yemilab area.  As a result, there is no need to be concerned about management of water in the area.  
Ditches along the edges of all the ramps and drifts provides a means of collecting any water that
may appear. However, except for accidental liquid spills, it is anticipated these will remain dry.

%% file: Ch4-RadiationProtection_v1.tex
\chapter{Radiation Protection\label{radpro}}

This chapter will address aspects of radiological protection related to the IsoDAR
experiment installed at the Yemilab
site.  
IsoDAR, and its high radiation environment must be isolated from the rest of the Laboratory
to the extent that background levels in the other deployed experiments are not increased over natural levels.

Natural backgrounds at Yemilab derive from uranium, thorium and potassium content in the rocks,
as well as products from reactions of the attenuated muon flux that does reach this depth.  
The uranium (0.8 ppm) and thorium (3.3 ppm) content of the rocks is unusually low, providing
a nice advantage for Yemilab over other sites.  Radon, whose genesis comes from these actinides, is not
as severe an issue as it is in other underground laboratory sites.  Potassium, on the other hand, 
at about 12,000 ppm, is notably higher than at other sites.  $^{40}$K, with a 1.3 billion year halflife,
will always be present at the same level.  Its beta decays (both + and -) are not significant in themselves,
however the positron produces annihilation radiation and there is a 1.4 MeV gamma transition.
With respect to muons, the depth of 2500
meter-water-equivalent provides an attenuation factor of about 10$^6$.  

Environmental isolation of IsoDAR from other experiments has been addressed in several other sections: 
in the last two sections of Chapter 2, as well as in the last sections of Chapter 4 addressing
water and air circulation.  For completeness, some of these topics will be covered again, from
the perspective of radiation protection.
The chapter will end with a brief discussion of personnel radiological protection, relating
to staff that require access to the site, and 
instrumentation to ensure adequate personnel protection.

Radiation around accelerators is highest when the beam is on.  X-rays from the RF systems, neutrons generated from beam loss, and prompt gamma radiation from beam particles stopping in material produce high radiation levels.  
As a consequence, there will be no personnel allowed inside the IsoDAR caverns
under these conditions. When the beam is off, and the RF system is not operating,
the only source of background radiation is from gammas emitted by radioactive
decay of isotopes produced by the neutrons generated by nuclear reactions from
beam particles slowing down and stopping in materials.  

\section{Shielding}

This section will address the shielding necessary to adequately attenuate this neutron flux to mitigate activation levels.  To evaluate this we should identify where the neutrons are produced: the ``source term'' describing the amount and distribution of proton loss from the beam.  Easiest will be the target itself, as almost 100\% of the beam is lost there, by design.
 As the beam loss in other areas, such as the cyclotron and transport lines, will be
minimized as much as possible, the evaluation of the ``source term'' for calculating induced radioactivity from these areas will be more complicated.

\begin{figure}[t]
\centering
\includegraphics[width=5in]{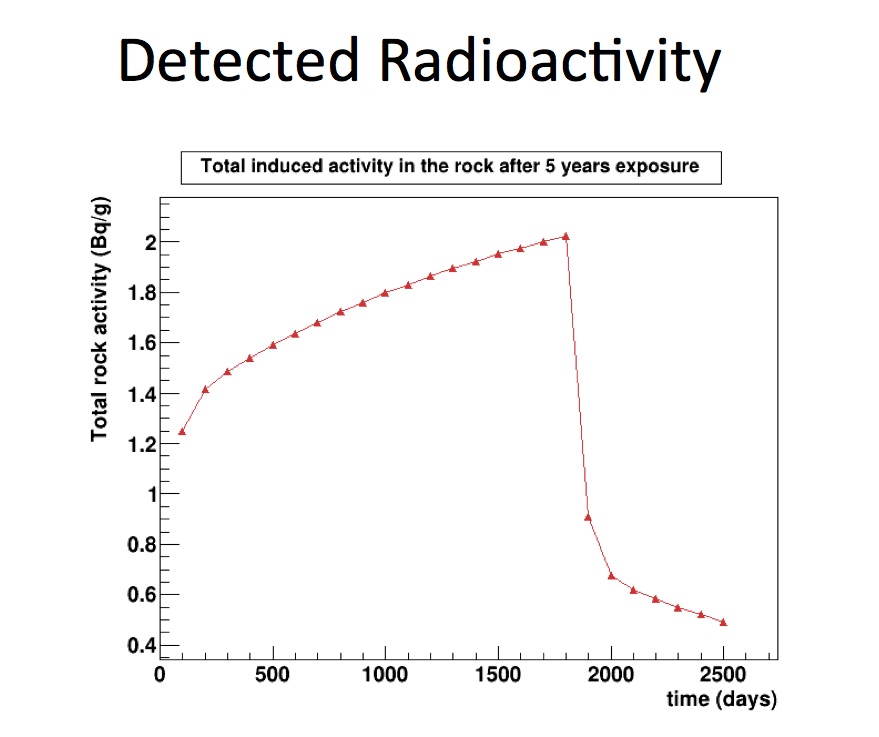}
\caption{{\footnotesize  The curve shows the buildup of activity in the surrounding 
rocks for the 5-year planned beam-on-target.  Once beam is shut off, rapid decay
of the many short-lived isotopes occurs, leaving only long-lived activities.  The 
``end of the experiment'' is declared at approximately day 2500, by which time the
activity level has dropped very significantly. The vertical scale should be ignored, the calculation was done for an area with different rock composition than what is found in Yemilab, however the generic shape of the curve is still very relevant.}
\label{decayCurve}}
\vspace{0.2in}
\end{figure}

In addition to the neutron flux, we must know the material the neutrons are going through, and the isotopes produced by absorption of the neutrons.  In an area such as Yemilab, this material is either brought in -- such as accelerator components, support equipment, or shielding itself; or is part of the environment, that is the rock in which the caverns are excavated.

In an underground laboratory, the question of rock activation is extremely important.  
Not only can this rock contribute to short-term background levels, but residual activation long after the experiment is over can be extremely deleterious. Components of the experiment that are brought in, and that might be activated, can be removed.  But removing activated rock is an entirely different matter.

In the following, we will study shielding requirements in two time regimes, starting with long term activation of the rock.

\subsection{Shielding for Long Term Rock Activation}

Figure~\ref{decayCurve} shows the buildup and decay of radioactivity in a high-intensity, high-energy beam area over the course of five years of an experiment's running, and how this radioactivity decreases after the beam is shut off for the last time.  The short-lived isotopes all decay away quickly after this shutoff, but remaining are those isotopes with half-lives measured in years that contribute to a lingering background of gamma rays.

To start, we should examine the half-lives of isotopes that might be formed in the rock.  And in particular, those with half-lives of the order of a few years.

Table~\ref{isotopes} lists long-lived isotopes typically produced in environments with neutron fluxes.  These isotopes are almost always found in shielding blocks around accelerators, and are the contributors to gamma radiation detected long after the neutron fluxes have ceased.
Of interest to us are the abundances of the parent isotopes (listed in parentheses in column 3) measured in the limestone in the Yemilab area.

\begin{table}[!t]
\caption{Long-lived isotopes produced in rock. 
\newline(*TNC = thermal neutron capture; Abundance assayed for Yemilab rock, na = abundance not measured)\label{isotopes}}
\centering
\renewcommand{\arraystretch}{1.25}
\begin{tabular}{lllll}
\hline
Radionuclide & Half life & Parent & Cross section & Produced by\\
	&	& (Abundance) & (barns)	&  \\
	\hline
$^{60}$Co & 5.3 years & $^{59}$Co ($<$10 ppm) & 37 & TNC* \\
$^{152}$Eu & 13.5 years 	& $^{151}$Eu ($<$0.5 ppm) & 4600	& TNC \\
$^{154}$Eu & 8.6 years & $^{153}$Eu ($<$0.5 ppm) & 310 & TNC \\
$^{134}$Cs & 2.1 years & $^{133}$Cs (na) & 29 & TNC \\
$^{46}$Sc & 84 days &  $^{45}$Sc (na) & 27 & TNC \\
$^{22}$Na & 2.6 years & $^{23}$Na (0.022\%) & 0.1 & (n,2n) $<$11 MeV\\
\hline
\end{tabular}	
\end{table}

Though cross sections are higher for the thermal neutron capture (TNC) isotopes, the very low concentration of these makes them less important than sodium, particularly if there is a reasonably high number of neutrons above the 11 MeV threshold.  As seen in Figure~\ref{neutron_spetrum}, the IsoDAR neutron energies reach 60 MeV, so $^{22}$Na indeed can be a major source of induced activity.
We are fortunate that the sodium concentration at Yemilab is so low, typical rock concentrations are in the 10\% or higher range.

\subsubsection{Requirements}

The problem with long-term activation relates to regulations for the handling
and containment of activated materials.  In Korea the regulation states that 
any material with a radioactivity level greater than 10 Bq/gm must be
contained in a ``controlled area for radiation protection."
During the years of operation the entire IsoDAR area must be classed
as such a ``controlled area.''  However, after the experiment is completed,
and fully decommissioned so the caverns can be repurposed for other programs, 
there must be no remaining ``controlled areas.''
In the case of IsoDAR, when beam is turned off for the final time, it is
expected that the area will be sealed off for about a year to allow all the hottest
areas (cyclotron interior, target and sleeve areas and inside of shielding structures)
to cool off.  Then another year's process will be dedicated for dismantling and removing the radioactive pieces.
So, after a 2 year period it would be expected the IsoDAR space would be ready for its next
mission.  At this time, a radiation assay will be performed, and the activity levels in any area
of the caverns must be 
below the 10 Bq/gm level.

Shielding must be designed to keep neutrons entering the rock to levels below where they
will show activities above the 10 Bq/gm level.  
The consequence of exceeding these levels will be that the surfaces of the rock, perhaps as much as 20-30 cm,
would need to be chiselled off.  This would be an expensive, and messy process.
It is important to do the shielding design properly, and conservatively.

It should be noted that 10 Bq/gm is the approximate level one finds from normal concentrations of uranium,
thorium and potassium in ordinary rocks.  So, basically, induced activities should not exceed the radioactivity
levels found in nature.

\subsubsection{Rock Activation Calculations}

To assess the effect of neutrons emerging from the shielding, and subsequent activation of rock layers,
a sophisticated GEANT4 calculation was performed~\cite{bungau:shielding}.
These studies should all be considered as {\it preliminary}, as they will need to be confirmed by use of
the code accepted in Korea for such calculations, most likely PHITS.

\begin{figure}[t]
\centering
\includegraphics[width=5in]{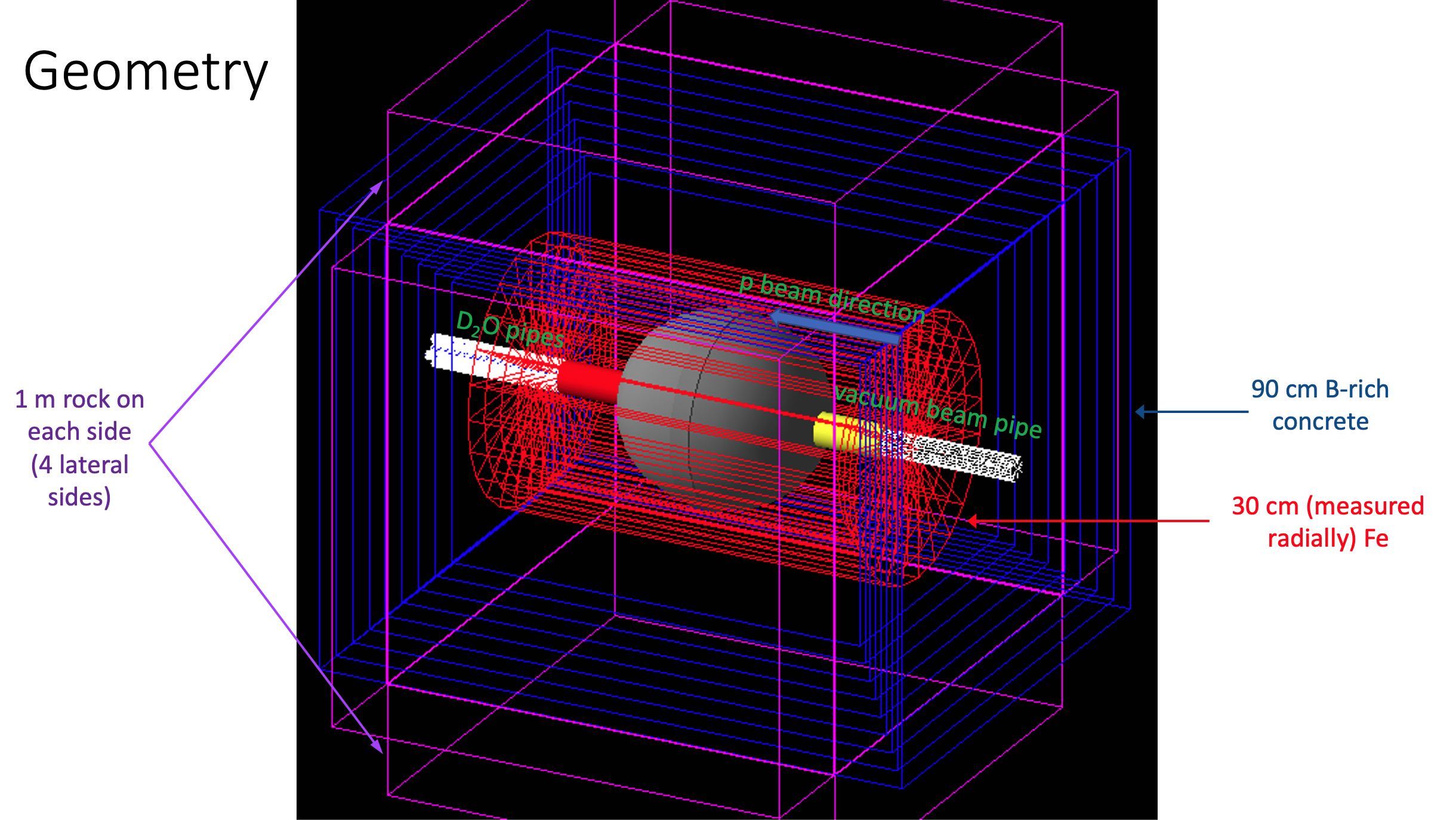}
\caption{{\footnotesize  Geometry entered into the GEANT4 Monte Carlo calculations
for neutrons emerging from the shielding and penetrating through modeled rock layers
outside the shield.  Protons enter from the right through the vacuum pipe, and stop in the target at 
the center of the sphere, representing the Be + Li sleeve.  The sleeve is surrounded by a 30 cm
iron cylinder, which in turn is enclosed in a 90 cm concrete rectangular structure. The concrete contains a very 
large amount of boron in its aggregate, developed at JLAB for absorption of slow neutrons~\cite{JLAB}.  Outside of the concrete,
the model includes 1 meter thick layers of rock, whose composition is pure calcium carbonate, but with the amounts
of contaminants identified in Table~\ref{isotopes}.}
\label{GEANT4}}
\vspace{0.2in}
\end{figure}

For these long-term rock-activation studies, GEANT4 geometry model used is shown in Fig.~\ref{GEANT4}. Following the target and shielding design described in 
section~\ref{target-section}, the target is modeled as the nested hemispheres of Be and D$_2$O, the sleeve as
the mixture of 75\% (by weight) of Be and 25\% of 99.99\% enriched $^7$Li, the sleeve is a steel (modeled as iron) 
sphere with inner
radius of 74 cm and outer radius of 81.5 cm.  The sleeve is placed in a 30-cm thick iron cylinder, with 30 cm thick iron
endcaps.  This cylindrical assembly is enclosed in a rectangular block of solid concrete, with a minimum
thickness, in the orthogonal directions and at the ends, of 90 cm.  This concrete is a special blend developed
at JLAB, with the aggregate material being pure boron carbide, providing efficient absorption of slow neutrons~\cite{JLAB}.

\begin{figure}[t]
\centering
\includegraphics[width=5in]{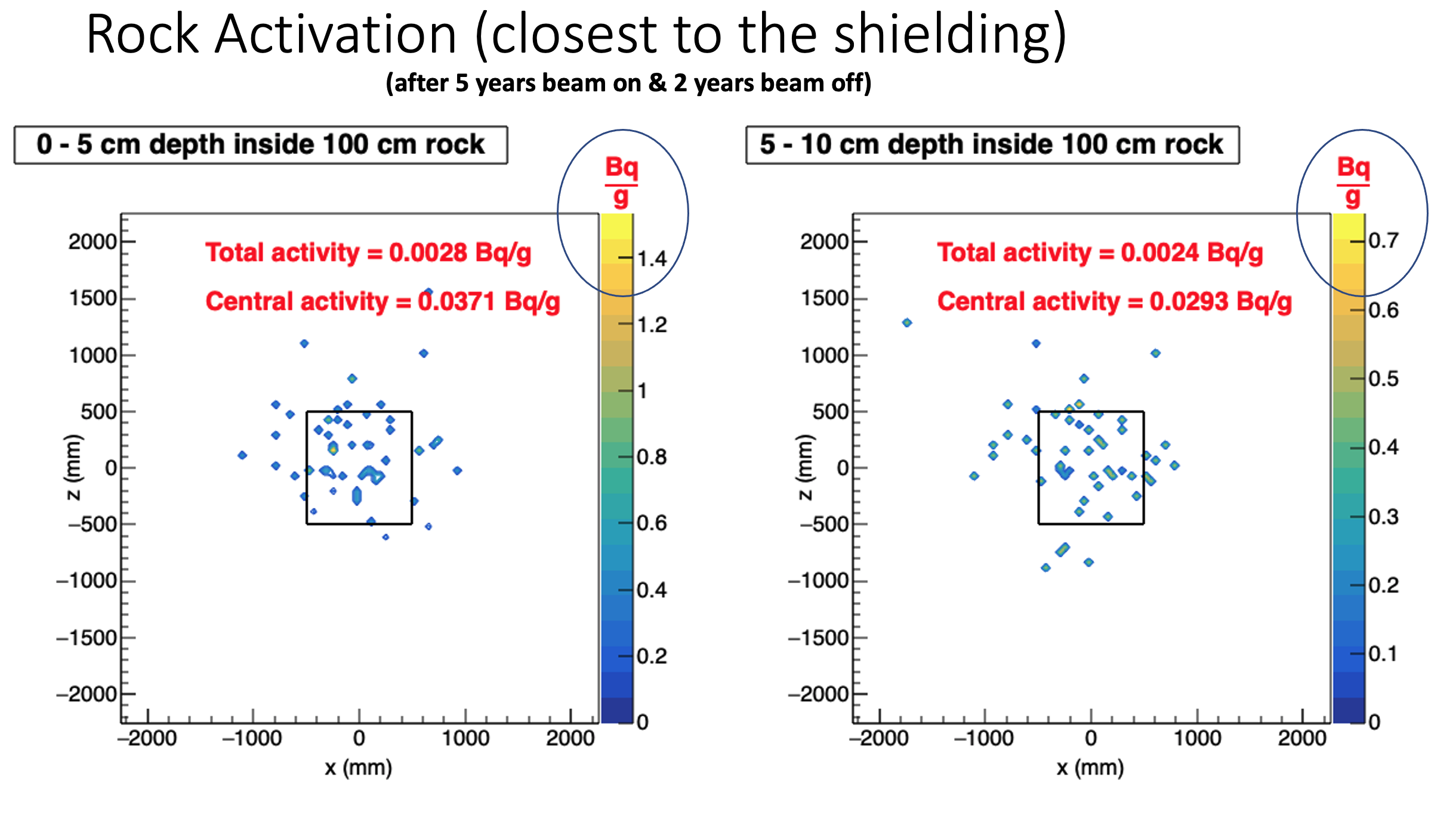}
\caption{{\footnotesize Total induced activity in the rock layers closest to the surface
of the shielding.  This is an (x,y) plot, the beam is coming from the top of the figure 
heading downwards.  The highest concentration of activity is at the very center, which
is the 90\degree direction from the target.
 It is clear to see that the specified configuration of shielding, namely 30 cm
of iron and 90 cm of boron-loaded concrete, easily meets the 10 Bq/gm activation requirement.}
\label{activation}}
\vspace{0.2in}
\end{figure}

Fig.~\ref{activation} shows the results from total activation (including all the isotopes being tracked)
in the Yemilab rock as measured in the two (5 cm thick) layers of rock closest to the outer edge of the concrete.
This is measured two years after the end of the 5 year run.  
The results show that the activation levels are better than a factor of 100 below the regulatory mandates.  So this level of shielding is perfectly adequate for long-term rock activation at Yemilab.  
Note, however, that if the sodium concentration in the rock were closer to what is found in other labs, the picture would be quite different, and substantially more shielding would be required.

\subsection{Short Term Activation Issues}

Access into the cavern as quickly as possible after beam is interrupted is important for the
overall efficiency of the experiment.  As a result
one would like repairs or corrections that resulted in the beam shutoff to be taken care of as
quickly as possible.  This means that there should be as little delay as possible before entry
is allowed to the caverns to correct the situation.

An environmental check by a Radiation Technician to assess
the background present is mandetory.  It is the discretion of this individual to allow, or not, and the conditions
for access to the
caverns, based on radiation-safety requirements.  

Calculating the 
flux of neutrons entering the environment of the cavern is straightforward, but the activation
these neutrons cause is dependent on the material present, often on trace amounts of material, such 
as sodium in concrete, wall coverings, or flooring.  As indicated before, sodium content of the base
rocks has been established to be low, but material brought in to the cavern must receive a similar amount of scrutiny.

At present, we believe that sodium will be the primary source of short-term background as well, due to the
15 hour half-life of $^{24}$Na (slow neutron capture on parent $^{23}$Na), and the wide-spread distribution of sodium throughout the cavern environment. 12-hour $^{64}$Cu is another well-known radioactivity in accelerator environments. This material is found in the electric coils of magnets.  These are quite localized, though, and could be adequately shielded.  We are researching whether there are other possible isotopes that could contribute to this short-term background.

\subsubsection{Shielding to Minimize Short-Term Activation}

\begin{figure}[t]
\centering
\includegraphics[width=5in]{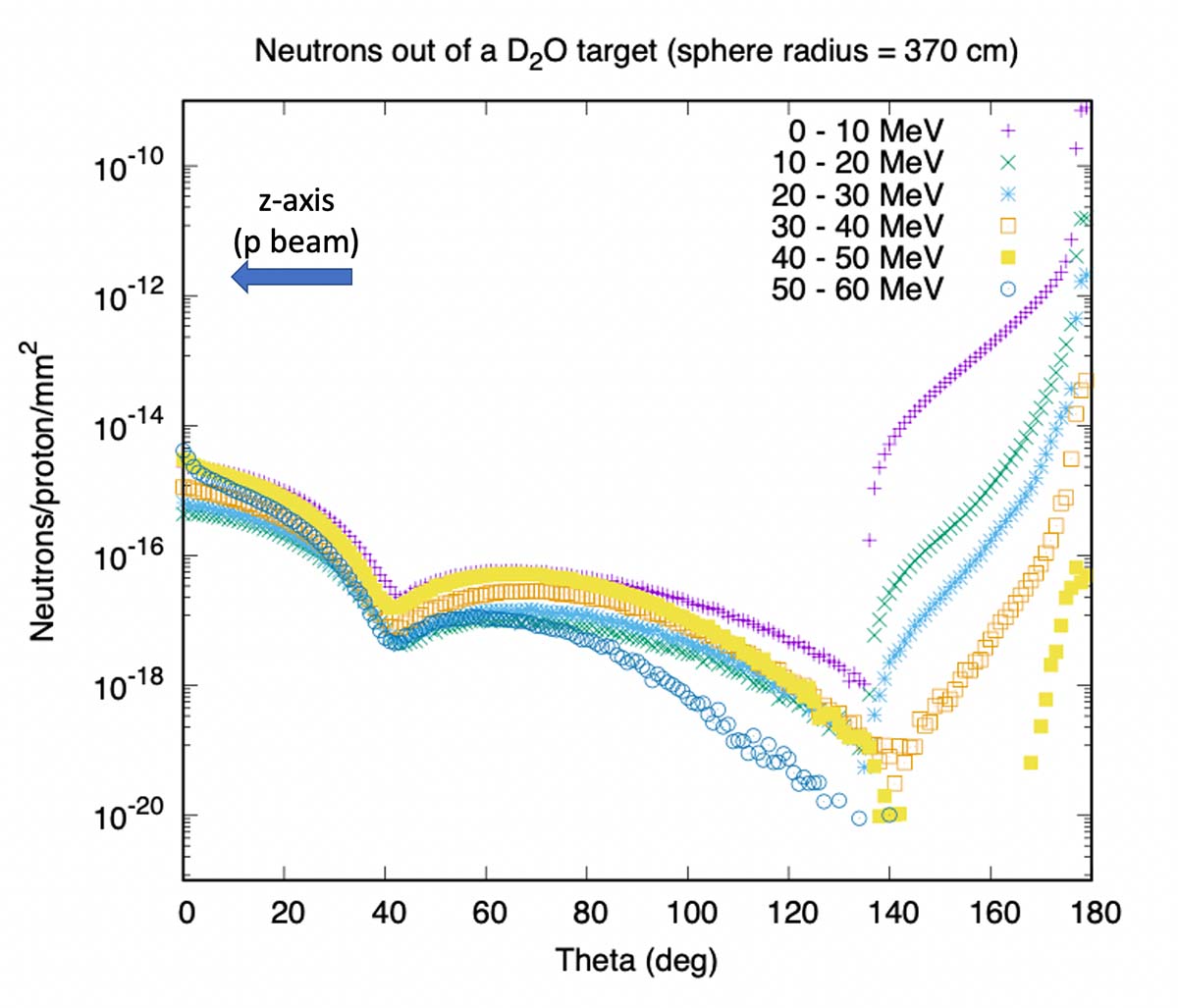}
\caption{{\footnotesize Neutron flux emerging from original shielding design.  While flux in the perpendicular direction is low, the end-caps are clearly less efficient for shielding neutrons.  In the backward direction (along the direction of the incident beam) the vacuum pipe allowing the beam to reach the target, is a clear path for neutrons.  In the forward direction the greater flux of higher-energy neutrons, along with the low-density water pipes also lead to greater neutron leakage.  To mitigate these, extra shielding is needed in these areas. Note that an attenuation of 10$^{16}$} corresponds to only a few neutrons per mm$^2$ per second.
\label{Adriana0}}
\vspace{0.2in}
\end{figure}

The challenge is to minimize the total neutrons penetrating through the shielding, reaching areas where activation can occur.  
In the previous section we saw that neutron flux in the perpendicular direction, could be adequately attenuated with the proposed shield of 30 cm of steel and 90 cm of borated concrete.  
However if we look at the total neutron flux leaving this shielding design, shown in Fig.~\ref{Adriana0} we can see more neutrons emerge from the ends.

\begin{figure}[t]
\centering
\includegraphics[width=5in]{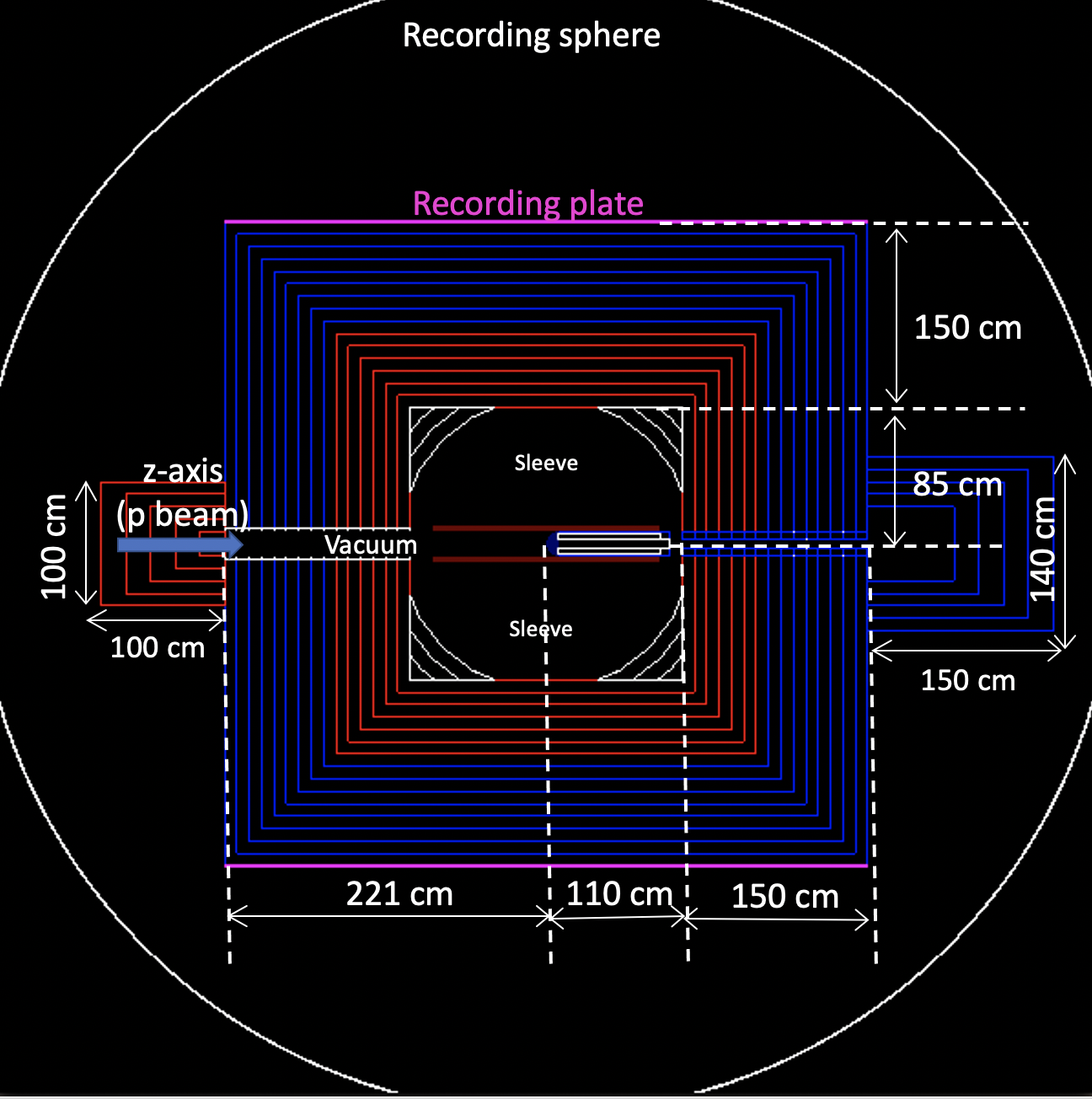}
\caption{{\footnotesize First iteration on shielding geometry.  Steel thickness of cylinder and endcaps has been increased from 30 to 60 cm (red and white areas), a concrete cap placed over the exiting water pipes, and the 90\degree beam bending magnet modeled by a 1 cubic meter block of steel.}
\label{Adriana1-Geom}}
\vspace{0.2in}
\end{figure}

While we do not yet have a good measure of the conversion of the flux of emergent neutrons into the expected background radiation level immediately after beam shut-off, we have continued the GEANT4 studies to beat down the neutrons penetrating through the shielding.  Figure~\ref{Adriana1-Geom} is a first attempt to address this.  The steel cylinder surrounding the sleeve and the steel end plates have been increased from 30 to 60 cm, a concrete endcap has been added to cover the water outlets from the torpedo, and the 90\degree entrance magnet has been modeled by a 1 cubic meter block of steel.

\begin{figure}[t]
\centering
\includegraphics[width=5in]{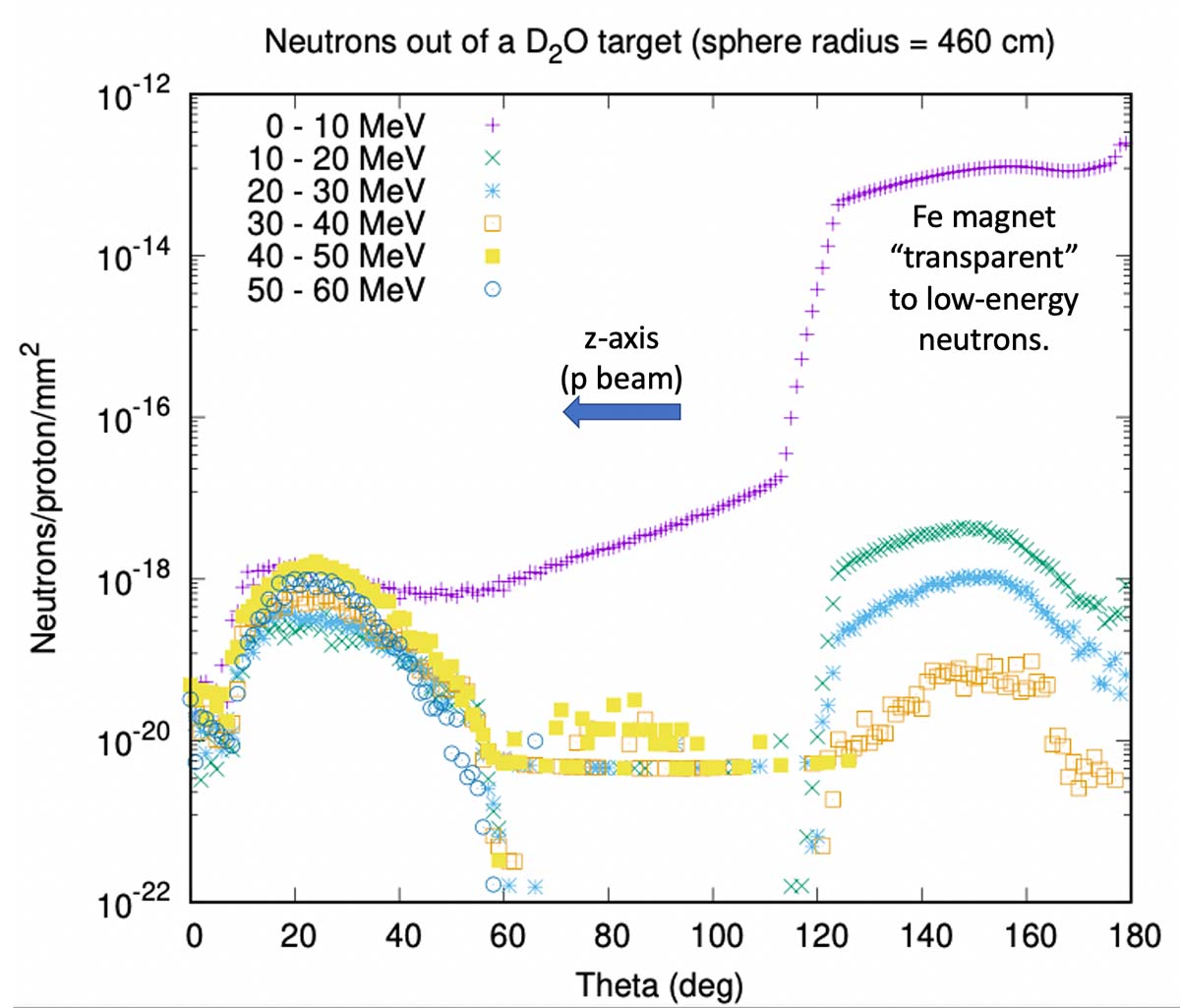}
\caption{{\footnotesize Neutron spectral distribution from the geometry shown in the last figure.  The 30 cm of steel serves to bring the neutron level a factor of one thousand below previous levels, and the additional shielding on the ends begins to bring neutron levels down significantly.  However it is clear that attention must be given to attenuating neutrons below 10 MeV. }
\label{Adriana1}}
\vspace{0.2in}
\end{figure}

The neutron spectrum calculated for this configuration is shown in 
Fig.~\ref{Adriana1}.  We see that the higher-energy neutrons emerging in the perpendicular direction are brought down another four orders of magnitude (to a few hundred neutrons per \textit{square meter}). 
The forward neutrons are also brought down three orders of magnitude, and the steel covering the beam pipe helps quite a bit.  
Additional design work is necessary to attenuate neutrons below 10 MeV.  This involves introducing more concrete shielding.  Further design studies are now underway.

\subsection{Neutrons into the Detector}


Background in the 2.5 kiloton liquid scintillation counter (LSC) affects its sensitivity to the desired physics measurements.  Studies of natural backgrounds that could mimic a 3 MeV single light flash show a possible rate of a few events per year.
Our goal then should be to keep the flux of neutrons above 3 MeV that penetrate through the thick steel shielding block between the target and the detector, to less than this amount.  

\begin{figure}[t]
\centering
\includegraphics[width=5in]{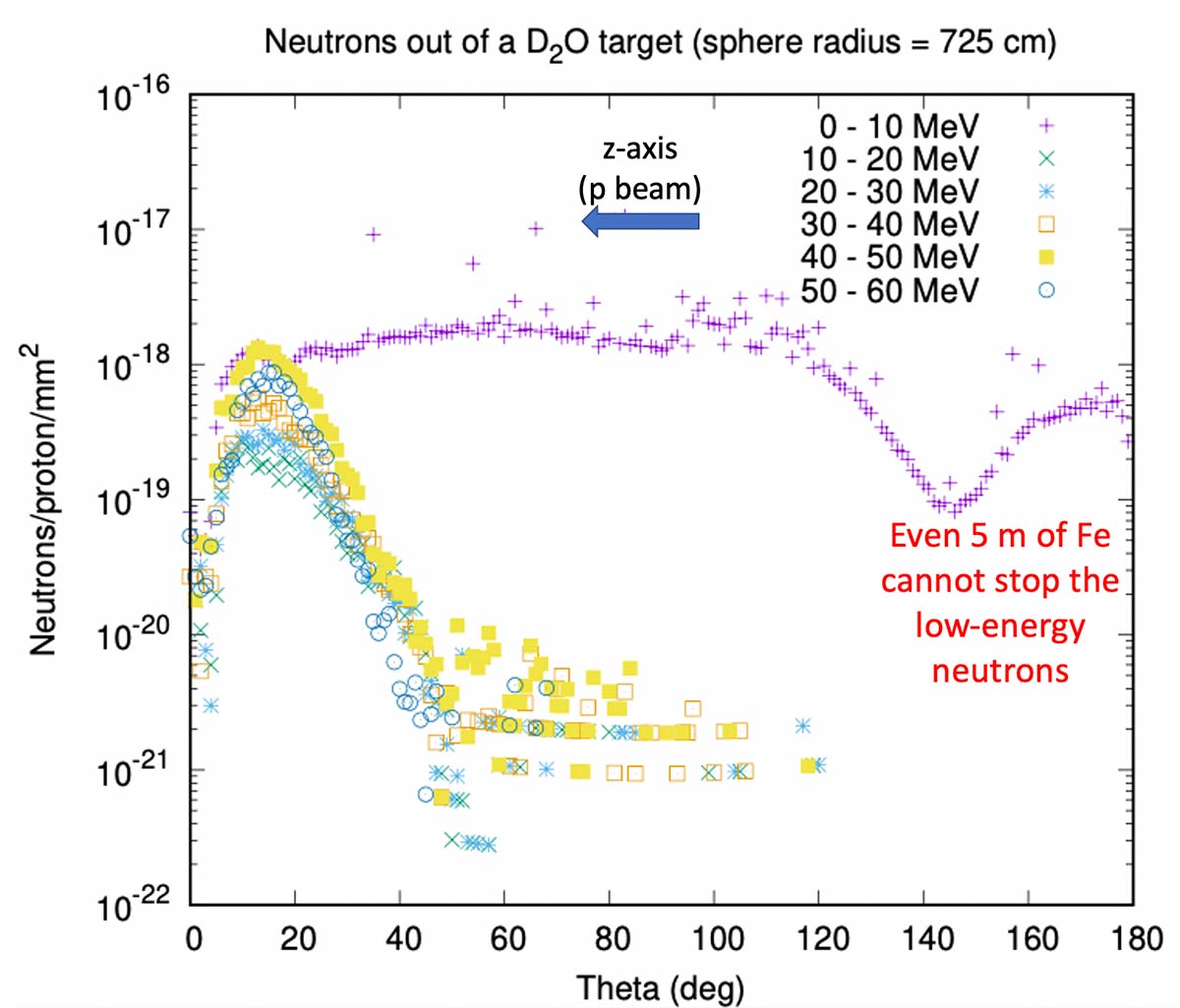}
\caption{{\footnotesize Neutron spectral distribution obtained by adding an additional large block of iron that is 4 meters thick to the geometry shown in Fig.~\ref{Adriana1-Geom}.  The calculation is not sensitive enough to evaluate the true attenuation of high energy neutrons, much more sensitivity is needed to probe the ``neutrons/year'' range.  However, the study does point out that steel alone is inadequate to shield the low energy neutrons.  How many of these are above 3 MeV is currently being studied. }
\label{Adriana2}}
\vspace{0.2in}
\end{figure}

A first GEANT4 run is shown in Fig.~\ref{Adriana2}.  In this study an additional four meters of steel (iron) were placed behind the 1 cubic meter simulating the bending magnet.  The study shows that neutrons above 10 MeV are attenuated to less than 10$^{-22}$, or somewhere around 10 neutrons per year (per square mm).  To evaluate a lower rate, and to probe the real efficiency for this thickness of steel, a different method of calculation will be needed.  However, one obvious result of this calculation is that steel alone is totally inadequate to attenuate all the neutron energies.  As the 0-10 MeV cut is broad, it is not possible to say how many of these neutrons are above 3 MeV.  We are refining the energy bins to evaluate whether the flux above 3 MeV is high, or whether the penetrating neutrons are all much lower energies.  In any event, mixing concrete or boron layers in the steel may be a good design choice.

\subsection{Shielding:  Cyclotron}

As mentioned above, the ``source term'' for neutrons from the cyclotron is more difficult to evaluate.
Neutrons all come from beam loss, which will be minimized as much as possible.
However, such losses cannot be brought to zero.

A practical and commonly used guideline to keeps the beam losses inside the cyclotron vault
and transport lines to less than 200 watts.  This number, developed at PSI,
results in activation levels of components that still allow hands-on maintenance.
For calibration, 1 watt of beam loss will come from 1 microamp of beam at 1 MeV.
Most of the beam loss we will experience that lead to neutron activation
will be at the full energy of the cyclotron,
60 MeV, so 200 watts will be about 3 microamps or 1.8 x 10$^{13}$ protons/second.
As the full beam is 10 milliamps,
this represents a loss of 1 part in 3000 of the primary beam.  This very low level of beam loss
is in fact achieved at the 590 MeV cyclotron at PSI.   We are
working closely with the PSI group, to benefit from their experience.

As indicated before, losses can be characterized as ``controlled'' and ``uncontrolled.''
``Uncontrolled'' losses arise from beam interacting with residual gas in the accelerator
and transport lines, from beam halo, namely trajectories that take the
beam particles outside of the normal stable orbits, or any other mechanism that causes particles to
strike surfaces inside the vacuum tanks.
The radiation resulting from this type of loss is distributed fairly uniformly around
the entire inner surfaces of the vacuum enclosures, so is difficult to shield.

``Controlled'' losses, on the other hand, are expected, and are channelled into 
well-shielded areas.  In our case, it is inevitable that some beam will be lost
on the extraction septum.  This loss occurs
because it is impossible to have completely clean separation between turns (orbits) 
in the cyclotron, so a small number of particles can be found in the space 
between the last turn to circulate, and the turn that enters the extraction channel.
This channel is defined by a thin septum, about a millimeter thick of graphite, usually.
This septum is one electrode of a high-voltage plate system that provides a kick
to the beam so it exits the cyclotron.

As shown in Fig.~\ref{extract}, we are planning on protecting the septum with a narrow stripper
foil that converts H$_2^+$ ions that would strike the septum, into protons, that can be
directed out of the cyclotron into a well-shielded dump.  This will certainly help.
It is anticipated that this ``waste" beam could even be utilized
for production of radioisotopes, though this is outside the scope of
this CDR.

\begin{figure}[t]
\centering
\includegraphics[width=5in]{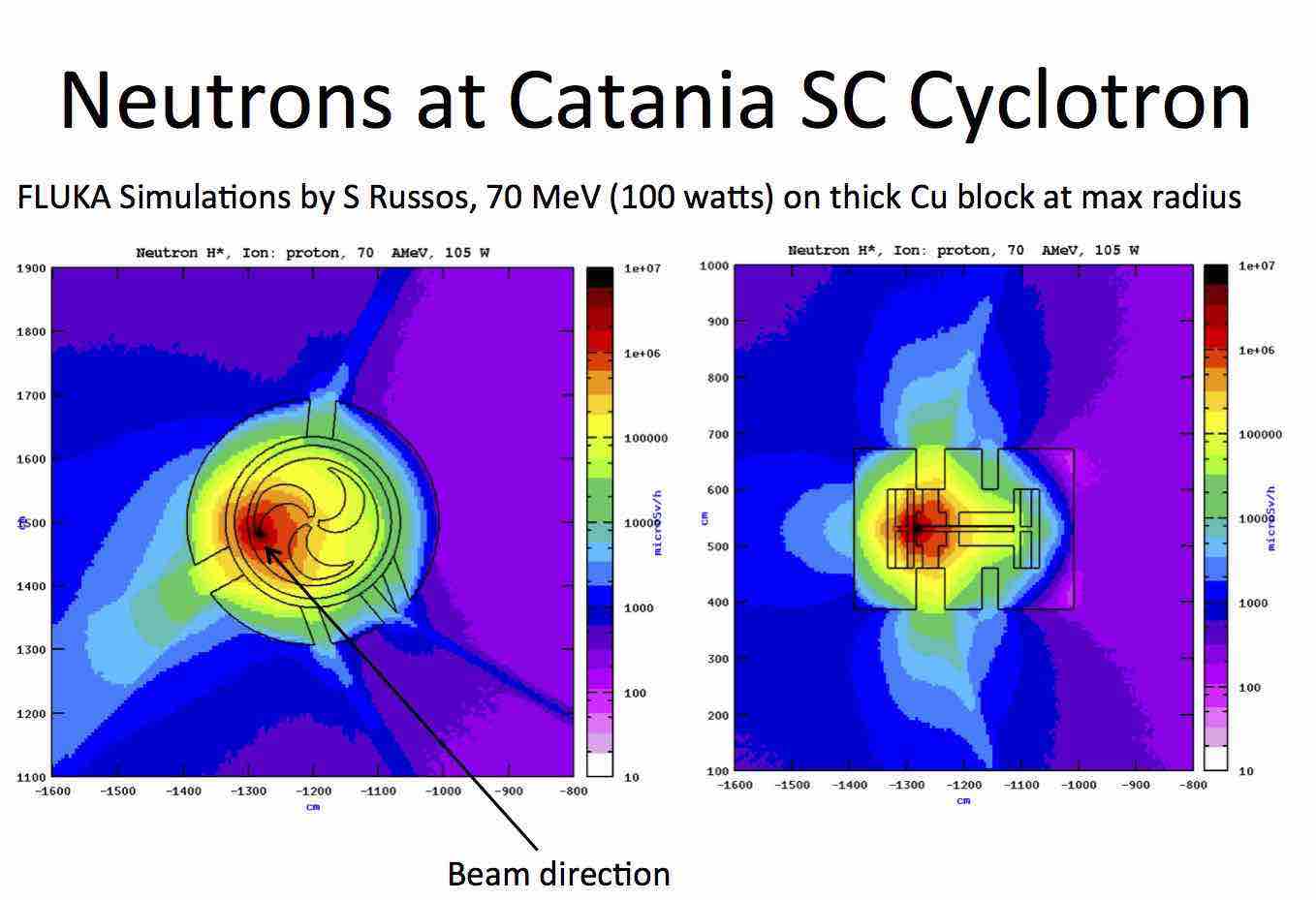}
\caption{{\footnotesize FLUKA simulation of neutron flux around the Catania 70 MeV 
superconducting cyclotron.   Left is the top view, right is the side view.  The simulation
assumed 100 watts of 70 MeV protons strike a copper block, at about the location of
the extraction septum.  One can see that the bulk of the steel of the cyclotron contains
most of the neutrons, however areas where the steel is thin: extraction port (about 6 o'clock in 
left figure) and the RF dee stems (above and below the plane in right figure) do show significant
neutron dose levels. }
\label{cyclotron-neutrons}}
\vspace{0.2in}
\end{figure}

On the whole, characterizing and quantifying uncontrolled losses is a difficult problem, but
simulations can be performed. 
Figure~\ref{cyclotron-neutrons}
shows one such calculation, done with FLUKA for the geometry of the Catania
superconducting 70 MeV cyclotron.  This figure demonstrates the complexity of
the problem, but does show a significant concentration of neutrons emanating
from the extraction septum.  The very large amount of steel in these machines provides a 
significant amount of self-shielding, however there are thinner parts, particularly 
for RF dee structures as well as injection and extraction ports, that provide paths for neutrons to escape.
One can perform similar calculations for the designed geometry
of our cyclotron, and doing these will give us a good idea of where neutron concentrations are high.  
These calculations have not yet been initiated.
But when performed, we should get an indication of localized shielding that can 
be deployed around the cyclotron to mitigate the radiation fields. 

In addition, once the cyclotron completed and during commissioning, radiation leakages can be measured, and
at that time more shielding can be added in areas where it is deemed necessary.

However, the low sodium content in the rocks again confirms that long-term activation from any
neutron flux emanating from the cyclotron, which will be orders of magnitude lower than what is present
in the target room, does not require that the cyclotron itself be enclosed in its own vault of concrete and steel.

\subsection{Shielding:  Beam Transport Line (MEBT)}

Beam is extracted from the cyclotron as H$_2^+$, but we choose to strip it as soon as
possible into protons.  This will minimize uncontrolled losses in the MEBT, which would mostly
come from stripping of the fragile H$_2^+$ molecules by residual gas in the transport line.

The stripper is located a few meters from the extraction port.  The stripper is followed by an 
analysis magnet and beam dump for unstripped beam, and halo-cleansing collimators.  Shielding will 
be designed to absorb neutrons produced in this area.

Transport of the beam to the target is a very straightforward matter, however beam losses along
this path cannot be avoided.  In principle these will be extremely small, and mitigating measures
can be put in place to minimize radiation from these beam losses.
Scattering of protons on residual gas could result in immediate loss of the particle.  However, the
highest cross sections will be for scattering at very small angles, this will result in a halo around
the primary beam.  
Halo is also a normal effect seen in transport of high-intensity beams.  It can be well calculated
with appropriate simulation codes.  
To mitigate this halo, one normally places collimators that scrape this halo off, allowing the core
of the beam to continue through.  These collimators are placed inside shielding blocks, containing
the neutrons from this ``controlled loss.'' 
More than one such collimation section may be needed in the beamline, as halo from all sources
is regenerated down the line.

\section{Control of Activated Materials}

As in any accelerator facility that produces ion beams with energies above the Coulomb barrier, there is always a chance
of  activating any pieces that may be exposed to the beam or neutrons produced by the beam.  
The measures for handling such materials are well-established, and will be adopted
for our experiment.
One item of great importance will be monitoring for leaks in the primary cooling circuit for the target.
Because of the intense neutron flux that this water is exposed to, it will contain
activated material.  Primary among these will be tritium.  However, other isotopes, such
as $^{15}$O, $^7$Be, and several others will undoubtedly be present.  

Design of 
water circuits to operate in this environment are well understood, they are called RAW 
(RadioActive Water)
systems.  Good experience exists at Fermilab with such systems, and no doubt experienced engineers
at J-PARC and KEK have designed and operated such systems as well.  
Procedures for the design and maintenance of our RAW cooling circuit will be adapted from the other
laboratories with experience with such systems, these will render
highly unlikely, or in the worst case, contain any leaks that might occur.

Should a leak develop during normal operations, system instrumentation would detect a large leak
through water pressure drops.  Small leaks can be detected through air monitoring for trace levels of tritium or other products.  
The systems put in place to monitor air exiting the high-radiation areas would pick up this signal.
This will be discussed in the next section.
But in any event, the impermeable polyethylene lining of the caverns would contain any leaked
water, so there would be no penetration into the rock surfaces.  Cleanup would occur
using standard decontamination procedures.

Other activated pieces will come from maintenance activities on accelerator components, 
in particular areas inside the cyclotron that will need to be replaced.  Best examples are: the
extraction septum, the stripper foil assemblies, both internal and external.  Other pieces, such as 
the spiral inflector, internal cyclotron probes, and beam line instrumentation are among those 
that will undoubtedly require service.
(Beam lost in the central region of the cyclotron will be at too low an energy to cause activation, 
but the spiral inflector and other central region components will be directly exposed to neutrons
produced at the outer extremities of the cyclotron, so are likely to be activated.)

Handling of these activated pieces will also follow standard procedures of accelerator
laboratories:  localized shielding if levels are too high, limiting time for personnel to work
in these areas, and well-shielded lockers or containers to store and transport the
activated pieces for proper disposal.

Low levels of radioactivity are likely to be seen in vacuum pump oil.
When servicing
these pumps, special monitoring and handling procedures will be followed.

Personnel working in all of these areas will follow the usual procedures of clean suits, face masks,
protective eyewear, helmets, gloves, hair nets, and will always check out through
radiation detectors suitable for the isotopes likely to be encountered, including tritium.

\section{Exhaust Air Control -- Radiological Aspects}

A small chance exists that neutrons escaping from the shielding may undergo nuclear
reactions with surrounding air in the caverns.  
In addition, airborne contaminants might exist: from discharge of the vacuum pumps
that might collect material from the inside of the vacuum chamber of the cyclotron and
beam lines; or from evaporation of accidental leaks or discharges from the primary 
coolant circuit.  Tritium would be the most likely volatile product of such an event.

The primary mitigation measure will be to maintain the atmosphere in the caverns at
a slightly negative pressure.  This will ensure that air is always flowing into the enclosure, 
and the exhaust, channelled through blowers maintaining the negative pressure, 
can be properly monitored for radioactivity prior to exhaust outside the mine.

Isotopes normally found in high-energy, high-intensity accelerator environments 
are $^{11}$C, $^{13}$N, $^{15}$O and $^7$Be, the first three produced by (n,2n) reactions,
 the latter from spallation reactions of high-energy neutrons, all on air constituents.  
 These reactions all require high-energy neutrons,  which we make great effort to
 minimize through good shielding.  
 The first three all have short halflives, longest is $^{11}$C (20 minutes), so presence of these
 can be mitigated by storing air a short time before releasing it.
 $^7$Be is not volatile, it can be filtered, however, as discussed below, special care should be 
 taken to ensure it is not spread.
 Contaminants from vacuum pump exhausts are also not volatile, and can also
 be filtered.
 Tritium is a more severe problem, however the tolerance levels for release are 
 substantially higher than for other isotopes, and detection of any amount of
 tritium will immediately signal a water leak in the primary coolant circuit which
 will cause immediate shutdown of the system for repair.  This shutdown should
 occur prior to reaching the allowed limits for tritium release.
 
 The strategy followed in other high-power accelerator centers that have similar beam
 characteristics to ours (Catania, Legnaro, PSI), is to maintain their accelerator
 vaults at a negative pressure, monitor the air, run it through several stages of
 filtration, following which it is immediately released, assuming no activity is detected.
  We could follow the same procedures, with good filtration, but then as an added
 measure, conduct the air through a flexible pipe to the surface for discharge in
 an area that is suitably remote and removed from human occupancy.  The long distance
 of travel prior to release will provide further time for any short-lived products to decay.
 
 It is important, then, that the air handling system in the IsoDAR area does not allow
 any of the air from the IsoDAR area to reach other areas of the Yemilab complex.  As
 stated earlier, air circulation in an underground environment usually pipes fresh air into 
 the farthest reach of the complex, and allows air to flow back through the spaces to the outside,
 The air system for IsoDAR should be handled differently:  air should be actively pumped out
 of the IsoDAR caverns and piped directly to the outside.  Also, as there is only one ingress
 into the IsoDAR area, one can insure that there is always positive airflow INTO the IsoDAR area.
 This is the best way of ensuring no possible contamination will occur from any activation 
 products inside the IsoDAR area.
 
 Also as mentioned before, $^7$Be is a natural product of neutron spallation with nitrogen and oxygen 
 in the air. This isotope settles out on surfaces, and can be picked up on clothing, shoes, hair, hands.
 It will be very important to establish, from the start of operations, and on a continuing basis, how
 much of a hazard this isotope will present.  A thorough and regular monitoring program should
 be put in place to establish whether or not special mitigation and cleaning programs are needed.
 Until the level of hazard is established, personnel should be provided with proper gowns, masks, hair
 nets, booties, etc. to be put on prior to entry to the area, and should be removed and disposed upon
 leaving the area.  If it is determined that the hazard is not severe, a graded approach can be taken
 on PPE requirements.

\section{Radiological Aspects of Ground Water Management}

The very dry nature of the limestone environment of Yemilab points to this as
a problem which should be negligible.  However, in the event
any ground water becomes evident, measures should be in place for
ensuring it is properly handled.

The mitigation strategy would be to collect all this water, and monitor it for contamination prior to releasing it.
To accomplish this, the caverns will be carefully surveyed to determine the entry points of
water, and plot its course through each cavern.  Also, exit points should be mapped, 
and sealed to prevent uncontrolled escape of this water.
All of the caverns in the Yemilab area have  channels along their edges that will catch this water,
so it can be collected and moved to a containment vessel for analysis and proper disposal.

\section{Personnel Access Control}

\subsection{Procedures}

From the start of commissioning
of IsoDAR until the end of the experiment, 
areas around the cyclotron, beam transport and target must be declared
as ``Controlled Areas for Radiation Protection.''
Access to these areas will be strictly controlled and monitored.    

During operations, personnel will be excluded from these areas due to high radiation levels present.  Sources of radiation include 
x-rays from the high-voltage and cyclotron RF systems;
and neutrons and gamma rays from beam loss in the accelerator
and transport lines.  

During beam-off periods, radiation comes mainly from the decay of components activated by 
losses of high-energy protons. 
While these levels may be initially high, these will drop off quickly as the short-lived isotopes decay.  

Should personnel require access to the Controlled Areas, the following procedures should be followed.
A Radiation Safety Officer will survey of the areas of requested access,
establishing that radiation levels are sufficiently low to permit access.  
This officer will also cordon off areas of high activity for which there can be no access, as well as
establish limits on the work-period at any given location inside 
the Controlled Area.

Access will be carefully monitored.  All personnel working in controlled areas must have certified training and must carry radiation-monitoring devices (badges).  Those entering must be
properly logged in and out of the controlled areas.

\subsection{Controlled Areas}

When beam is off, there are several types of Controlled Areas,  distinguished by the 
level of radiation.  
The highest levels will likely be found
 inside the cyclotron shielding vault, and
the target vault.  The radiation in these areas will consist primarily of gammas. In particular, 
the cyclotron extraction region and the filters of the primary coolant circuit
in the target will require the highest level of access control.
The layout is designed such that 
other areas, such as the electrical substation and power supply areas around the
cyclotron and the beam transport line will have much lower radiation
levels.

\section{Monitoring Instrumentation and Interlock Systems}

Access will be controlled with lock-out, tag out procedures interlocked to to the beam producing sub-systems of the cyclotron.   This assures that entrance can only occur off-beam.

We will make use of well-established radiation-safety instrumentation.
There are many excellent models world-wide and at accelerator facilities in Korea for the systems that
need to be implemented at IsoDAR.

%% file: Ch6-conclusion_v1.tex
\chapter{Conclusion}

This Conceptual Design Report addresses the present plans for deploying the IsoDAR experiment at the site of the large (2.5 kiloton) liquid scintillator planned for the new Korean Yemilab facility, under the auspices of the Institute for Basic Sciences, Center for Underground Physics.  

In keeping with the nature of a Conceptual Design Report, the designs
and solutions that are presented here not necessarily final, but represent means of accomplishing the goal of 
installing and operating the Isotope Decay at Rest Experiment.   With that said,
we have identified all the major issues that must be faced in placing this type of experiment in a
low-background underground laboratory
and conclude that the task is feasible.    In particular, 
the onset of construction of the cyclotron hall, in accordance with this design report, allows IsoDAR to proceed without 
disrupting the body of low-background, high-sensitivity
experiments that are 
planned for deployment at Yemilab.

The process of this study, which incorporated input from US and IBS-CUP scientists, has established the strength of the IsoDAR$@$Yemilab collaboration.   The editors are grateful for all input from the parties involved.